\documentclass[prb,twocolumn]{revtex4}
\usepackage[dvips]{graphicx}
\usepackage{latexsym}
\begin{document}

\newcommand{\fig}[2]{\includegraphics[width=#1]{#2}}

\title{Thermal fluctuations in pinned elastic systems: field theory of
rare events and droplets}

\author{Leon Balents}
\affiliation{Department of  Physics, University of California,
Santa Barbara, CA 93106--4030}
\author{Pierre Le Doussal}
\affiliation{CNRS-Laboratoire de Physique Th{\'e}orique de
l'Ecole Normale Sup{\'e}rieure, 24 rue Lhomond 75231 Paris
France}

\date{\today}

\begin{abstract}
  
  Using the functional renormalization group (FRG) we study the
  thermal fluctuations of elastic objects (displacement field $u$,
  internal dimension $d$) pinned by a random potential at low
  temperature $T$, as prototypes for glasses. A challenge is how the
  field theory can describe both {\it typical} (minimum energy $T=0$)
  configurations, as well as thermal averages which, at any non-zero
  $T$ as in the phenomenological droplet picture, are dominated by
  {\it rare} degeneracies between low lying minima. We show that this
  occurs through an essentially non-perturbative {\it thermal boundary
    layer} (TBL) in the (running) effective action $\Gamma[u]$ at
  $T>0$ for which we find a consistent scaling ansatz to all orders.
  The TBL describes how temperature smoothes the singularities of the
  $T=0$ theory and contains the physics of rare thermal excitations
  (droplets). The formal structure of this TBL, which involves all
  cumulants of the coarse grained disorder, is first explored around
  $d=4$ using a one loop Wilson RG. Next, a more systematic Exact RG
  (ERG) method is employed, and first tested on $d=0$ models where
  where it can be pushed quite far.  There we obtain precise relations
  between TBL quantities and droplet probabilities (those are
  constrained by {\it exact identities} which are then checked against
  recent exact results).  Our analysis is then extended to higher $d$,
  where we illustrate how the TBL scaling remains consistent to all
  orders in the ERG and how droplet picture results can be retrieved.
  Since correlations are determined deep in the TBL (by derivatives of
  $\Gamma[u]$ at $u=0$), it remains to be understood (in any $d$) how
  they can be retrieved (as $u=0^+$ limits in the non-analytic $T=0$
  effective action), i.e. how to recover a $T=0$ critical theory.
  This formidable ``matching problem'' is solved in detail for $d=0$,
  $N=1$ by studying the (partial) TBL structure of higher cumulants
  when points are brought together. We thereby obtain the
  $\beta$-function at $T=0$, {\it all ambiguities removed}, displayed
  here up to four loops. A discussion of the $d>4$ case and an exact
  solution at large $d$ are also provided.
\end{abstract}

\maketitle

\section{Introduction}

Complex and disordered materials are often dominated, in their static
properties, by rare but large fluctuations, and in their dynamics by
ultra-slow relaxations. Such behavior occurs in a variety of systems
involving many interacting degrees of freedom, from supercooled
liquids to spin glasses \cite{reviewsg}. Understanding the nature of
the low lying excitations in these highly complex systems and even
more so, obtaining an analytic theory of the low temperature behaviour
in realistic dimensions, is a formidable task.  A class of problems
which shares some of the basic physics and seems more amenable to
analytic approaches are elastic manifolds pinned by quenched disorder.
They are of interest for numerous experimental systems, such as
interfaces in magnets \cite{reviewrf,creepexp}, contact lines of
fluids wetting a rough substrate \cite{MoulinetGuthmannRolley2002},
charge density waves \cite{gruner_revue_cdw,quantum_creep} (CDW),
Coulomb glasses \cite{coulomb_glass} and Wigner crystals
\cite{wigner,chitra} in heterojunctions and on the surface of helium
\cite{helium}, or vortex lattices in superconductors
\cite{blatter,bglass,natter}. It seems likely that a broader class of
problems, including random field systems, can be understood by similar
methods.

In this paper we will thus focus on the low temperature equilibrium
behaviour of elastic manifolds pinned by a random potential. Such
systems are usually described by a displacement field $u({\bf r})$ which
measures deformations away from a reference ordered state in the absence
of any thermal, disorder induced, or quantum fluctuations. In their
classical version they are typically described by a model Hamiltonian
\begin{equation}
  H[u] = \int \! d^d{\bf r} \left[ {1 \over 2}|\nabla u|^2 +
    W(u({\bf r}),{\bf r}) \right], \label{eq:ham}
\end{equation}
where the random potential $W(u,r)$ can be chosen, for concreteness,
Gaussian with zero-mean and variance $\overline{W(u,r) W(u',r')} =
R(u-u') \delta^{(d)}(r-r')$.  We indicate here and in the following a
disorder (ensemble) average by an overline.  At equilibrium and low
temperature the elastic manifold is usually pinned by disorder into non
trivial rough configurations.  In these so called glass phases,
deformations scale with size $L$ as $u \sim L^\zeta$ with roughness
exponent $\zeta$, and are dominated by the competition between elastic
and disorder energies.  Different physical cases are described by
varying longitudinal (i.e.  internal) dimensionality ($d$), transverse
(i.e. embedding) dimensionality ($N=$ the number of vector components of
$u$ -- we focus on $N=1$ for simplicity here), and $R(u)$. For instance
$R(u)$ is a periodic function in the case of CDW or lattices, while in
the case of interfaces in a ferromagnet it is short range for random
bond disorder, and long range ($R(u) \sim |u|$) for random field
disorder.

As in other energy dominated glass phases in related systems,
in the pinned elastic medium temperature is formally irrelevant
in the RG sense as the ratio of typical temperature to
energy scales flows to zero
as $T_L \sim L^{- \theta}$.
Here, the exponent $\theta=d-2 + 2 \zeta$ ($>0$) as a
consequence of the so called statistical tilt symmetry
\cite{SchulzVillainBrezinOrland1988,HwaFisher1994b}
(STS), which
expresses that the probability
distribution of the disorder is translationally invariant. However,
because of the multiplicity of low energy configurations,
thermal effects are rather subtle and do in fact dominate
the behaviour of some observables at small but non zero temperature.
The basic mechanism is that even if local quasidegeneracies of ground
states may occur only rarely, they can induce large deformations.

A minimal physical description of such thermal
excitations in disordered glasses is provided by
the phenomenological droplet picture \cite{droplets0,droplets}. In its simplest form, it
supposes the existence, at each length scale $L$, of a small number of
excitations of size $\delta u \sim L^\zeta$ above a ground state,
drawn from an energy distribution of width $\delta E \sim L^\theta$
with constant weight near $\delta E=0$. This means that
with probability $\sim T/L^\theta$ this excitation can be
thermally active with $\delta E \sim T$. The consequence is that
static thermal fluctuations at a given scale are dominated by such rare
samples/regions with two nearly degenerate minima.
For instance, the
$(2n)^{\rm th}$ moment of $u$ fluctuations is expected to behave as
\begin{equation}
  \overline{(\langle u^2\rangle - \langle u\rangle^2)^n} \sim
  c_n (T/L^\theta) L^{2n\zeta}. \label{eq:twowell}
\end{equation}
as a straightforward consequence (angular brackets denote the thermal
average with respect to the canonical distribution). This simple picture
seems to describe models such as Eq.~(\ref{eq:ham}) relatively well, at
least in low dimensions,\cite{dropevid,dpfisher,mezard}. The statistics
of low energy excitations has been much studied numerically recently in
other systems such as low dimensional spin glasses \cite{dropevid2}. The
droplet picture has been refined in many ways to account for more
complex systems, and it is still controversial if this scenario fully
describes systems such as spin glasses.  Issues such as the fractal
dimension $d_f$ of these droplet excitations (``fat'' droplets) and the
possible global multiplicity of true ground states are still debated.
In any case, it is clear that the basic mechanism of stochastic local
ground state quasi-degeneracies must be part of any realistic analytical
theory of such systems.

Similar physics occurs in the dynamics of these systems.
Although
we will not address it in details here, it is useful to study, in parallel
to the statics, equilibrium dynamics.\cite{us_dyn_long} The Langevin equation of
motion is
\begin{eqnarray}
  \eta \partial_t u_{rt} = \nabla^2_r u_{rt} + F(u_{rt},r) +
  \zeta(r,t), \label{eom}
\end{eqnarray}
with friction $\eta$, thermal noise $\zeta(r,t)$ with $< \zeta(r,t)
\zeta(r',t')  \rangle  = 2 \eta T \delta^d(r-r') \delta(t-t')$, and random force
$F(u,r) = - \partial V(u,r)/\partial u$ (noting $u(r,t)\equiv u_{rt}$)
with correlator denoted $\Delta(u) = - R''(u)$. Equilibrium dynamics at
non-zero temperature $T>0$ provides an equivalent way, through the
fluctuation dissipation relations, to study the statics.  Extension of
the droplet picture to the dynamics supposes that equilibrium dynamics
in the glass phase is dominated by thermal activation between the
thermally active low lying quasidegenerate minima, with typical barriers
scaling as $U_b \sim L^\psi$. Little is known about the distribution of
these barriers, but there is some evidence
\cite{drossel_barrier}
that $\psi \approx \theta$.

Despite the wide applicability and extensive theoretical studies of the
elastic model, Eqs.~(\ref{eq:ham}-\ref{eom}), there are few analytical
results for the ground state properties, and even less for thermal
excitations. There are exact solutions in the mean-field
($N\rightarrow\infty$) limit \cite{mezpar,leto1,leto2,frgN} and for
fully connected models \cite{infinited1,infinited2} (i.e. the
$d\rightarrow\infty$ limit).  These limits are interesting and probably
capture part of the physics but they also have peculiarities which are
still not fully understood. In both, thermal activation over divergent
barriers ($U_L \sim L^\theta$) is not included, and the infinite $N$
limit (or Cayley tree \cite{dpct,derridarsb}) leads to global state degeneracies
(with $\delta E \sim T$) -- replica symmetry breaking (RSB) \cite{beyond}
-- which may, or may not, be lifted at finite $N$. For finite $N$ or $d$
exact results exist only in two cases. There are some exact results for
the roughness and the free energy distributions of the $1+1$ dimensional
directed polymer problem ($d=1$ $N=1$ case)
\cite{henley,kardar,imbrie,derrida1d,johansson} and the related
Burgers turbulence, and KPZ growth or exclusion problems
\cite{kpz,spohn,exclusion}. The $d=0$ limit of a particle in a random
potential is simpler but also interesting as in some cases the full
structure of the zero temperature fixed point can be solved exactly and
many results obtained. This is the case for the $N=1$ Sinai or so called
toy model (i.e. RF) case ($\theta=2/3$): there the droplet picture is
``exact'' and the droplet probabilities are exactly known
\cite{toy,cecile}. Other interesting results have been obtained for any
$N$ in the marginal case ($\theta=0$) of a log correlated $R(u)$, where
RSB seems to occur \cite{carpentier}.

A potentially promising method is the field-theoretic Functional
Renormalization Group (FRG). It was originally developed at $T=0$ to
lowest order in an expansion in $\epsilon=4-d \ll 1$ (one loop) for the
statics \cite{fisher_functional_rg} and later for depinning
\cite{frg_dep} (around the upper critical dimension $d_{uc}=4$). It is
called ``functional'' since one must follow the RG flow of the full
correlator $R(u)$ which becomes marginal in $d=4$, as $u$ becomes
dimensionless.  It is shown that $R(u)$ becomes non-analytic beyond the
Larkin length as the random force correlator $\Delta(u) = -R''(u)$
develops a cusp singularity at $u=0$. Physically this cusp is associated
to the existence of many metastable states and its meaning has been
discussed in \cite{fisher_functional_rg,balents_rsb_frg,frgN}.  At the
depinning transition it yields a non-zero threshold force as $f_c \sim
|\Delta'(0^+)|$.  Most importantly, the non-analyticity of the effective
action seems to allow to evade the so called dimensional reduction phenomenon
which renders perturbation theory in an analytic $R(u)$ trivial and
yields the incorrect result $\zeta=\epsilon/2$.  Instead, in the space
of non analytic functions, well behaved fixed points do exist, yielding
reasonable values for the exponents ($\zeta=\epsilon/3$ for random field
disorder, $\zeta=0.283 \epsilon$ for random bonds,
\cite{fisher_functional_rg} and a logarithmic growth of displacement for
CDW and Bragg glass \cite{bglass}).

An outstanding but crucial question is the inherent consistency of this
FRG method given this non-analyticity, a highly anomalous occurence in
conventional field theory.  In particular, the occurence of
non-analyticity in the effective action at {\sl finite scale} usually
does not occur, because fluctuations average out any singularities
associated with multiple local minima.  Here, the {\sl zero temperature}
field theory contains only disorder (sample to sample) and not thermal
fluctuations, and these considerations do not apply. Instead
it is tempting to compare the integration over fast modes to some (iterative) minimization
procedure, which can then produce non-analytic energy landscape for
fast modes (''shocks''). This picture can be made precise in simple
single mode $d=0$ ''toy RG'' models
\cite{balents_rsb_frg}. The emergence of such shocks
is well known in the equivalent Burgers-KPZ system in $d=1$.
In any case, the
usual justification of perturbative renormalization group calculations
(i.e. $\epsilon$ expansions) must be re-examined. We will comment
further on the physics associated with this structure below.

A sign of the difficulties inherent in justifying the approach occurs
already in attempts to extend this FRG method beyond the one loop
approximation, and also at non-zero temperature. Formidable difficulties
appear already at two loop order for the beta function at $T=0$
(and even in the one-loop corrections to the two point correlation
function). Because
the effective action is non-analytic, the renormalized perturbation
theory is {\it ambiguous} \cite{frg2loop}. An exact solution of the FRG
at large $N$ to any number of loops shows that special care has to be
given to how any specific correlation is precisely defined, i.e. how the
system is prepared \cite{frgN}. For $N=1$ quasistatic depinning the
ambiguity can be fully lifted, as at any steady state velocity $v>0$,
the elastic manifold always moves forward.  The resulting field theory
is found renormalizable to two loop \cite{frgdep2loop}, and the
predictions fully supported by numerical simulations. A solution of the
ambiguity for the statics, based on a requested (and hoped for)
renormalizability property, was proposed in \cite{frg2loop}. An
alternative route, technically more complex but more transparent
physically, which we will follow here, is to study the system at
non-zero temperature $T>0$ where the action remains analytic and no such
ambiguities appear.

Extensions of the FRG to non-zero temperature have been studied
previously to one loop \cite{balents_loc,chauve_creep_epl,chauve_creep,ergchauve}.  Since
the renormalized temperature $T_L \sim L^{-\theta}$ tends to zero, one
expects that the asymptotic fixed point is still the $T=0$ one. This
convergence was found to be highly non-uniform.  $ $From the FRG flow
equation {\it truncated to follow the second cumulant alone ignoring
  feedback from higher cumulants}, it was found that $\Delta(u) = -
R''(u)$ indeed remains analytic and differs from the $T=0$ fixed point
function only in a {\it thermal boundary layer} $u \sim T_L$ where the
cusp singularity is rounded.  A highly desirable consequence of this
early one loop analysis \cite{balents_loc,chauve_creep_epl,chauve_creep} was that the
correct scaling for creep barriers $U_b \sim L^\theta$ was recovered as
a consequence of the boundary layer, from the divergence of $- R''''(0)
\sim 1/T_L$, the quantity which enters the friction (ie the time scale)
renormalization.  This very same divergence however, is also very
troublesome in considering any extension to higher number of loops (as
was found in \cite{ergchauve}) since one easily realizes that adding one
loop extracts one power of $1/T_L$, thus adding one new divergence.
Until now the status of the boundary layer beyond one loop thus seemed
rather unhealthy.

The occurrence of these technical issues is not surprising when one
reflects upon the non-trivial physics that the field theory must
describe.  On the one hand, typical samples and their ground states --
which control static correlations at $T=0$ -- are critical in a fairly
conventional sense.  For instance, they are described by scaling
exponents ($\zeta,\theta,\ldots$) and amplitudes, which apparently
(hopefully!)  can be computed systematically in an $\epsilon$ expansion.
On the other hand, thermal fluctuations of the manifold are dominated by
rare low-lying metastable configurations, which play no role in the
$T=0$ theory.  The distribution (over disorder) of the thermally
fluctuating part of the correlations (e.g. $\langle (u-\langle
u\rangle)^2\rangle$) is thus extremely broad, and must be encoded
somehow also in the field theory at $T>0$.  Both types of quantities are
formally defined extremely similarly as derivatives of the effective
action (e.g of the renormalized $R(u)$) at $u=0$.  Nevertheless, they
scale very differently and represent rather different physics.  The
coexistence of these two features strongly argues for an unconventional
structure of the field theory.

In this paper we solve these issues in $d=0$ and propose a consistent
form for the effective action of the field theory in any dimension to
any number of loops. We establish the clear connection between the
formal thermal boundary layer and the physical droplet picture in any $d$ (see
\cite{ergchauve} for an early calculation). Thus we obtain a ``field
theory of droplets'', i.e. we show that the present extension of the
field theoretic FRG captures correctly the basic feature of glasses,
rare events due to quasi-ground-state degeneracies.  The mechanism is
quite non-trivial. A short account of these results has been given in
\cite{us_short}, together with a discussion of the thermal boundary
layer structure which also arises in the dynamics and yields there to
fluctuations of barriers, as discussed in detail in \cite{us_dyn_long}.

First, in Section~\ref{sec:bound-layer-with}, within a Wilson
formulation, we derive and carefully analyze the full one loop FRG
equations. We discover that the thermal boundary layer extends to {\it
  all cumulants} of the disorder. Thus the truncation to the second
cumulant performed in previous studies is not strictly consistent at
$T>0$. It does however give the correct overall scaling, since we find
that a consistent ansatz can be found for all cumulants. This results in
a hierarchy of FRG equations for the scaling forms of each cumulant or
the random potential, with good properties.  Unfortunately, there is no
longer a small parameter ($\epsilon$ has disappeared as a small parameter!)
which prevents a perturbative solution. We must thus {\it assume} that there
exists a global solution, which we call the {\it Thermal Boundary Layer
  Ansatz} (TBLA).

To further explore this novel theory we turn to the Exact RG method
(ERG), explained in Section~\ref{sec:exact-rg-general}.  The ERG follows
the full (non local) effective action {\it functional} under coarse
graining.  To better understand the physics, we start in
Section~\ref{sec:zero-dimension:-erg} with a detailed analysis of the
$d=0$ case, where the effective action becomes a (replica) function
rather than a functional, and can be studied without any truncation.
There, the droplet picture is exact, and a variety of exact results are
available for comparison with the methods developed here. From the TBLA
and the ERG, we derive exact relations for droplet probabilities.  These
are tested on the exact solution of the toy model, and are found to pass
this non-trivial check.  In Section~\ref{sec:matching}, we introduce and
discuss the problem of {\sl matching} the zero temperature and boundary
layer regimes.  In particular, there are a variety of correlation
functions describing the critical (typical) correlations of the ground
state, and as such should be determined physically by the $T=0$
non-analytic field theory.  Formally, however, they are obtained as
indicated above from derivatives of the effective action at $u=0$, deep
within the TBL. Thus, some information should be transmitted
through the TBL. We show how these correlations are indeed determined
from particular limits of the (non-analytic) zero temperature ``outer''
solution.  This is accomplished through a careful consideration of the
{\sl Partial Boundary Layer} (PBL) structure that obtains when any two
relative replica displacements ($u_a-u_b$) become of order $T_L$. Thus
we will describe the {\it full solution of the ambiguity problem} (for $d=0$
and $N=1$) in the zero temperature beta function.  As an example of the power of
this prescription, we calculate the beta function and various
correlations as a series in $R$ up to {\it four loop order}.

With this physical understanding of the FRG equations in the thermal
boundary layer regime in hand, we then turn to $d>0$ in
Section~\ref{sec:conclusion}.  We show that a fully consistent boundary
layer ansatz can be found for the whole non-local effective action, and
is self-consistent when applied to the ERG equations. This result
provides a non-trivial ''solution'' to the above-mentioned disease of
proliferation of $1/T_L$ divergences by resumming them order by order
into a finite result.  We show explicitly that the results from the
droplet picture are recovered.  Perturbative control (i.e the $\epsilon$
expansion) is na\"ively expected in the outer non-analytic solution.
The necessary extension of the matching understood in $d=0$ to obtain
physical quantities (even zero temperature ones) remains a challenge.
We discuss this issue and some perspectives on future work at the end of
this section. Whether the method could account for other scenarios,
more complex than the initial droplet picture, is also addressed.

A panoply of appendices give various technical details and a
description of related models. In particular the case $d>4$ is
examined in details: it is shown, through analysis of the one loop
FRG, exact solution at large $d$ and qualitative arguments, that a
phase with non-analytic effective action exists at $T=0$ and
strong enough disorder, but that the TBL remains of finite width,
at variance with $d<4$.

\section{Boundary Layer within the Wilson FRG }
\label{sec:bound-layer-with}

In this Section we establish the Wilson one loop FRG equations,
which are a convenient starting point
for our investigation of the thermal boundary layer. We include from
the start all cumulants of the random potential since we do find,
a posteriori, that they must all be considered.

\subsection{model, properties and notations}

$ $From now on we consider the equilibrium statics at $T>0$, defined through the
equilibrium Hamiltonian in (\ref{eq:ham}).  Formally, all static
quantities of interest can be obtained from the replicated generating
(partition) function
\begin{equation}
  Z[h_a] = \overline{\prod_{a=1}^n \int Du_a \exp \left[ - \frac{H[u_a]}{T} +
    \int_r \! h_a(r) u_a(r) \right]}.
\end{equation}
The number of replicas $n$ can be taken to zero, or kept
arbitrary as a bookeeping device. Since some of
our calculations are independent of $n$, we will explicitly
indicate when the limit $n=0$ is considered.

The connected cumulants of the random potential $W(u,r)$
in  Eq.~\ref{eq:ham} are denoted ($k \ge 2$):
\begin{eqnarray}
&& \overline{W(u_1,r_1) .. W(u_k,r_k)}^c \nonumber \\
&& = (-)^k S^{(k)}(u_1,\cdots,u_k) \delta^d(r_1,\cdots,r_k)
\label{conncumdis}
\end{eqnarray}
with $S^{(2)}(u,u') \equiv R(u-u')$. The cumulants higher than second
are generated by coarse graining, and are thus included here from the
start. We ignore for the moment correlations between different points in
internal space: this choice and its extensions are discussed below. The
probability distribution of the random potential is chosen
translationally invariant, so the cumulants satisfy
\begin{eqnarray}
  S^{(k)}(u_1+\lambda,\cdots,u_k+\lambda) = S^{(k)}(u_1,\cdots,u_k).
\label{translationalinv}
\end{eqnarray}
They are also even so that $S^{(k)}(- u_1,\cdots, - u_k) = S^{(k)}(u_1,\cdots,u_k)$,
the $S^{(k)}$ being by construction fully symmetric functions of
their arguments.

Performing the disorder average one obtains:
\begin{eqnarray}
&& Z[\vec{h}] = \int \!\!\!D\vec{u} e^{ - {\cal S}[u] +
    \int_r  \vec{h} \vec u }, \\
&& {\cal S}[u] = \int d^d r ( \sum_a {1
        \over {2T}} |\nabla u^a|^2 - V[\vec{u}] ) , \label{eq:staticPF}
\end{eqnarray}
where we denote vectors in replica space as $\vec{u}=(u_1,\cdots,u_n)$, and we define
the characteristic function of the pinning disorder
\begin{eqnarray}
  V[\vec{u}] = \sum_{k=2}^\infty {1 \over T^k k!} \sum_{a_1
\cdots a_k}
    S^{(k)}[u_{a_1},\cdots,u_{a_k}], \label{eq:Vexpand}
\end{eqnarray}
which is a fully symmetric, translationally-invariant function of the
replica vector $\vec{u}$. The property (\ref{translationalinv}) results
in STS, e.g.  in Fourier space $\ln Z[\vec{h}=h] = \ln
Z[\vec{h}=0] - T n h(q) h(-q)/(2 q^2)$ resulting in infinite sets of
relations between correlation functions and other important consequences
\cite{SchulzVillainBrezinOrland1988,HwaFisher1994b}
, to which we will return later. One consequence is that the
``one replica'' part of the action in (\ref{eq:staticPF}) is not
corrected by disorder under coarse graining (which is also why one can
set the elastic coefficient to unity).

Let us give some useful notations and properties. For
notational simplicity we will often write:
\begin{eqnarray}
&& S^{(k)}(u_{12\ldots k}) \equiv S^{(k)}(u_{1},\cdots,u_{k})
\end{eqnarray}
for any function, and for a function of one argument $R(u_{12}) \equiv
R(u_{1}-u_{2})$.  Since any function antisymmetric in any two replicas
gives zero contribution to the action when summed over replicas, such
functions may be added freely to any cumulant without physical
consequences.  This freedom will be used to simplify some expressions.
Usually the cumulants $S^{(k)}$ will be taken fully symmetric in their
arguments.  It is then useful to define the symmetrization with
respect to a given set of arguments:
\begin{eqnarray}
&& {\rm sym}_{1\cdots k} \phi(u_{1\ldots k}) = \frac{1}{k!} \sum_{\pi \in P_k}
\phi(u_{\pi(1)\cdots\pi(k)})
\end{eqnarray}
$P_k$ being the set of permutations of $k$ objects. Next we
note two useful properties of the above decomposition of the
action in cumulants, i.e. in number of replica sums (\ref{eq:Vexpand}).
Translational invariance (\ref{translationalinv}) implies relations between derivatives,
e.g.:
\begin{eqnarray}
&&  (\sum_{a=1}^k \partial_a) ~ S^{(k)}[u_{1},\cdots,u_{k}] = 0.
\label{prop1}
\end{eqnarray}
This implies properties such as:
\begin{eqnarray}
&& {\rm sym} S^{(k)}_{10..0}[u_{1\cdots k}] = 0, \nonumber \\
&& {\rm sym} S^{(k)}_{20..0}[u_{1\cdots k}] = - (k-1) {\rm sym}
S^{(k)}_{11..0}[u_{1\cdots k}], \label{STSrels}
\end{eqnarray}
where here ${\rm sym}={\rm sym}_{1\cdots k}$, and we denote everywhere
partial derivatives $S^{(k)}_{n_1 n_2..n_k}$ for $n_i$ derivatives in
the $i$-th argument.

Finally, there is a ``gauge invariance'' property: the change
\begin{eqnarray}
&& S^{(k)}[u_{1\cdots k}] \to S^{(k)}[u_{1\cdots k}] + {\rm
  sym}_{1\cdots k}
\phi(u_{1\cdots k-1})
\end{eqnarray}
yields an extra term in (\ref{eq:staticPF}) proportional to an explicit
$n$. Thus, if one considers $n=0$, one can add freely to each $k$
replica term any function of $p<k$ replicas.  Physically, this gauge
freedom corresponds to the fact that the random potential may be shifted
by a $u$-independent random function of $x$, changing only the free
energy in a trivial way but leaving all displacement correlation
functions unchanged.  Explicitly, this shift is of the form
\begin{eqnarray}
  \label{eq:gaugemeaning}
&&   W(u,x) \rightarrow W(u,x) + w(x),
\end{eqnarray}
where $w(x)$ can have arbitrary cross-correlations with $W$, i.e.
\begin{eqnarray}
  \label{eq:constcross}
&&   \overline{W(u_1,x_1)\cdots W(u_k,x_k) w(x_{k+1})\cdots w(x_p)}
\nonumber \\ && =
  \delta_{x_1\cdots x_p} \Phi^{(p,k)}(u_1,\cdots,u_k),
\end{eqnarray}
with arbitrary $\Phi^{(p,k)}$.

\subsection{conventional scaling and one loop FRG equations}

Let us first summarize the conventional view on the behaviour
of the above action (\ref{eq:staticPF}) under coarse graining
and RG as it has emerged from previous works
\cite{fisher_functional_rg,balents_loc,chauve_creep,ergchauve}. In the usual
zero-temperature power counting, namely $T\rightarrow
b^\theta T$, $x\rightarrow b x$, and $u\rightarrow b^\zeta u$, all the
cumulants $S^{(k)}$ for $k>2$ are irrelevant, and the second
cumulant $R(u-u')$ is marginally relevant just below $d=4$. Similarly,
terms describing correlations between different internal space points are
strongly irrelevant, which justifies to restrict to (\ref{conncumdis})
with $k=2$ (in the field theory language the only needed counterterms
are for the local part of the second cumulant of the disorder).

More precisely, within the Wilson approach with a running
UV momentum cut-off $\Lambda_l = \Lambda e^{-l}$, one makes the
above action dimensionless ${\cal S}[u] = A_d \tilde{{\cal S}}[u \Lambda_l^\zeta] $ as:
\begin{equation}
\tilde{\cal S}[u] =  \int d^d \tilde{r} ( \sum_a {1
        \over {2 \tilde{T}_l}} |\nabla_{\tilde{r}} u^a|^2 - \tilde V[\vec{u}] ) ,
\label{eq:staticPF2}
\end{equation}
with $\tilde{r} = r \Lambda_l$.
We have introduced
the ratio of thermal energy $T$ to pinning energy $L^\theta$ which defines the running
temperature:
\begin{eqnarray}
&& \tilde{T}_l = A_d T \Lambda_l^\theta, \\
&& \theta=d-2+2\zeta > 0.
\end{eqnarray}
The rescaled temperature $\tilde{T}_l$ flows to zero,
indicating the dominance of energy over entropy as
appropriate for a pinned glass phase. For later convenience we
also introduced $A_d = S_d/(2 \pi)^d = 1/(2^{d-1} \pi^{d/2} \Gamma(d/2))$.
The rescaled generating function reads:
\begin{eqnarray}
&& V[\vec{u}] = A_d \Lambda_l^d \tilde V[\vec{u} \Lambda_l^\zeta], \label{eq:Vexpand0} \\
&& \tilde V[\vec{u}] = \sum_{k=2}^\infty {1 \over \tilde{T}_l^k k!} \sum_{a_1
\cdots a_k}
    \tilde S^{(k)}[u_{a_1},\cdots,u_{a_k}], \label{eq:Vexpand2}
\end{eqnarray}
and corresponds to the rescaled
dimensionless local cumulants:
\begin{eqnarray}
&& R(u) = A_d^{-1} \Lambda_l^{4 - d - 4 \zeta} \tilde{R}(u \Lambda_l^\zeta), \label{resc} \\
&& S^{(k)}[u_{a_1},\cdots,u_{a_k}] = A_d^{1-k} \Lambda_l^{d - k \theta}
\tilde S^{(k)}[u_{a_1} \Lambda_l^\zeta ,\cdots,u_{a_k} \Lambda_l^\zeta].
\nonumber
\end{eqnarray}
These equations embody the conventional scaling. Within the Wilson formulation
one finds that for $\epsilon=4-d \ll 1$ there are well defined
fixed point functions $\tilde{R}^*(u) \sim
\epsilon \ll 1$ associated to $\zeta$ also of $O(\epsilon)$.
One also expects
that for $k>2$ the fixed point value of the functions $\tilde{S}^{(k)*}(\vec{u}) \ll \tilde{R}^*(u)$
(more specifically $\tilde{S}^{(k)*} \sim \epsilon^k \ll  \epsilon$ for
$k>2$). This smallness has been argued to
justify the truncation $S^{(k)}\approx 0$ for $k>2$ at one-loop, to
leading order in $\epsilon$ at $T=0$. Prior investigations of thermal
effects have presumed this truncation
remains valid at $T>0$ \cite{balents_loc,chauve_creep,ergchauve} also to obtain the
leading order in $\epsilon$.  We now check this
assumption.

To capture this leading-order behavior, we therefore employ a
one loop FRG analysis.  A general one loop FRG scheme contains in fact
much more than the minimal physics needed -- in principle -- for a consistent
$O(\epsilon)$ analysis by the usual power-counting.  However, we will
pursue it in the general form in order to have a single calculation
which encompasses both this simple regime and the thermal boundary
layer. We emphasizes that at this stage it has mainly
a heuristic value, since, as we find
later, the restriction to one loop is not necessarily
justified within the boundary layer. Similarly, the consideration
of non-local terms in internal space for the disorder cumulants
will also, in the end, become necessary. We ignore for now all of these
further complications, which
will be captured by the more rigorous (but heavier) exact RG scheme
studied below. The present much simpler one loop Wilson RG will give us
a first good idea of what is happening in this theory.

Expanding the action (\ref{eq:staticPF}) to
quadratic order in the ``fast'' modes with Fourier component
non-zero only in the shell $\Lambda_l < q < \Lambda_{l+dl}$ and performing the resulting
Gaussian functional integral one obtains the differential flow equation for
the function $V(\vec{u})$ as:
\begin{eqnarray}
&& \partial_l V(\vec{u}) = - {1 \over 2} A_d \Lambda_l^d {\rm Tr}
  \ln \left[ \delta^{ab} - {T \over \Lambda_l^2} \partial_a \partial_b
    V(\vec{u}) \right] \label{eq:Wilson1}  \\
&& = A_d \Lambda_l^d \sum_{p \ge 1} c_p \left[{T \over \Lambda_l^2}\right]^p Tr((V'')^p)
\label{eq:Wilsoncp}
\end{eqnarray}
where the $Tr$ denotes a trace in replica space only.
For convenience we have introduced the parameters $c_p=1/(2 p)$ since an
almost identical equation will be derived below for the
exact RG in $d=0$ (with $c_p=1$).

We can now use the expansion in (\ref{eq:Vexpand}) and collect terms
with a given number of replica sums on either side of
(\ref{eq:Wilson1}). The general form of this equation is derived in
Appendix~\ref{sec:gener-struct-frg}. We give here only the result for
the projection of the FRG equation (\ref{eq:Wilson1}) onto two and
three replica components:
\begin{widetext}
\begin{eqnarray}
  && \partial_l \tilde{R}(u) =
  (\epsilon-4\zeta+\zeta u \partial_u)\tilde{R}(u) +
  \tilde {T}_l \tilde{R}''(u) + \tilde{S}^{(3)}_{110}(0,0,u) +
  \frac{1}{2} \tilde{R}''(u)^2 - \tilde{R}''(0) \tilde{R}''(u)
  \label{Wilson2cum}
  \\
  && \partial_l \tilde{S}^{(3)}(u_{123}) = (2
  \epsilon-2-6\zeta+\zeta u_i\partial_{u_i})
  \tilde{S}^{(3)}(u_{123}) +
   \,{\rm
    sym}\left(\frac{3\tilde{T} _l}{2}\tilde{S}^{(3)}_{200}(u_{123})
    + \frac{3\tilde{T} _l}{2}{\sf R}''(u_{13})
    {\sf R}''(u_{23}) \right)     \nonumber \\
  && + 3 \, {\rm sym}\left( {\sf R}''(u_{12}) (
    \tilde{S}^{(3)}_{110}(u_{113}) -
    \tilde{S}^{(3)}_{110}(u_{123}))
    +\gamma {\sf R}''(u_{12}){\sf R}''(u_{13})^2  -
    \frac{\gamma}{3} {\sf R}''(u_{12}) {\sf R}''(u_{23})
  {\sf R}''(u_{31}) \right), \nonumber \\
&& + \frac{3}{2} {\rm sym} \tilde{S}^{(4)}_{1100}(u_{1123})  \label{Wilson3cum}
\end{eqnarray}
\end{widetext}
\noindent where here ${\rm sym}={\rm sym}_{123}$, and $\gamma=1$. Since $\tilde{R}''(0)$
does not feed into higher cumulants (for $n=0$) we define:
\begin{eqnarray}
&& {\sf R}(u) = \tilde{R}(u) - \tilde{R}''(0) u^2/2
\end{eqnarray}
Note that since additive ($u$- independent) constants in (\ref{Wilson2cum})
are pure gauge (for $n=0$ in the sense discussed above) we will
in general ignore them below. Equations,
with similar structure exist for higher cumulants (see Appendix~\ref{sec:gener-struct-frg}
for details).

\subsection{analysis within conventional scaling}

One first notes that (\ref{Wilson2cum}, \ref{Wilson3cum}) does not
close, since there is a feedback of cumulant $k+1$ in cumulant $k$,
though only in a single term.  At zero temperature, in the conventional
scaling discussed above this should not be a problem. Let us first
discuss that case and formally set $\tilde{T}_l=0$ in
(\ref{Wilson2cum},\ref{Wilson3cum}).  We will call the solutions of the
resulting equation the ``outer'' solution.  This terminology is
motivated by the thermal boundary layer analysis which is performed
below.

\subsubsection{zero temperature}

For $T=0$ the solution $\tilde{R}_l(u)$ of (\ref{Wilson2cum})
(neglecting the third cumulant term $S^{(3)}$)
converges to a zero temperature fixed point function \cite{fisher_functional_rg} $\tilde{R}^*(u)$,
with a small set of universality
classes \cite{bglass,balents_loc}. For, e.g. $W(u,x)$ periodic on
the interval $0<u<1$, one finds $\zeta=0$ and the force-force correlator
$\tilde \Delta^*(u)=- \tilde{R}^{* \prime \prime}(u)={\rm Min}_{n\in {\cal Z}}
{\epsilon\over 6}[(u-n-1/2)^2-1/12]$.  Its {\sl non-analytic} behavior
at small $u$,
\begin{eqnarray}
&& \tilde \Delta^*(u) = \tilde \Delta^*(0) - \chi |u| + O(u^2),
\label{cusp}
\end{eqnarray}
is, however, {\sl super-universal}, i.e.  the same for all disordered
elastic models.  We denote $\chi\equiv|\tilde\Delta'(0^+)|= \epsilon
\tilde{\chi}$, where $\tilde{\chi}$ is a $O(1)$ constant whose numerical
value depends on the model.  The physical significance of this cusp has
been discussed by several authors
\cite{fisher_functional_rg,balents_rsb_frg,frgN}. It is related to the
existence of multiple metastable minima in the effective potential,
since these are implied by the divergence of the mean squared curvature of the
pinning potential, $-\tilde\Delta''_{T=0}(0)=\overline{(\partial_u^2
  V(u))^2}=+\infty$.  We will see further consequences below. Note that
the second derivative of (\ref{Wilson2cum}) at $u=0$ yields the fixed
point constraint
\begin{eqnarray}
&& - \tilde{R}^{* \prime \prime}(0) = \tilde \Delta^*(0) = \chi^2/(\epsilon - 2 \zeta).
\label{secondder}
\end{eqnarray}
Note that this will be the precise definition of the parameter $\chi$ in the remainder
of this paper. The relation between $\chi$ and the first derivative
given above then holds only within the one-loop Wilson approximation, and is
corrected at higher order.

The neglect of $\tilde{S}^{(3)}$ then appears consistent.
Given that the second cumulant
flows to a fixed point ${\sf R}^* \sim \epsilon$ then (\ref{Wilson3cum})
shows that the third cumulant also flows to the fixed point:
\begin{eqnarray}
&& \tilde{S}^{(3)*}(u_{123}) =
\frac{3}{2} \gamma \, {\rm sym} ( {\sf R}^{* \prime \prime} (u_{12}) {\sf R}^{* \prime \prime}(u_{13})^2
\nonumber \\
&& - \frac{1}{3} {\sf R}^{* \prime \prime}(u_{12}) {\sf R}^{* \prime \prime}(u_{23})
{\sf R}^{* \prime \prime}(u_{31}) ).
\label{fp}
\end{eqnarray}
Thus, at least formally, $\tilde{S}_{T=0}^{(3)} \sim \epsilon^3$. The other terms, including
the fourth cumulant $\tilde{S}_{T=0}^{(4)} \sim \epsilon^4$ (and higher which have fixed points
\cite{ergchauve} similar to (\ref{fp}) ) can be neglected self consistently
in that picture, as well $\tilde{S}^{(3)}$ in the equation for $\tilde{R}(u)$, as they
yield only higher-order corrections in $\epsilon$.

Although this picture is consistent at $T=0$ to the lowest order in $\epsilon$,
closer examination shows possible ambiguities in the feeding term $\tilde{S}^{(3)}_{110}(0,0,u)$
in Eq. (\ref{Wilson2cum})
if one uses (\ref{fp}) which make the next order in $\epsilon$ more problematic.
In particular, differentiating (\ref{fp}) and taking $v=u_1-u_2$ small yields:
\begin{eqnarray}
&& \tilde{S}^{(3)}_{110}(v,0,u) \approx
\gamma {\sf R}^{* \prime \prime}(u) (
{\sf R}^{* \prime \prime \prime}(u)^2 - {\sf R}^{* \prime \prime \prime}(v)^2)
\nonumber \\
&& - \frac{\gamma}{2} {\sf R}^{* \prime \prime}(v) {\sf R}^{* \prime \prime \prime}(u)^2
- \gamma {\sf R}^{* \prime \prime}(v) {\sf R}^{* \prime \prime \prime \prime}(v)
{\sf R}^{* \prime \prime}(u). \label{lim}
\end{eqnarray}
It turns out that in this expression has well defined and coinciding
limits as $v \to 0^{\pm}$.  However, in the formal FRG equation
(\ref{Wilson2cum}) $\tilde{S}^{(3)}_{110}(v,0,u)$ is evaluated exactly
at $v=0$ and generally the replacement of that expression by either
limit $v \to 0^{+}$ or $v \to 0^{-}$ may not be valid in particular
renormalisation scheme. More general ambiguities arise at higher orders
and cannot be so easily resolved regardless of the scheme. Even in the
present Wilsonian formulation, treatment of this term alone is not
consistent since other two-loop contributions arise at order
$\epsilon^3$, thus it is not clear that the smoothness of this limit is
sufficient to guarantee a well behaved $\epsilon$-expansion even to this
order. Studies beyond Wilson\cite{bucheli_frg_secondorder} are
discussed elsewhere,\cite{frg2loop} so we will not address them here. Instead,
we follow a different route and focus on $T>0$, where no ambiguity appears,
and first summarize the naive $T>0$ analysis still to lowest order in
$\epsilon$.  A systematic study of the resolution of these ambiguities
by connecting zero and non-zero temperature quantities will be performed
for the case of $d=0$ in Section~\ref{sec:zero-dimension:-erg}.

\subsubsection{non-zero temperature}

Reinstating the $\tilde{T}_l$ terms in
(\ref{Wilson2cum},\ref{Wilson3cum}), the naive $T>0$ analysis {\it still
  assumes} that $\tilde{S}^{(3)} \ll \epsilon \tilde{R}$ and converts
(\ref{Wilson2cum}) into a decoupled flow equation for $\tilde{R}_l(u)$.
Since $\tilde{T}_l = e^{- \theta l}$ flows rapidly to zero,
$\tilde{R}_l(u)$ does converge to the $T=0$ fixed point. However, since
at $T>0$ the function remains analytic around $u=0$ the convergence to
the non analytic $T=0$ fixed point function is highly non-uniform. One
finds \cite{balents_loc,chauve_creep_epl} that for $u \alt \tilde{T}
_l/\epsilon$ the cusp singularity is rounded in a {\sl thermal
  boundary-layer} (TBL).  Its detailed solution yields two regimes at
large $l$ (precisely $\tilde{T} _l \ll \epsilon^2$):
\begin{eqnarray}
&& \tilde{R}_l(u) \to \tilde{R}^*(u) \quad \text{for} \quad u \sim O(1)
\gg \tilde{T} _l/\epsilon, \\
&& \tilde{R}_l(u) = \frac{1}{2} \tilde{R}^{* \prime \prime}(0) u^2
+ \tilde{T} _l^3 (\epsilon \tilde \chi)^{-2} r(\tilde{u}),\nonumber  \\
&&
\quad \text{for} \quad  \tilde{u} = \frac{\epsilon \tilde{\chi}
  u}{\tilde{T} _l} \sim O(1),   \label{bl2}
\end{eqnarray}
where we introduced the TBL scaling variable $\tilde{u}$ and scaling
function $r(\tilde{u})$.  The matching of the thermal boundary layer
(\ref{bl2}) to the zero temperature behaviour outside (\ref{cusp})
requires that:
\begin{eqnarray}
&& r''(\tilde{u}) \sim |\tilde{u}|,  \qquad {\rm for}\, |\tilde{u}| \gg 1.
\label{match}
\end{eqnarray}
Substituting the TBL form (\ref{bl2}) in (\ref{Wilson2cum})
one finds that (up to additive constants) the leading terms are
$O(\tilde{T} _l^2)$ yielding the constraint that
\begin{eqnarray}
&& (\epsilon - 2 \zeta) \frac{\tilde{R}^{* \prime \prime}(0)}{\chi^2 } \frac{u^2}{2}
+ \frac{1}{2} (r''(\tilde{u}) - r''(0))^2 + r''(\tilde{u}) \nonumber
\\
&& \label{blr1}
\end{eqnarray}
is a constant, which must thus equals $r''(0)$.  Note that the
$\partial_l \tilde{R}$ term is $O(\tilde{T} _l^3/\epsilon^2)$ and
the rescaling term (apart from the one acting on the purely
quadratic part of $\tilde{R}$) is even smaller as $O(\tilde{T}
_l^3/\epsilon)$ so both can be safely neglected at large $l$ such
that $\tilde{T} _l \ll \epsilon^2$ \cite{f4}.  Using
(\ref{secondder}), (\ref{blr1}) yields \cite{chauve_creep}
\begin{eqnarray}
&& r^{\prime\prime}(\tilde{u}) - r''(0) = \sqrt{1+\tilde{u}^2}-1,
\label{solu}
\end{eqnarray}
which is smooth for $|\tilde{u}|$ of $O(1)$ and
has the desired matching property (\ref{match}).
Note that the flow of $\tilde{R}''_l(0)$ is not
fully specified by the solution. However the existence
of the above single TBL together with the matching implies
that the value to which it converges must be equal to the $T=0$
fixed point value, as can be seen from (\ref{blr1}) (a different
asymptotic value would require a more complex TBL scenario, not
supported here).

Thus the problem of {\it matching} is successfully solved within this
approach.  The convergence to (\ref{bl2}) was studied in
\cite{chauve_creep}, where it was found that an expansion in higher
powers of $\tilde T_l$ could be solved order by order. In
\cite{ergchauve} it was found that the form (\ref{solu}) was particular
to Wilson, as other one loop schemes (with the same truncation to second
cumulant) give other analytical forms. Although $r(u)$ was found non
universal, its large $u$ (\ref{match}) and small $u$ behaviour must be.
For instance, the divergence of the mean curvature is constrained to be
\begin{eqnarray}
&& \tilde T_l \tilde R_l''''(0) \to \chi^2,
\label{fourth}
\end{eqnarray}
(i.e. $r''''(0)= 1$). This simple consequence of the TBL was found to
yield for susceptibility fluctuations the same scaling as predicted by
droplets \cite{ergchauve}.  Also, since a similar analysis (and
truncation) in the dynamics \cite{chauve_creep} yields that the friction
grows as $\partial_l \ln \eta_l = R_l''''(0)$ the TBL nicely recovers
the scaling of barriers $U_b \sim L^\theta$ expected in the creep
motion.  Thus there must be some truth in this TBL scaling, although
this simple analysis is far from being sufficient, as we now show.

Before doing so, we note that {\sl above four dimensions}, the situation
is rather different, and discussed in
Appendix~\ref{sec:above-four-dimens}, where this case is analyzed by FRG
and large $d$ methods.  For a weak random potential with smooth
correlations, no cusp arises even at $T=0$, and there is therefore no
TBL.  For stronger disorder or a sufficiently non-smooth potential, the
zero temperature cusp does arise.  In this case, however, the TBL has a
width $\sim T$ instead of $\tilde T_l$, and so does not scale to zero
with the infra-red cutoff.  The physics of this difference -- which we
believe is associated to the {\sl short-scale} nature of metastable
states for $d>4$ -- is discussed in Appendix~\ref{sec:above-four-dimens}.

\subsection{thermal boundary layer: the systematic analysis}

We need now to estimate the magnitude of the term
$\tilde{S}^{(3)}_{110}$ in the second cumulant equation (\ref{Wilson2cum}), {\it in
the TBL region } $u \sim T_l/\epsilon$, which has not been done
before. For that we must carefully reexamine the equation for the
third cumulant (\ref{Wilson3cum}). First one sees that $\tilde S^{(3)}(u_{123})$ will also
have a TBL form for $u_i \sim T_l/\epsilon$ since the $O[(\tilde {\sf
R}'')^3]$ terms in (\ref{Wilson3cum}) feed their boundary-layer
scaling forms into $\tilde S^{(3)}$. Naively balancing these terms with
the rescaling part ($\approx -2 \tilde{S}^{(3)}$) suggests
$\tilde{S}^{(3)} \sim \tilde{T} _l^3$. This is, however,
inconsistent, since the remaining terms linear in
$\tilde{S}^{(3)}$ would then be $O(\tilde{T} _l^2)$.  Instead, the
only consistent TBL ansatz for the third cumulant is:
\begin{eqnarray}
\tilde{S}^{(3)}(u_1,u_2,u_3) = (\epsilon\tilde\chi)^{-2}
\tilde{T}_l^4 s^{(3)}(\tilde u_1,\tilde u_2,\tilde u_3)
\label{bl3}
\end{eqnarray}
The second derivative term feeding into Eq.~(\ref{Wilson2cum}) then
reads:
\begin{eqnarray}
\tilde{S}^{(3)}_{110}(0,0,u) = \tilde{T}_l^2
s^{(3)}_{110}(0,0,\tilde u) \quad \text{for} \quad \tilde u = O(1)
\end{eqnarray}
and is thus of the same order $O(\tilde{T}_l^2)$ than the other
surviving terms in the boundary layer for $R(u)$. Thus the above
described result for $r(\tilde u)$ appears as an
(uncontrolled) approximation, since we now have two coupled
equations for the TBL scaling functions $r$ and $s^{(3)}$. The equation
(\ref{Wilson2cum}) now leads to:
\begin{eqnarray}
&& 0 = - \frac{\tilde{u}^2}{2} + \frac{1}{2}
(r''(\tilde{u})-r''(0))^2 \nonumber \\
&& + r''(\tilde{u}) - r''(0) +
s^{(3)}_{110}(0,0,\tilde u),
\label{blr1new}
\end{eqnarray}
while $s^{(3)}(\tilde u_1,\tilde u_2,\tilde u_3)$ satisfies a TBL
equation simply obtained from (\ref{Wilson3cum})
by setting $\tilde{S}^{(3)} \to
s^{(3)}$, ${\sf R} \to r$, $u_i \to \tilde{u}_i$, $\tilde T_l \to
1$, suppressing the linear rescaling term and setting the l.h.s.
to zero.

Of course we must now worry about the fourth cumulant term $\tilde S^{(4)}_{1100}$
in the equation for $\tilde S^{(3)}$, so in fact there is an
infinite set of coupled equations and the scaling forms in
(\ref{bl2}-\ref{bl3}) must be extended to {\sl all}
cumulants. We will now establish the consistent form of all
cumulants and derive the TBL equation for all of them. There are a
few subtleties in that construction.

Let us first write the one loop FRG equation for the $k$-th
cumulant in a very schematic form:

\begin{widetext}

\begin{eqnarray}
&& \partial_l \tilde S^{(k)} = {\cal O} \tilde S^{(k)} + \tilde S^{(k+1)
\prime \prime} +
\tilde{T}_l \tilde S^{(k) \prime \prime}
+ \sum_{m=2}^{k} \sum_{\sum k_i=k+m } \tilde S^{(k_1)
\prime \prime}\cdots \tilde S^{(k_m) \prime \prime} + \tilde{T}_l
\sum_{m=2}^{k-1} \sum_{\sum k_i=k-1+m } \tilde S^{(k_1) \prime
\prime}\cdots \tilde S^{(k_m) \prime \prime} \label{general}
\end{eqnarray}

\end{widetext}
where ${\cal O}$ denotes the rescaling linear operator, the two next
terms are tadpoles, and each of the last two terms is a loop with $m \geq
2$ disorder vertices. The counting of $T$ factors results from there being
$m$ propagators per loop, hence a factor
$T^m$, and a factor $T^{- k_i}$ associated to each $k_i$-replica vertex
$\tilde S^{(k_i)}$. One easily sees that only these terms can appear. This
property and their detailed structure is worked out
in Appendix~\ref{sec:gener-struct-frg}, but is irrelevant for the present discussion.
Among them there is a feeding term of $O[({\sf R}'')^k]$.

A look at (\ref{general}) suggests that the natural generalization
of (\ref{bl2}-\ref{bl3}), namely
\begin{eqnarray}
&&  \tilde S^{(k)}(u_1, \cdots , u_k) \stackrel{?}{=} ( \tilde \chi \epsilon)^{-2} T_l^{k+1}
s^{(k)}(\tilde u_1, \cdots \tilde u_k)
\label{blodd}
\end{eqnarray}
should produce a consistent TBL equation with all terms in
(\ref{general}) of order $O(\tilde T_l^k)$, except the rescaling and
left hand side which are $O(\tilde T_l^{k+1}/\epsilon^2)$ and thus
disappear from the TBL equation as $T_l \ll \epsilon^2$.  This is
essentially correct, apart from one simple, but important, modification.
In the TBL equation for $r$ (\ref{blr1}, \ref{blr1new}) one term
survives in the rescaling part, coming from the quadratic part of
$R(u)$, and this is because this term scales naturally as $\epsilon
\tilde R^{*,\prime \prime}(0) u^2 \sim \tilde T_l^2$ (while all other
even monomials scale as larger, hence subdominant, powers of $\tilde
T_l$). One can then ask whether one should consider similar terms in the
TBL equation for $s^{(k)}$, which is $O(\tilde T_l^k)$.  The
construction of the $k$-replica terms is detailed in
Appendix~\ref{sec:structure-cumulants}. One easily sees that for $k=2 p$
even, the monomial of {\it lowest} degree one can construct which is a
genuine $k$-replica term is indeed of order $u^k$ and thus scales as
$\tilde{T}_l^k$.  It reads:
\begin{eqnarray}
&& q_{2p} u_1 u_2.. u_{2p}, \label{even}
\end{eqnarray}
where $q_{2p} = \tilde{S}^{(2p)}_{11..1}(0,\cdots,0)$, and is fully
symmetric, even as it must be, and translationally invariant (see
Appendix~\ref{sec:structure-cumulants}). A $k$-replica monomial must
contain at least one of each replica, i.e. the product (\ref{even}),
otherwise it is a pure gauge. No such term exist for $k=2p+1$ odd, and
one shows that the term of smallest order in that case is of order
$u^{k+3}$. For instance for the third cumulant, one checks that the term
of order $u^4$, namely
\begin{eqnarray}
&& \text{sym}_{123} u_1^2 u_2 u_3
\end{eqnarray}
is forbidden by translational invariance (see
Appendix~\ref{sec:structure-cumulants}). Thus the third cumulant starts
at higher order $S^{(3)} \sim u^6$.  The conclusion is thus that the TBL
form (\ref{blodd}) is correct for {\it odd} cumulants but for {\it even}
cumulants it should be replaced by the correct TBL form:
\begin{eqnarray}
&&  \tilde S^{(2 p)}(u_1\cdots u_k) = q_{2p} u_1 u_2.. u_{2p}
\\
&& + (\tilde \chi \epsilon)^{-2}
T_l^{2 p +1} s^{(2 p)}(\tilde u_1 \cdots
\tilde u_{2p}). \nonumber
\label{bleven}
\end{eqnarray}

Having determined the scaling for all cumulants within the TBL, we
briefly return to the third cumulant TBL equation.  Using
(\ref{bleven}), we can express the feeding terms from the fourth
cumulant in their appropriate TBL form.  Remarkably, as for the second
cumulant, a formal analytic solution for $s^{(3)}$ is possible (and this
appears to generalize to all cumulants).  Using the gauge freedom and
STS relations (\ref{STSrels}) one directly rearranges the third cumulant
equation into the form
\begin{widetext}
\begin{eqnarray}
&&
    s^{(3)}_{110}(\tilde u_{123}) - s^{(3)}_{110}(\tilde u_{113}) =
    \frac{1}{1+r''(\tilde u_{12})} \Big\{ G(\tilde{u}_{123}) + a(\tilde
    u_{13}) + a(\tilde u_{23})+  \,{\rm sym}_{123}\,\Big[
    \frac{1}{2}
    r''(\tilde u_{13})
    r''(\tilde u_{23}) \nonumber \\ &&   + \gamma r''(\tilde u_{12})
    r''(\tilde u_{13})^2  -
    \frac{\gamma}{3} r''(\tilde u_{12}) r''(\tilde u_{23})r''(\tilde u_{13})
    +\frac{1}{2}s^{(4)}_{1100}(\tilde u_{1123}) \Big] \Big\},
    \label{eq:formals3}
\end{eqnarray}
\end{widetext}
where $G(\tilde{u}_{123})=G^{(1)}(\tilde{u}_{123})+
G^{(2)}(\tilde{u}_{123})$, where $G^{(1)},G^{(2)}$ are translationally
invariant functions symmetric in $1\leftrightarrow 2$ and respectively
antisymmetric in $1\leftrightarrow 3$, $2\leftrightarrow 3$, and
$a(\tilde u)$ is an even function.  These are constrained by the
requirement that $s^{(3)}$, or equivalently $s^{(3)}_{111}$ is a
symmetric function, and in the latter case, fully gauge invariant.
Although this has been written as a formal solution for $s^{(3)}$, it
clearly involves the fourth cumulant quantity $s^{(4)}_{1100}(\tilde
u_{1123})$.  Thus ``solutions'' of this form do not close, and form a
coupled hierarchy.  Indeed, it is equally valid to view
(\ref{eq:formals3}) as expressing the fourth cumulant feeding term in
terms of lower cumulants.  This and other similar relations are used
later in the paper to simplify some expressions.

Let us write the TBL scaling form in a more compact way in terms of the
{\it rescaled characteristic function } defined in (\ref{eq:Vexpand2}).
It takes the form:
\begin{eqnarray}
&& \tilde{V}_l(\vec{u}) = \sum_{p \geq 1} f_{2p} \sum_{a_1,..,a_{2p}}
\tilde u_{a_1} \cdots \tilde u_{a_{2p}} +
  \frac{\tilde{T} _l}{(\epsilon\tilde\chi)^2}  v( \vec{\tilde u} ), \nonumber \\
&& \label{blgen}
\end{eqnarray}
where we have defined the boundary-layer replica vector, and scaling function:
\begin{eqnarray}
&& \vec{\tilde u} = \epsilon \tilde{\chi} \vec{u}/\tilde{T} _l \\
&& v( \vec{\tilde u} ) =
\sum_{k=2}^\infty {1 \over k!} \sum_{a_1 \cdots a_k}
s^{(k)}[\tilde u_{a_1},\cdots, \tilde u_{a_k}], \label{eq:Vexpand3}
\end{eqnarray}
Note that all dependence in $\tilde{T} _l$, which has been made apparent
in (\ref{blgen}), factors in front of all cumulants and that the TBL characteristic function
$v( \vec{\tilde u} )$ approaches a fixed point form at large $l$. Below we will
discuss the physical
meaning of the set of ($l$-dependent) constants,
\begin{eqnarray}
&& f_{2p} = \tilde{S}^{(2 p)}_{11..1}(0, \cdots, 0)/(\tilde\chi\epsilon)^{2p},
\label{f2p}
\end{eqnarray}
with $f_{2} = - \tilde{R}''(0)/(\tilde\chi\epsilon)^2$ (remember that $u_a u_b \equiv - \frac{1}{2}
(u_a - u_b)^2$ up to gauge terms).

Inserting the $l$ dependent form (\ref{blgen}) into (\ref{eq:Wilson1}), taking into
account all $l$ dependent rescaling factors, and discarding terms small as
$\tilde T_l \ll \epsilon^2$, one gets the TBL-FRG equations which
must be satisfied by the constants $f_{2p}$ and the function $v$ at
their fixed point. The $Tr \ln$ term in (\ref{eq:Wilson1}) takes the form:
\begin{eqnarray}
&&
{\rm Tr}  \ln \left[ \delta^{ab} - \alpha M_{ab} - \partial_a \partial_b
v(\vec{\tilde u}) \right],  \label{trln} \\
&&
M_{ab} = \partial_a \partial_b [ \sum_{p \geq 1} f_{2p} \sum_{a_1,..,a_{2p}}
\tilde u_{a_1} \cdots \tilde u_{a_{2p}} ],
\end{eqnarray}
with $\alpha=(\epsilon \tilde{\chi})^2/\tilde T_l$, and $\partial_a =
\partial/\partial \tilde{u}_a$ here.

However we see that because the monomials associated to each $f_{2p}$
is such that one replica appears only once, the matrix $M_{ab}$
is independent of replica $a$ and $b$. Thus in the expansion of
(\ref{trln}) to any order in $M$, any power of $M$ or any
product of the form:
\begin{eqnarray}
&& \sum_b M_{ab} \partial_b \partial_c  v(\vec{\tilde u}) =
M_{ab} \sum_b \partial_b \partial_c  v(\vec{\tilde u}) = 0
\end{eqnarray}
vanishes when traced, because of translational invariance
(\ref{prop1}). Thus $M$ can be set to zero (for
$n=0$) in (\ref{trln}) and we finally obtain the
non-trivial one loop Wilson fixed point TBL equation
in a very compact form:
\begin{equation}
\!\! \sum_{p \geq 1} x_{2p} f_{2p} \sum_{a_1,..,a_{2p}}  \!\!
\tilde u_{a_1} \cdots \tilde u_{a_{2p}} \!\!
- \frac{1}{2} {\rm Tr} \!
  \ln \left[ \delta^{ab} -  \partial_a \partial_b
    v(\vec{\tilde u}) \right] = 0, \label{eq:BL2}
\end{equation}
where all $l$ dependence has disappeared, the
trace being over replica indices. The first term is,
as discussed above, the only one surviving from the
rescaling part and contains the scaling eigenvalues:
\begin{eqnarray}
&& x_2 = \epsilon-2 \zeta = O(\epsilon), \\
&& x_{2p} = d- 2 p(\theta-\zeta) = O(1) \quad p>1.
\end{eqnarray}
(\ref{eq:BL2}) is a highly compact form which,
expanded in the number of replica sums, yields coupled
boundary layer equations for $r$ and $s^{(3)}$, and all higher
cumulants. It is useful for later purpose to also give the
schematic structure of these TBL equation for the coupled
function $s^{(k)}(u_1,\cdots,u_k)$:
\begin{widetext}

\begin{eqnarray}
&& 0 = \delta_{k,\text{even}} x_{k} f_{k}
  \tilde u_1 \!\cdots \tilde u_{k} +  s^{(k+1)
\prime \prime} + s^{(k) \prime \prime}
+ \sum_{m=2}^{k} \sum_{\sum k_i=k+m } s^{(k_1)
\prime \prime} .. s^{(k_m) \prime \prime} +
\sum_{m=2}^{k-1} \sum_{\sum k_i=k-1+m } s^{(k_1) \prime
\prime} .. s^{(k_m) \prime \prime} \label{generals}
\end{eqnarray}

\end{widetext}
for any $k \geq 2$, where, as in (\ref{general}) all coefficients and
detailed structure has been suppressed (they are identical to the one in
(\ref{general}) and given in Appendix~\ref{sec:gener-struct-frg}). Note
that the coefficients $f_k$ appear only in the first term and have no
feedback in the others.  Similarly, the naively first non-gauge term in
$s^{(2p)}$ proportional to $\tilde u_1 \cdots \tilde u_{2p}$
(subdominant to $f_{2p}$) drops out of the equation completely. Thus
this term can be taken as zero, and $s^{(2p)}$ thus is chosen to start
as $\tilde u^{2p+2}$.

\subsection{physical consequences, correlations, droplets}

Let us now pause and analyze what has been achieved. We have derived,
within the one loop Wilson FRG, coupled flow equations for all cumulants
of the disorder. We have found a scaling form for its solution at large
$l$ (i.e. small $\tilde T_l$) which exhibits a thermal boundary layer,
consistent for all cumulants.  It is expressed in terms of an infinite
set of unknown parameters $f_{2p}$ and functions $s^{(p)}$. These encode
for the probability distribution of the coarse grained disorder, and
obey the reduced fixed point equations (\ref{eq:BL2} , \ref{generals}).

The first observation is that {\it there is no remaining small
  parameter} in Eqs.~(\ref{eq:BL2},\ref{generals}). It is clear in the
$Tr \ln$ term. For the other terms it also holds {\it if conventional
  (zero temperature) scaling holds}. Indeed although $f_2 \sim
1/\epsilon$, the product $x_2 f_2 = 1$, and in the conventional zero $T$
scaling $\tilde{S}^{(2 p)}_{11..1}(0, \cdots, 0) \sim \epsilon^{2 p}$,
thus $x_{2p} f_{2p} \sim O(1)$. The problem has thus become fully {\it
  non-perturbative} and one is thus in the uncomfortable situation where
a full solution of this fixed point equation (\ref{eq:BL2} ,
\ref{generals}) is required simultaneously for all cumulants of the
disorder distribution, an infinite number of functions.  Although some
formal algebraic relations between cumulants can be obtained in the TBL,
as discussed above for $s^{(3)}$, no general solution of the full
hierarchy appears possible.  Thus at this stage we have no choice but to
{\it assume the existence} of a global, well behaved solution to the
TBL-FRG equation and explore some of the consequences of this {\it
  thermal boundary layer ansatz} (TBLA).  As we show below, it implies
that a smooth zero temperature limit exists for displacement
correlations determined by the $f_{2k}$ coefficients, i.e.  a well
defined limit exists as the bare temperature $T \to 0^+$ \cite{f3}
The above definition of the $f_{2k}$  further implies that  {\it
conventional scaling holds} for all $T=0$ correlations, i.e.
$\overline{u^{2k}}^c_{T=0} \sim \epsilon^{2k}$ for $k>1$, as we
now show.

\subsubsection{zero temperature limit: evading
dimensional reduction}

Let us explore this further, one of the aim being to obtain
zero temperature correlation functions (as well as $T>0$
ones). The coefficients $f_{2p}$ (\ref{f2p})
have an interesting physical meaning, linked to the so called
``dimensional reduction'' (DR) property. They are the rescaled cumulants of the
so called {\it random force}, precisely:
\begin{eqnarray}
&& \overline{F(0,r_1).. F(0,r_k)} = S^{(2 p)}_{11..1}(0, \cdots, 0)
\delta^d(r_1,\cdots ,r_k).
\end{eqnarray}
The model (\ref{eq:ham}), as other disordered systems such as random
field spin models, has the amazing DR exact property that if one
computes {\it any observable} (e.g. correlation functions) at $T=0$ in
perturbation theory in {\it analytic} disorder cumulants, the result is
equal {\it to all orders in perturbation theory} to the result in the
simple model with only a random force, introduced by Larkin to model
physics at short scales:
\begin{eqnarray}
&& H_{rf}[u] = \int \! d^d{\bf r} \left[ {1 \over 2} |\nabla u|^2 - u({\bf r}) \cdot
F(0,r) \right], \label{eq:hamrf}
\end{eqnarray}
which is simply, of course (in Fourier)
\begin{eqnarray}
&& u(q) = F(0,q)/q^2
\label{solurf}
\end{eqnarray}
we will see that the $f_{2p}$ do indeed determine the correlations at
$T=0$ (connected with respect to disorder), simply as expected from
(\ref{solurf}):
\begin{eqnarray}
&& \overline{ u(q_1) .. u(q_{2 p}) }^c|_{T=0}
= (2 \pi)^d \delta^d(\sum_i q_i) C_{T=0}(q_1,\cdots,q_{2 p}) \nonumber \\
&& C_{T=0}(q_1,\cdots,q_{2 p}) =  q_1^{-2}.. q_{2 p}^{-2} S^{(2 p)}_{11..1,l}(0, \cdots, 0),
\label{corrw}
\end{eqnarray}
except that from the TBLA these cumulants are $l$ dependent and
flow under RG, yielding a non trivial result, different from the
``'naive DR estimate'' which assumes them constant and is incorrect. Thus the
structure (\ref{blgen}) has nice properties: it does obey correctly
the DR property, since setting naively $T=0$ by simply removing
the term proportional to $\tilde T_l$ in (\ref{blgen}) does
indeed leave us only with a random force. At the same time,
though, the TBL provides a novel and non-trivial
mechanism to evade the ``naive DR estimate'' since it gives a non-trivial
renormalization of the random force cumulants. It is particularly
interesting since only two other mechanisms have been proposed
previously to evade DR, namely (i) replica symmetry breaking,
only convincingly demonstrated in mean field models
(ii) the cusp and non analyticity of the disorder cumulants in the FRG,
demonstrated in the statics only to one loop (although up to two loops
at depinning). Here we have constructed another one,
as apparent from the structure (\ref{blgen}).

Specifically the zero temperature correlations can be computed
by restoring the $l$ dependence, using from (\ref{resc}) that
$S^{(2 p)}_{11..1,l}(0,\cdots,0) = A_d^{1- 2 p} \Lambda_l^{d - 2 p \theta + 2 p \zeta}
\tilde{S}^{(2 p)}_{11..1,l}(0,\cdots,0)$ and the TBLA property (\ref{f2p})
that $\tilde{S}^{(2 p),l}_{11..1}(0,\cdots,0)$
converges to its fixed point value $\epsilon^{2 p} \tilde \chi^{2 p} f_{2 p}$.
We can then simply estimate the correlation when e.g. all
momenta have similar magnitude $q_i^2 \sim q^2$. Setting
$l=\ln (q/\Lambda_0)$, one finds from (\ref{corrw}):
\begin{eqnarray}
&& C_{T=0}(q_1,\cdots,q_{2 p})|_{q_i^2 \sim q^2} \sim  b_p f_{2 p} \epsilon^{2 p}
q^{- 2 p (d - 1 + \zeta)}, \nonumber \\
&& \label{eq:dropcorr}
\end{eqnarray}
with $b_p=A_d^{1- 2 p} \tilde \chi^{2 p}$. Taking into account the
Fourier transform, this is the form expected from the scaling $u \sim
L^\zeta$, i.e. real space correlations $\overline{u_{r_1}... u_{r_{2
      p}}}^{T=0} \sim L^{2 p \zeta}$ ($L$ the common scale). The
prefactors of the correlations however, and thus the full probability
distribution, are determined by the constants $f_{2 p}$. It is thus
likely (but at this stage not established) that they should be universal
(for long range and periodic disorder, and up to a common non universal scale
for short range disorder). Similar correlations were computed recently for
depinning \cite{distrib}.  It is worth noting that, as was
the case for $R''(0)$, the terms involving the higher cumulants of the
random force $f_{2p}$ do not feed back into the renormalization of the
nonlinear part of pinning force (and thus do not appear in the $tr
\ln$), as was the case for $R''(0)$.  This is also to be expected on
physical grounds, since they only result in the shift (\ref{eq:hamrf})
and can thus be eliminated by a translation.

\subsubsection{non-zero $T$, droplets}

In addition to the terms containing $f_{2p}$ which describe the zero
temperature limit, (\ref{blgen}) also contains a term proportional
to $\tilde{T}_l$ which describes the first correction proportional
to $T$ in the correlation functions. Remarkably, its content is
precisely the one of the droplet picture. Let us indicate here
how it works, remaining very schematic for now.  A more
precise and detailed formulation will be given in
Section~\ref{sec:zero-dimension:-erg}.

Let us consider an arbitrary correlation amongst replicas (connected
with respect to disorder). Since same replicas correspond to same thermal average,
it corresponds to some disorder average of some product of thermal averages
(itself connected \cite{footnote2}), and we write schematically:
\begin{eqnarray}
&& \langle u^{a_1}_{q_1} .. u^{a_{2p}}_{q_{2p}}  \rangle _c =
(2 \pi)^d \delta^d(\sum_i q_i) C_{q_1,..q_{2p}}^{a_1 .. a_{2p}} \\
&& = \overline{ \langle u ..u \rangle  .. \langle u.. u \rangle  }.
\end{eqnarray}
Lowest order perturbation theory gives
\begin{eqnarray}
&& C_{q_1,..q_{2p}}^{a_1 .. a_{2p}} =
\frac{T^{2 p}}{q_1^2 .. q_{2 p}^2} V^{(2p)}(0),
\end{eqnarray}
where here $V^{(2p)}(0)$ denotes a $2p$-th derivative at
$\vec{u}=0$, the exact replica index structure of which we
ignore here. Substituting the rescaled form for the
characteristic function yields
$V^{(2p)}_l(0) = A_d \Lambda_l^{d + 2 p \zeta} \tilde{V}^{(2p)}_l(0)$.
Inserting now the TBLA (\ref{blgen}) gives
\begin{eqnarray}
&& C_{q_1,..q_{2p}}^{a_1 .. a_{2p}} = b_p \epsilon^{2 p}
\frac{\Lambda_l^{d + 2 p \zeta - 2 p \theta}}{q_1^2 .. q_{2 p}^2}
(f_{2p} + \frac{\tilde{T}_l}{(\epsilon \tilde{\chi})^{2}} v^{(2p)}(0) )
\nonumber \\
&&
\end{eqnarray}
Thus we find that at $T \to 0$, as expected:
\begin{eqnarray}
&&
\overline{ \langle u ..u \rangle  .. \langle u.. u \rangle  } \to \overline{ u(q_1) .. u(q_{2 p}) }^c|_{T=0}
\end{eqnarray}
obtained above (\ref{corrw},\ref{eq:dropcorr}) (since at $T=0$ the
correlations become independent of the replica index -- in the absence
of RSB). The contribution of thermal excitations, to lowest order takes
the form
\begin{eqnarray}
&& C - C_{T=0} =  (T q^\theta) \epsilon^{2 p - 2} q^{- 2 p (d - 1 + \zeta)} b'_p v^{(2p)}(0),
\end{eqnarray}
with $b'_p=A_d^{1- 2 p}  \tilde \chi^{2 p-2}$. This is exactly the scaling
expected from the droplet picture at low temperature. In real space:
\begin{eqnarray}
&& C - C_{T=0} \sim T L^{- \theta} L^{2 p \zeta}
\end{eqnarray}
with $C=\overline{\langle u_{r_1}.. \rangle ...\langle.. u_{r_{2 p}} \rangle }^c$. The complicated
set of TBL fixed point coefficients thus encode for the distribution of
low energy excitation.
This equivalence will be further explored below.

Another question is what fixes the coefficients $f_{2p}$ ?
Let us examine the FRG equation (\ref{ergS4}) for the fourth cumulant
(see Appendix~\ref{sec:gener-struct-frg}). We need to take four derivatives at zero.
Since ${\sf R}''(u)$ starts at $u^2$,
$S^{(3)}_{110}$ as $u^4$, $S^{(5)}_{110}$ as $u^6$,
only one term in addition to the rescaling term contributes.
The analysis is identical for any $2p$-th cumulant and
one finds
\begin{eqnarray} \label{rel2p}
&& \partial_l \tilde S^{(2p)}_{1..1}(0,..0)  =
(d - 2 p (\theta - \zeta)) S^{(2p)}_{1..1}(0,..0)\nonumber
\\
&& + p \tilde T_l \tilde S^{(2p)}_{311..1}(0,..0),
\end{eqnarray}
a simple generalization of the relation (\ref{fourth}) for the second cumulant.
Thus the $f_{2p}$ are determined from the hierarchy,
and from the function $v$ itself.

A further hypothesis is that these $f_{2p}$ {\sl match} the behavior of
the outer solution, in particular,
\begin{eqnarray}
  f_{2p} \stackrel{?}{=} \lim_{u_i\rightarrow 0} \lim_{\tilde{T}_l
    \rightarrow 0} \tilde{S}^{(2p)}_{11..1}(u_1,\cdots,u_{2p}) \equiv
  \tilde{S}^{(2p)}_{11..1}(0^+,\cdots,0^+). \nonumber \\
  &&   \label{eq:matchf2p}
\end{eqnarray}
Here we have introduced the notation $0^+$ for the small argument limit
of the outer solution.  Given the existence of the TBL, this is not a
trivial statement.  It is a particular case of a more general question
of how (and which) zero temperature quantities match to $T\rightarrow 0$
objects within the TBL.  The appealing result of the above postulate is
that the $f_{2p}$ coefficients are truly properties of the system at
zero temperature, despite being defined from (\ref{f2p}) deep inside
the TBL region ($u=0$ rather than $u=0^+$).  This is, however, only an
hypothesis.  The structure and consistency of matching is explored and
verified in some detail in $d=0$ in Section~\ref{sec:matching}.

If one attempts to generalize the TBLA to match further and further
(nonanalytic) derivatives of the outer solution, it seems natural to
expect that:
\begin{eqnarray}
&& \tilde{V}_l(\vec{u}) = \tilde{V}^{T=0}_l(\vec{u})
+  \frac{\tilde{T}_l}{(\epsilon\tilde\chi)^2}  v( \vec{\tilde u} )
+ \left[\frac{\tilde{T}_l}{(\epsilon\tilde\chi)^2}\right]^2  v_2( \vec{\tilde u} ) + \cdots
 \nonumber \\
&& \label{blexp}
\end{eqnarray}

Another consequence of the TBL ansatz is that the {\sl scaling} of
moments of the pinning force derivatives ($\tilde{S}_{n_1\cdots
  n_{2k}}^{(2k)}(0,\cdots,0)$ for $k=1,2,\ldots$) is not modified from the
naive results of Refs.~\onlinecite{balents_loc,chauve_creep}.  Thus, in
particular, the interpretation of the boundary layer as representing
thermal activation, and moreover the equality of the barrier exponent
with $\theta$ remains unchanged.  It is somewhat less clear, but we
think likely, that the general {\sl form} of the velocity-force
characteristic\cite{balents_loc,chauve_creep}\ (in the low velocity
limit) should be unmodified, as this relies only on the scaling
assumption.  We leave this question open to future analysis (see also
the parallel work on the dynamics in \cite{us_short,us_dyn_long})

However, although the {\sl scaling} of moments is preserved, the
prefactors are certainly modified by inclusion of higher cumulants.
Indeed, we have shown that the $\tilde{S}^{(3)}(0,0,u)$ term in
(\ref{Wilson2cum}) modifies $\tilde{R}^{(4)}(0)$ (and all higher
derivatives) if the full hierarchy is solved self-consistently.  This
value appears explicitly in calculations of susceptibility fluctuations
\cite{ergchauve}\ and in the velocity-force characteristic
\cite{chauve_creep}, so these results may be modified by the
non-Gaussian distribution encoded in the higher-replica terms.  Note
however that $s$ in (\ref{blr1new}) starts as $u^6$ and thus it cannot
change the result $r''''(0)= 1$, thus (\ref{fourth}) remains true.  It
would obviously be desirable to solve the hierarchy encoded in
(\ref{eq:BL2}) to obtain these and other physical quantities.  This
is an important open problem.

\section{Exact RG and general identities in any dimension}
\label{sec:exact-rg-general}

The Wilson one loop approach was helpful in exploring the new physics of
the thermal boundary layer.  Since the non perturbative nature of the
problem has been unveiled, the validity of the approximations implicit
in the one loop Wilson scheme, which is known to work in perturbative
situations, are here difficult to assess.  In particular the assumption
of uncorrelated space points does not remain correct, and a more
controlled and general technique is required to perform a detailed
analysis.  We therefore turn now to a more powerful Exact RG (ERG)
method.  Within the ERG, we will still not obtain a complete solution of
the TBL hierarchy, but we will obtain some concrete results.  For
instance,  in $d=0$, up to a change in
some coefficients ($c_p$ defined above) the one loop FRG equations are
{\it exact}. This is very interesting both as a new way to study $d=0$
but even more so as a way to confirm the structure found here in an
exactly solvable case, since in some cases exact solutions exist in
$d=0$.  This paves the way for future ERG studies in $d>0$, as discussed
in Section~\ref{sec:conclusion}.

In this section we review the ERG method, and obtain useful exact
identities valid in any dimension. We will turn to its specific
application in $d=0$ in the following section.  As a general method, the
ERG was introduced long ago \cite{wilson_kogut,polchinski} and much studied since,
but not so often for
perturbative calculations, and rarely in the context of disordered
systems.  In recent studies however a systematic multilocal expansion
scheme was developed \cite{ergchauve,erg} and applied to perturbative
study of pinned systems and the FRG. We will find useful the version
developed directly on the effective action functional \cite{erg}, and
will borrow from it.  In parallel, we will also derive simpler, but
powerful, exact relations, directly for correlation functions using from
scale invariance and STS. These will teach us much about the physics
associated with the thermal boundary layer.

\subsection{effective action}

We are interested in calculating correlation functions in the
replica theory, typically:
\begin{eqnarray}
&& \langle u^a_x u^b_y  \rangle  \quad , \quad  \langle  u^a_x u^b_y u^c_z u^d_t  \rangle  \quad , ...
\end{eqnarray}
as we recall correlations with an odd number of $u$
fields vanish. A useful object in doing so is the effective
action functional $\Gamma[u]$. We recall its definition as
a Legendre transform:
\begin{eqnarray}
&& \Gamma[u] = \int_r h^a_r u^a_r - W[h] \\
&& \partial_{h^a_r} W[h] = u^a_r \\
&& W[h] =  \ln Z[h]
\end{eqnarray}
where $Z[h]$ was defined in (\ref{eq:staticPF}). It thus has a
physical interpretation in terms of a free energy
with imposed expectation values for the fields
(e.g. probability distribution of the order parameter).

Let us consider the perturbation theory of the
replicated problem defined by $Z[0]$ in (\ref{eq:staticPF}),
the free propagator being:
\begin{eqnarray}
&&  \langle u^a_q u^b_{-q} \rangle_0 = \frac{T}{q^2 + m^2} \delta_{ab}
\end{eqnarray}
and the interaction vertices come from the disorder cumulants.
We have added a small mass which confines the manifold in a finite
region in $u$, and
provides an infrared cutoff. Other IR cutoff, such as finite size
$L$, or momentum IR cutoff $\Lambda' \leq \Lambda$ can also be considered
(the UV cutoff is still denoted $\Lambda$).

We recall that $W[h]$ is the generating function of {\it connected}
correlations, noted $ \langle .. \rangle _c$, which are sums of
connected graphs. $\Gamma[u]$ is the generating function of {\it one
  particle irreducible} (1PI) graphs (which cannot be disconnected by
cutting one propagator) and thus contain loops. The interest of the
effective action functional is that since it resums the loops,
correlations are simply obtained from it as sums of tree diagrams.

There are two useful expansions of the functional $\Gamma[u]$, the {\it polynomial
expansion}, schematically:
\begin{eqnarray}
&&  \Gamma[u] = \Gamma[0] + \frac{1}{2}
\sum_{ab} \int_{r r'} \Gamma^{(2) a b}_{r,r'} u^a_r u^b_{r'} +
\frac{1}{4!} \Gamma^{(4)} u u u u + .. \nonumber
\end{eqnarray}
where the ``proper vertices''
$\Gamma^{(2 p) a_1..a_{2 p}}_{q_1,..q_{2p}}$ are the $2 p$-th functional derivative
of $\Gamma[u]$ in $u=0$ and are the sums of all 1PI graphs with $2 p$
external legs (note here $2 p$ does not refer to the number of replicas, but to
the number of external legs). A different expansion is the {\it multilocal
expansion}:
\begin{eqnarray}
&& \Gamma[u] = \Gamma_0 + \frac{1}{2 T} \int_q (q^2 + m^2) u^a_q u^a_{-q} - {\cal V}[u], \\
&& \! \! \! \! {\cal V}[u] = \int_r V(\vec u_r) + \int_{r,r'}
V^{(2)}(\vec u_r,\vec u_{r'},r-r') + \ldots \nonumber \\
& & \label{eq:multilocalexp}
\end{eqnarray}
with $\Gamma_0= - \frac{1}{2} n L^d \int_q \ln(q^2 + m^2)$, where we
have decomposed the interaction part, noted $- {\cal V}[u]$, into a
local part defined by the function $V(\vec u)$ of a replica vector, a bilocal
part $V^{(2)}(\vec u_r,\vec u_{r'},r-r')$ defined as a function of two replica
vectors and one space argument, with by definition $\int_r
V^{(2)}(\vec u,\vec v,r)=0$, trilocal etc.. Each of these terms can then also be
expanded in sums over different numbers of replicas, as was done in
(\ref{eq:Vexpand}). Because of STS, $V$, $V^{(2)}$,\ldots contain only
sums over $k \geq 2$ replicas (the single replica part is not corrected)
and thus represent the ``renormalized disorder'' ($V$ encodes all
cumulants of the ``renormalized random potential'' at the same space
points, $V^{(2)}$ at two different space points, etc.).

$ $From this Section on, we will make a convenient change of notation.
We will denote the original action (i.e. the bare model) studied
here as:
\begin{eqnarray}
&& {\cal S}[u] =  \frac{1}{2 T} \int_q (q^2 + m^2) u^a_q u^a_{-q}
- \int_r V_0[\vec{u}_r],
\label{eq:staticPFnew}
\end{eqnarray}
where the index $0$ denotes the bare model. The function $V(u)$
thus defines the ``renormalized disorder'' and is expanded
as in (\ref{eq:Vexpand}) in terms of the ``renormalized cumulants'' $R$,$S^{(k)}$
while $V_0(u)$ represent the bare disorder, and bare
cumulants $R_0$,$S^{(k)}_0$. A similar expansion for the
bilocal terms, and higher, will be considered below.

\subsection{exact RG}

The aim is to compute $\Gamma[u]$, e.g. the renormalized cumulants,
as a function of the IR cutoff since some kind of fixed point
behaviour with universality and scaling is expected in that
limit. Remarkably, the effective action satisfies the
{\it exact functional equation}:
\begin{eqnarray}
&& - m \partial_m {\cal V}[u] = m^2 \int_x \sum_a [ G^{-1}_{xa,xa}
- (- \nabla^2 + m^2)^{-1}_{xx}], \nonumber \\
&&  G_{xa,yb} = (- \nabla^2_x + m^2) \delta_{ab} \delta_{xy}
- T \frac{\delta^2 {\cal V}[u] }{\delta u^a_{x} \delta u^b_{y}}
\label{ergeq}
\end{eqnarray}
where the inversion is both in replica and in internal space. This
equation is supplemented by an initial condition, namely in the absence
of any fluctuation (with all loops suppressed) the action and effective
action are equal. For instance, if one takes $m=\infty$, or $\Lambda' =
\Lambda$, in that case:
\begin{eqnarray}
&& \Gamma[u] = {\cal S}[u]  \quad , \quad
{\cal V}[u] = \int_r V_0[\vec{u}_r],
\label{init}
\end{eqnarray}
i.e. the renormalized disorder is equal to the bare one
($V=V_0$, $V^{(2)}=0$, ..).
Thus solving the Exact RG equation (\ref{ergeq}) from
the initial condition (\ref{init}) allows, in principle
to compute $\Gamma[u]$ in the small $m$ limit.

\subsection{correlations from the effective action}

Once the effective action functional $\Gamma[u]$ is known, any
connected correlation function can be computed, since it is the
sum of all connected tree diagrams made with the proper vertices.
This means that it can be computed from expanding
\begin{eqnarray}
&&  \langle u_1 u_2 .. u_{2 p}  \rangle _c =  \langle u_1 u_2 .. u_{2 p}
e^{- \Gamma[u]}  \rangle_c^{\rm tree} \label{eq:tree}
\end{eqnarray}
(indices here mean both space and replicas) and using Wick's theorem,
but (as indicated by the ``tree'' superscript) keeping only the connected tree
diagrams.  The two point function is thus:
\begin{eqnarray}
&&  \langle u^a_{q} u^b_{-q}  \rangle  =  \langle u^a_{q} u^b_{-q}  \rangle _c
= \Gamma^{(2)}(q)^{-1}_{ab} \\
&& = T \frac{1}{q^2 + m^2} \delta_{ab} +
\frac{1}{(q^2 + m^2)^2} \Gamma^{(2)}(q),
\end{eqnarray}
where $\Gamma^{(2)}(q)$ is the off-diagonal disorder at finite momentum. At zero
momentum it is exactly given by the local part $V$ of the renormalized
disorder
\begin{eqnarray}
&& \Gamma^{(2)}(q=0) = -  R''(0)
\end{eqnarray}
as one easily sees that
only two legs entering the $S^{(k)}$ vertex leaves $k-2$ free
replica sums, i.e. a factor $n^{k-2}$. However $\Gamma^{(2)}(q=0) - \Gamma^{(2)}(0)$
at finite momentum is given by the two replica part of the
bilocal part $V^{(2)}$ and higher.
These can be computed order by order in the local part (useful if
this one is of order $\epsilon$). In perturbation theory it contains all 1PI graphs with
a flowing momentum. It is thus in general simpler to
compute only correlations at $q=0$ in presence of a mass
since then one needs only the effective action at $q=0$, i.e the local part
of the renormalized disorder.

The higher correlations can be computed in a similar way.
For instance, schematically one has for a four point function:
\begin{eqnarray}
&&  \langle u_1 u_2 u_3 u_4 \rangle  - (  \langle u_1 u_2 \rangle
\langle u_3 u_4 \rangle  + \text{2 perm} ) \nonumber \\
&&
= \int_{1',2',3',4'}
 \langle u_1 u_{1'} \rangle   \langle u_2 u_{1'} \rangle   \langle u_3
 u_{1'} \rangle   \langle u_4 u_{1'} \rangle  \nonumber \\
&& \times \Gamma^{(4)}_{1',2',3',4'}.
\end{eqnarray}
More generally in expanding (\ref{eq:tree}) there is thus
a $(\Gamma^{(2)})^{-1}$ on each line of these tree diagrams
and the the vertices can be only $\Gamma^{(4)}, \Gamma^{(6)}, ..$.

Analyzing (and even more so, solving) the ERG equation is not a trivial
task since one has to deal with a functional. One way to analyze it in
$d>0$ is the multilocal expansion scheme (\ref{eq:multilocalexp}), which
we will explore in Section~\ref{sec:conclusion}. We first study it in
$d=0$ since there it reduces to a flow equation for a function.
Interestingly this flow equation is very similar to the Wilson FRG
equation studied in the previous Section. We emphasize though,
that the relation between the cumulants $R$, $S^{(k)}$ (i.e. the {\it
  local} part of the effective action) and the correlation functions
{\it at zero momentum} are identical in any dimension. It is thus
particularly interesting to study the model with a mass, i.e. a
confining well for the interface, and study the correlations of its
center of mass $u_{cdm}$ since many of the $d=0$ formulae then apply.
These exhibit scaling with $m$ and amplitudes which have good limits as
$d \to 0$ (this is particularly interesting for the random field
interface).

\subsection{identities from exact RG and STS}

Before we do so we first establish {\it directly on the correlation
functions} two sets of simple and general identities.

The first is an exact RG identity. The exact RG can indeed
also be formulated directly on correlation functions. Usually it
leads to complicated hierarchies whose meaning is unclear. Here
it yields interesting constraints on the structure of the
low energy excitations. Let us consider the average of any observable $O[u]$
which does not contain the mass $m$ explicitly:
\begin{eqnarray}
&&  \langle  O[u]  \rangle  = \frac{ \int \prod_{a=1}^n Du^a O[u] e^{- {\cal S}[u] } }{
\int \prod_{a=1}^n Du^a e^{- {\cal S}[u] } }
\label{average}
\end{eqnarray}
where ${\cal S}[u]$ defined in (\ref{eq:staticPFnew}) has a simple explicit
$m$ dependence. Taking a derivative yields:
\begin{eqnarray}
&& - m \partial_m   \langle  O[u]  \rangle  = \frac{m^2}{T} \int_x \sum_{f} (  \langle  O[u] u^f_{x} u^f_{x}  \rangle
\nonumber \\
&& -   \langle  O[u]  \rangle   \langle  u^f_{x} u^f_{x}  \rangle   ) \\
&& = \frac{m^2}{T} \int_x \sum_{f}  \langle  O[u] u^f_{x} u^f_{x}  \rangle  \label{ergid}
\end{eqnarray}
where the last equality holds for $n=0$. This identity can be used, e.g. to generate
relations between correlation with $2 p$ and $2 p + 2$ $u$ fields.

The second is the consequence of STS. Let us consider again
(\ref{average}) and
in the numerator let us shift integration by $u^a_x \to u^a_x + v_x$, this yields:
\begin{eqnarray}
&&  \langle  O[u]  \rangle  =  \langle  O[u + v] e^{- \frac{1}{T} \int_x v_x (m^2 - \nabla^2_x) \sum_a u^a_x}  \rangle  \nonumber \\
&& \times
e^{ - \frac{1}{2 T} n \int_x (\nabla v_x)^2 + m^2 v_x^2 }
\end{eqnarray}
To linear order this gives:
\begin{eqnarray}
&& T  \langle  \sum_c \partial_{u^c_x} O[u]  \rangle  = m^2  \langle  O[u] \sum_f u^f_x  \rangle  \nonumber \\
&& -  \langle  O[u] \sum_f \nabla^2_x u^f_x  \rangle  \label{stsid}
\end{eqnarray}
When applied to odd powers of $u$ it generates identities between even ones.

The consequences of these identities will be explored in the following
sections.

\section{Zero dimension: ERG, correlations, droplet
probabilities and thermal boundary layer}
\label{sec:zero-dimension:-erg}

In this section, we consider the special case of zero (internal)
dimension of the manifold ($d=0$).  For this simple limit, many exact
results are known by other methods, providing useful tests of our very
general TBL Ansatz.  Moreover, since the phenomenological droplet theory is
known to be correct in $d=0$, this limit provides a means of further
understanding how droplet physics emerges from the TBLA.  In particular,
we will consider the connection between the TBLA and droplets in detail
for low moment correlators, including tests of universal amplitude
ratios with exactly solvable limits.  Beyond the physical connection of
the TBLA to droplets, we will also use the $d=0$ model to understand the
{\sl matching} problem, more specifically to study the emergence of a
well-defined but {\sl non-analytic} zero temperature RG $\beta$-function
beyond one loop.  The form of this $T=0$ $\beta$-function was postulated
in Ref.~\onlinecite{frg2loop}, but can be confirmed from first
principles only by a matching calculation such as considered here for
$d=0$.

A manifold with zero internal dimensions corresponds to studying a point
particle in $N$ external dimensions.  Specifically, specializing here to
the case $N=1$, we consider the
partition function of a particle in one (external) dimension:
\begin{eqnarray}
&& Z_W = \int_{-\infty}^{+\infty} du e^{ - [\frac{1}{2} m^2 u^2 + W(u)]/T}.
\end{eqnarray}
The particle is confined by a quadratic well and feels a random
potential $W(u)$, often chosen gaussian. For a glass phase to
exist (as $m \to 0$), the correlations should be long range:
\begin{eqnarray}
&& \overline{(W(u_1) - W(u_2))^2} = K(u) \label{Ku} \\
&& = 2 (R_0(0) - R_0(u)) \sim K u^{2\alpha} \nonumber,
\end{eqnarray}
with $\alpha >0$. This can be seen as the $d=0$ limit of
an interface model with long range disorder and a similar mass term.
The following scaling relations are then obtained by replacing $L
\rightarrow 1/m$:
\begin{eqnarray}
&& u \sim m^{- \zeta} \quad , \quad W \sim m^{- \theta} \sim m^{- \alpha
  \zeta}, \\
&& \theta = 2 \alpha/(2 - \alpha), \\
&& \zeta = 2/(2 - \alpha),
\end{eqnarray}
using the STS relation $\theta=2 (\zeta - 1)$. For $\alpha>0$ ($\theta >
0$) one expects a single ground state to dominate and the droplet
picture to be exact. In the marginal case $\theta=0$ there can be a
transition and a glass phase with a Gibbs measure dominated by a few
sites $u_\alpha$, the physics of the Derrida's random energy model and
RSB. This was found for gaussian disorder with logarithmic correlations
$\alpha=0$, $K(u) \sim \ln u$, but also holds for uncorrelated disorder,
provided the distribution has exponential tails.

The case where $W(u)$ is a Brownian process, the so called ``toy model''
with $\alpha=1/2$, has been much studied. It corresponds to the $d=0$ limit of
an interface in random field disorder, with:
\begin{eqnarray}
&& \zeta = \epsilon/3 = 4/3 \quad , \quad \theta= 2/3.
\end{eqnarray}
In this case, powerful real space RG methods, or path integral
techniques, allow one to obtain the full solution for the zero temperature
glass fixed point and compute virtually any universal observable in
statics (ground state and low temperature behaviour) and dynamics (Sinai
model) \cite{sinai,toy,cecile} .  We will use these
exact results for $\alpha=1/2$ to test our methods.

\subsection{exact FRG in $d=0$}

In $d=0$ since there is no space, $\Gamma[u]$ is purely local,
${\cal V}[u] = V(\vec u)$ and thus the ERG equation becomes
\begin{eqnarray}
&& - m \partial_m V(\vec u)  =
n - Tr [\delta_{cd} - \frac{T}{m^2}
\partial_{u_c} \partial_{u_d} V(\vec u)]^{-1} . \nonumber \\
&&
\label{ergeq0d}
\end{eqnarray}
It simply describes the flow of a {\it function} of the replica vector,
the characteristic function $V(\vec u)$ of the renormalized disorder given by (\ref{eq:Vexpand})
in terms of its cumulants
(the trace and inversion
refer to replica indices). One notes that it
is almost identical to the Wilson FRG equation (\ref{eq:Wilsoncp} ), up to
the coefficients $c_p = 1$ here. Here, however, in $d=0$,
it is {\it exact}. There is thus no need for new calculations to obtain the FRG
equations for all cumulants, which are the same as obtained in
Appendix~\ref{sec:gener-struct-frg}.
Here we define the rescaled temperature and cumulants as:
\begin{eqnarray}
&& \tilde{T} = 2 T m^\theta , \label{eq:Trescale}\\
&& R(u) = \frac{1}{4} m^{- 2 \theta} \tilde R(u m^\zeta) , \label{eq:Rrescale}\\
&& S^{(k)}(u_1,\cdots,u_k) = \frac{1}{2^k} m^{- k \theta} \tilde
S^{(k)}(u_1 m^\zeta ,\cdots,u_k m^\zeta) . \label{eq:Srescale}
\nonumber \\
&& \label{relations}
\end{eqnarray}
Now $V(u) = \tilde{V}(u m^\zeta)$ instead of (\ref{eq:Vexpand0})
while $\tilde{V}(u)$ is still given by (\ref{eq:Vexpand3})
in terms of the rescaled cumulants. Then we find that
the ERG flow equations in $d=0$ for $\tilde R,\tilde S^{(3)}$
are exactly the same as (\ref{Wilson2cum} , \ref{Wilson3cum})
setting $\gamma=3/4$ and $\partial_l \to - m \partial_m$
(and for the fourth cumulant $\tilde S^{(4)}$ given in Appendix~\ref{sec:gener-struct-frg}
with, in addition $\gamma'=1/2$). The schematic form of the
FRG equations for all cumulants (\ref{general}), and the TBL form (\ref{generals})
are also valid here. The detailed coefficients can be retrieved
from Appendix~\ref{sec:gener-struct-frg}.

Following the analysis of Section~\ref{sec:bound-layer-with}, we also consider the
TBL form for the rescaled solution of the ERG equation:
\begin{eqnarray}
&& \tilde{V}_l(\vec{u}) = \sum_{p \geq 1} f_{2p} \sum_{a_1,..,a_{2p}}
\tilde u_{a_1} \cdots \tilde u_{a_{2p}} +
  \frac{\tilde{T} _l}{\chi^2}  v( \vec{\tilde u} ) +
O(\tilde{T} _l^2) \nonumber \\
&& \vec{\tilde u} = \chi \vec{u}/\tilde{T} _l
\end{eqnarray}
with the same definition (\ref{f2p}) for the $f_{2p}$
and for the TBL cumulants. The only difference is
that here we define $(\epsilon \chi)^2 = \chi^2
\equiv - (2 - \theta) \tilde R^{* \prime \prime}(0)$.
We expect a fixed point value for $R_l^{\prime \prime}(0)$
but here, in $d=0$, $\epsilon =4$ is not a small parameter
any more and there is no longer an obvious relation between
$\tilde R^{* \prime \prime}(0)$  and the $\tilde R^{* \prime \prime \prime}(0^+)$
in the $T=0$ non analytic theory. In fact in
(\ref{Wilson2cum}) we expect the term proportional to
$\tilde S^{(3)}$ to be important everywhere (while it
could be neglected $\sim O(\epsilon^3)$ outside of
the TBL). The exact TBL equations in $d=0$ read:
\begin{equation}
\!\! \sum_{p \geq 1} x_{2p} f_{2p} \sum_{a_1,..,a_{2p}}  \!\!
\tilde u_{a_1} \cdots \tilde u_{a_{2p}} \!\!
+ {\rm Tr} \!
  \left[ \delta^{ab} -  \frac{1}{2} \partial_a \partial_b
    v(\vec{\tilde u}) \right]^{-1} = 0, \label{eq:BL2d0}
\end{equation}
with $x_2 f_2 = 1$, $x_{2p}= - 2 p (\theta-\zeta)$ for $p>1$.

So we now explore the consequences of the ERG equation on the correlation
functions in $d=0$

\subsection{comparison with droplet predictions}

In order to compare with the droplet theory we first discuss briefly
how to calculate physical quantities in a low temperature expansion.
One assumes that the behaviour up to order $T$ can be obtained
by considering no more than two quasidegenerate minima in the rescaled
landscape separated by a distance $x = u m^{-\zeta} \sim O(1)$.
We first introduce, in terms of
the rescaled variables, the probability distribution $P(x_1)$ of the position
$x_1$ of the
absolute minimum, normalized as $\int dx_1 P(x_1) = 1$.
The probability density (conditional on the global minimum being $x_1$)
that the secondary minimum is
located at position $x_2$ with rescaled energy above the ground state
$\epsilon = E m^\theta$, is denoted $P(x_1,x_2,\epsilon)$. Thus one has
\begin{eqnarray}
&& \int dx_2 \int_0^\infty d \epsilon  P(x_1,x_2,\epsilon) = P(x_1) .
\end{eqnarray}
We define the ``droplet probability density''
\begin{eqnarray}
  \label{eq:drop}
  D(x_1,x_2) = P(x_1,x_2,\epsilon=0) .
\end{eqnarray}
Note the symmetry  $D(x_1,x_2) = D(x_2 , x_1)$. This is the
probability density that the absolute minimum is (almost)
degenerate, and does not obey a normalization condition. In the
droplet picture it is finite and non-zero \cite{footnotehat} Let
us illustrate the way to construct the low temperature expansion
(see e.g.  \cite{cecile} for systematics in the toy model). Let us
consider a generic disorder average of product of thermal averages
of any observables of the rescaled position. One starts with
restricting to two wells (more generally $N$):
\begin{eqnarray}
&& \overline{\prod_i  \langle  O_i(x)  \rangle } = \int_0^{+ \infty} dE  \int dx_1 dx_2 P(x_1,x_2,\epsilon) \nonumber \\
&& \times \prod_i (p O_i(x_1)
+ (1-p) 0_i(x_2) ) \\
&& p = \frac{1}{1 + e^{- \epsilon m^{-\theta}/T} }
\end{eqnarray}
$p$ and $1-p$ being the (Gibbs) thermal occupation probabilities
of the two wells. Then one adds and substracts the $T=0$ part, and
introduces $w=\epsilon m^{-\theta}/T$:
\begin{widetext}
\begin{eqnarray}
\label{dropcalc}
&&  \overline{\prod_i  \langle  O_i(x)  \rangle } = \int dx_1 P(x_1)  \prod_i O_i(x_1) \\
&& + T m^{\theta}
\int_0^{+ \infty} dw \int dx_1 dx_2 P(x_1,x_2, T m^{\theta} w)
( \prod_i (p O_i(x_1) + (1-p) 0_i(x_2) ) - \prod_i O_i(x_1)) + O(T^2
m^{2 \theta}) ,  \\
&& p = \frac{1}{1 + e^{-w} } .
\end{eqnarray}
\end{widetext}
We have assumed that the restriction to two wells is valid, that
the thermal width of the packet in each well is small compared to
the distance between the wells. This width results in subleading
corrections \cite{f2}.

For small $m$ one gets the lowest order droplet result:
\begin{widetext}
\begin{eqnarray}
&&  \overline{ \langle  O(x)  \rangle } = \int dx_1 P(x_1) \prod_i O_i(x_1) \\
&& + T m^{\theta}
\int_0^{+ \infty} dw \int dx_1 dx_2 D(x_1,x_2)
( \prod_i (p O_i(x_1) + (1-p) 0_i(x_2) ) - \prod_i O_i(x_1))  + O(T^2 m^{2 \theta})
\end{eqnarray}
\end{widetext}
provided the integral converges, but it usually does since the
one well part has been substracted. Generalization to $N$
wells is studied in \cite{cecile}.

A first interesting constraint on these droplet functions is obtained
from STS.  The identity from Section~\ref{sec:exact-rg-general}
specialized to $d=0$ gives in replicas
\begin{eqnarray}
&& T  \langle  \sum_c \partial_{u^c} O[u]  \rangle  = m^2  \langle
O[u] \sum_f u^f  \rangle   \label{stsid0}
\end{eqnarray}
where $O(u)$ is any odd function.  For the particular case where
$O(u_a)$ is a function of a single replica, one directly obtains (for
$n=0$)
\begin{eqnarray}
  \label{eq:STSonesamp}
  T  \overline{\langle \partial_{u} O[u]  \rangle}  = m^2  (\overline{\langle  O[u]
u  \rangle}-\overline{\langle O[u]\rangle \langle u\rangle}) .  \label{stsidone}
\end{eqnarray}
To $O(T)$, the expectation value on the left hand side can be evaluated
directly at $T=0$, while the right hand side is determined from
averaging with respect to $D(x_1,x_2)$.  This gives in rescaled variables
\begin{eqnarray}
  \label{eq:STS1a}
&&   \int dx_1 O'(x_1) P(x_1) = \\ && \overline{2p(1-p)} \int\! dx_1 dx_2\,
  (x_1-x_2)D(x_1,x_2) O(x_1), \nonumber
\end{eqnarray}
for any odd $O(x_1)$.  Thus one finds the implication of STS within the
droplet picture:
\begin{eqnarray}
  \label{eq:STSdrops}
  && - P'(x_1) = \int dx_2 (x_1 - x_2) D(x_1 , x_2).
\end{eqnarray}
As discussed in the Appendix, this is the most general relation
between droplet probabilities that one can extract from STS. Upon
multiplication by odd functions (e.g. powers) of $x_1$ and
integration it generates an infinite series of identities between
zero temperature observables and finite temperatures ones.

One may also obtain an RG equation for the droplet probabilities.  One
considers the energy difference
\begin{eqnarray}
  \label{eq:energydiff}
  E = \frac{m^2}{2}(u_2^2-u_1^2) + V(u_2)-V(u_1)
\end{eqnarray}
between the two minima.  Upon an infinitesimal variation of $m$,
the total derivative comes only from the explicit depedence on
$m$, since the contribution from the implicit variation of
$u_1,u_2$ with $m$ vanishes using the minimization condition.
Thus, assuming no creation or annihilation of additional minima,
one obtains taking into account rescaling
\begin{widetext}
\begin{eqnarray}
  \label{eq:RGpdf}
&&  -m\partial_m P(x_1,x_2,\epsilon) = \big[2\zeta+\theta + \zeta x_i
  \partial_{x_i} + \theta \epsilon\partial_\epsilon -
  (x_1^2-x_2^2)\partial_\epsilon \big]P(x_1,x_2,\epsilon).
\end{eqnarray}
\end{widetext}
for $\epsilon>0$. \cite{footnotehat} Integrating this relation
over $\epsilon$ and $x_2$, and assuming $m\partial_m P(x_1)=0$
gives
\begin{eqnarray}
  \label{eq:RGdrop}
  && - \zeta (P(x_1) + x_1 P'(x_1) ) = \int dx_2 (x_1^2 - x_2^2 ) D(x_1 , x_2). \nonumber \\
&&
\end{eqnarray}
This relation (\ref{eq:RGdrop}) obtained within the droplet
picture from the above RG equation, can also be obtained by a
similar method to that used above for STS, from the more general
ERG relation (not itself assuming droplets)
\begin{eqnarray}
&& - m \partial_m   \langle  O[u]  \rangle  = \frac{m^2}{T} \sum_{f}
\langle  O[u] u^f u^f  \rangle,  \label{ergid0}
\end{eqnarray}
a particular case of the ERG equations, written directly on the
correlation functions given in Section~\ref{sec:exact-rg-general}.
It also generates another infinite set of identities between zero
temperature observables and finite temperature ones, also
discussed in the Appendix.

We now compare the droplet predictions for the correlation functions to
that of the ERG.  We begin with those of the droplet model.

\subsubsection{correlation functions from the droplet theory}

At zero temperature we are concerned with the sample to sample
fluctuations of the location of the global minimum.  The simplest
such quantity is the variance:
\begin{eqnarray}
  \label{eq:zerot2pt}
  && \overline{  u^2 }_{T=0} = m^{-2\zeta} \langle x_1^2\rangle_P,
\end{eqnarray}
where we defined the average
\begin{eqnarray}
  \label{eq:Pavg}
&&  \langle O(x_1)\rangle_P = \int dx_1\, P(x_1) O(x_1).
\end{eqnarray}
The non-Gaussian nature of the distribution $P(x_1)$ is probed by the
four-point correlator
\begin{eqnarray}
  \label{eq:zerot4pt}
  && \overline{u^4} - 3 \overline{u^2}^2  = m^{-4\zeta} (\langle
  x_1^4\rangle_P -3\langle x_1^2\rangle_P^2).
\end{eqnarray}

Other correlation functions probe the thermal fluctuations of the
particle. In particular one can define the sample dependent
susceptibility $\chi_s = \langle u^2 \rangle  - \langle u \rangle
^2$ and its moments. Although it is negligible in a typical
disorder configuration, it becomes large when there are two
quasi-degenerate minima with $\chi_s =m^{-2 \zeta} p(1-p)
(x_1-x_2)^2$. Its disorder average is the second cumulant of
thermal fluctuations
\begin{eqnarray}
&& \overline{  \langle u^2 \rangle  -  \langle u \rangle ^2 }
\simeq \frac{T}{2}
m^\theta m^{-2\zeta} \int\! dx_1 dx_2\, (x_1-x_2)^2 D(x_1,x_2) \nonumber \\
&& = T/m^2 \label{u2thermal}
\end{eqnarray}
We used (\ref{dropcalc}) and $\overline{p(1-p)}=\frac{1}{2}$ (see
Appendix~\ref{sec:details-droplets}). The symbol $\simeq$ (here
and throughout this section) indicates that the first term on the
right hand side of the first line is intended to give correctly
the amplitude of the $T$-linear term. In fact, as is well known,
this correlator is fully constrained by STS to give exactly the
free propagator result. Within the droplet calculation this can be
obtained from (\ref{eq:STSdrops}) upon multiplication by $x_1$ and
integration. For the present model (confined by a harmonic well)
it is exact at any $T$ (i.e. all higher order corrections cancel).

At quartic order one may define several thermal correlation
functions. One of them is the second moment of $\chi$:
\begin{eqnarray}
&& \overline{ (  \langle u^2 \rangle  -  \langle u \rangle ^2 )^2}
\simeq \frac{T}{12} m^{\theta-4\zeta} \int\! dx_1 dx_2\,
(x_1-x_2)^4 D(x_1,x_2) \nonumber
\end{eqnarray}
(see Appendix~\ref{sec:details-droplets} for details). There is
one other combination
\begin{eqnarray}
&& \overline{\langle (u -  \langle u \rangle  )^4 \rangle } \simeq
\frac{T}{4} m^{\theta-4\zeta} \int\! dx_1 dx_2\, (x_1-x_2)^4
\nonumber D(x_1,x_2)
\end{eqnarray}
from the droplet calculation. These two combinations are however,
related by STS which implies that:
\begin{eqnarray} \label{stsu4}
&& \overline{ \langle (u -  \langle u \rangle )^4 \rangle } = 3
\overline{ ( \langle u^2 \rangle  -  \langle u \rangle ^2 )^2 }
\end{eqnarray}
which again is an exact relation valid for all $T$ (see
Appendix~\ref{sec:details-droplets} and \cite{cecile}). Although
it cannot be derived from (\ref{eq:STSdrops}), the droplet theory
is clearly compatible with it. It in fact entails a property
\cite{f1} of the various moments of $p$.

Remarkably the RG, combined with STS, allows to relate these
quartic $T>0$ observables to quadratic ones at $T=0$. Here we can
see it from the droplet theory. Indeed multiplying
(\ref{eq:STSdrops}) by $8 x_1^3$, (\ref{eq:RGdrop}) by $6 x_1^2$
and subtracting we obtain:
\begin{equation} \label{relation}
\frac{1}{12} \int dx_1 dx_2 (x_1-x_2)^4 D(x_1,x_2)=
\frac{2-\theta}{2} \langle x_1^2 \rangle_P
\end{equation}
A fixed point has been assumed. Similar relations exist at the
fixed point between $2 n$ and $2 n-2$ observables for any $n \geq
2$.

Finally, it is interesting to look for a combination which should
have contributions only from three wells. Within the droplet
theory the coefficient of $T$ should vanish and thus this
correlation can be used to test droplet theory. The simplest one
occurs at sixth order and reads:
\begin{eqnarray} \label{d3}
&& D_3 = \frac{1}{6}  \langle (u_a - u_b)^2 (u_b - u_c)^2 (u_c - u_a)^2 \rangle  \\
&& = 2 \overline{ \langle u^3 \rangle \langle u^2 \rangle \langle
u \rangle} - \overline{ \langle u^2 \rangle^3 }  + \overline{
\langle u^4 \rangle ( \langle u^2 \rangle - \langle u \rangle^2 )
} - \overline{ \langle u^3 \rangle^2} \nonumber
\end{eqnarray}
this is the lowest order necessary to see $3$ well physics, since
the combination $ \langle (u_a - u_b)^2 (u_a - u_c) (u_b - u_c)
\rangle $ is zero by symmetrization (here $a,b,c$ denote any
triplet of distinct replicas). Thus the prediction of the droplet
theory is that:
\begin{eqnarray}
&& D_3  \simeq O(m^{2 \theta} T^2)
\end{eqnarray}

\subsection{correlation functions from the ERG}

In this section we give the relation between a set of correlation
functions of $u$ and the $m$-dependent cumulants $R$,$S^{(k)}$,
using the properties of the effective action summarized above in
Section~\ref{sec:exact-rg-general}. Details can be found in the
Appendix \ref{sec:corr-funct}. We stress, as discussed above, that
the relations given here between the $R$,$S^{(k)}$ and the
correlations are {\it exact}, not only in $d=0$, but in any
dimension (if one considers zero momentum $u \to u_{q=0}$, i.e.
center of mass). The rescaling and, of course, the fixed point
values, does however depend on $d$ (see
Sections~\ref{sec:bound-layer-with} and \ref{sec:conclusion}).
Next, we insert the expected asymptotic flow of the cumulant for
$m \to 0$ using the TBL ansatz and discuss consequences. The
resulting relations are then also expected to be ''exact'' but
with the following disclaimers (i) they are subjected to the TBL
assumptions and (ii) there are accurate up to subleading
corrections in $m$. For them we will use the symbol $\simeq$ and
will, here, reserve the symbol $=$ for relations which are exact
by construction

\subsubsection{two point function}

We start with the two point function:
\begin{eqnarray} \label{twopoint}
&&  \langle u^a u^b  \rangle  = T \frac{1}{m^2} \delta_{ab} - \frac{1}{m^4}  R''(0)
\end{eqnarray}
which contains a zero temperature part, and a thermal connected
part which form is dictated by STS. Higher correlations will have
similar properties and we will explicitly write them up to six-th
order. To classify them it is convenient, from now on in this
Section, to use the convention that {\it different replica indices
mean non equal replicas}. Then one has
\begin{eqnarray}
&&  \langle  u_a^2  \rangle  = \overline{  \langle u^2 \rangle  } = T \frac{1}{m^2} - \frac{1}{m^4}  R''(0) \label{2pt1} \\
&&  \langle  u_a u_b  \rangle  = \overline{  \langle u \rangle ^2 }  = - \frac{1}{m^4}  R''(0) \label{2pt2}
\end{eqnarray}
The disorder cancels between the two lines and the second cumulant
of thermal fluctuations is thus exactly given by (\ref{u2thermal})
above.

The TBLA gives, using Eq.~(\ref{eq:Rrescale}),
\begin{eqnarray}
&& \overline{  \langle u \rangle ^2 } \simeq  - \frac{1}{4} m^{- 2 \zeta} \tilde{R}^{*
  \prime \prime}(0).
\end{eqnarray}
This gives the second moment of the position of the absolute minimum.

\subsubsection{four point function}

The calculation of the four point function requires only one
$\Gamma^{(4)}$ vertex. Since $S^{(3)}$ starts as $u^6$ only $R$
and $S^{(4)}$ can give a contribution. The calculation is
performed in Appendix \ref{sec:corr-funct}. There are five
possible replica monomials. Their connected expectation values
read:
\begin{eqnarray} \label{conn4}
&& \! \! \!  \langle  u_a^4  \rangle _c = -  \frac{T^2}{m^8} R^{(4)}(0)
+ \frac{F_4}{m^8}  \\
&& \! \! \!   \langle  u_a^3 u_b  \rangle _c = - \frac{T^2}{m^8} R^{(4)}(0)
+   \frac{F_4}{m^8} \\
&& \! \! \!   \langle  u_a^2 u_b^2  \rangle _c =  \frac{T^2}{m^8} R^{(4)}(0)
+  \frac{F_4}{m^8} \\
&& \! \! \!   \langle  u_a^2 u_b u_c  \rangle _c =  \frac{F_4}{m^8} \\
&&  \! \! \!  \langle u_a u_b u_c u_d  \rangle _c = \frac{F_4}{m^8}
\end{eqnarray}
we denote the fourth cumulant of the random force $F_4 =
S^{(4)}_{1111}(0,0,0,0)$. To obtain their full expectation values
$ \langle  O  \rangle  =  \langle  O  \rangle _c +  \langle  O
\rangle _{disc}$ one must add their disconnected parts, given in
(\ref{disc4}) in terms of the two points averages (\ref{2pt2}),
and which contain only $T$ and $R''(0)$.

One sees that these five connected correlations depend only on two
independent quantities $F_4$ and and $R''''(0)$. The first one
measures the fourth order cumulant of the displacement at zero
temperature. Since in that limit all replica averages in
(\ref{conn4}) ( and in (\ref{disc4})) coincide
\begin{eqnarray}
&& \overline{u^4} - 3 \overline{u^2}^2 \simeq_{T=0} m^{- 4 \zeta}
\frac{1}{16} \tilde{S}^{(4)*}_{1111}(0,0,0,0) \label{4conn}
\end{eqnarray}
in terms of the rescaled quantities, assuming a fixed point.
Comparing with (\ref{eq:zerot4pt}) and generalizing, one finds
that the full distribution $P(x)$ of the position of the absolute
minimum, is retrieved from the ERG fixed point from its cumulants:
\begin{eqnarray}
&& \langle x^{2p} \rangle_P^c = \frac{1}{2^p} \tilde{S}^{(2p)*}_{1
\cdots 1}(0,\cdots,0)
\end{eqnarray}

At $T>0$ from (\ref{conn4}) there is only one independent
additional observable (as discussed in the Appendix
\ref{sec:corr-funct} this can be equivalently seen from the
STS-ERG relations). One finds:
\begin{eqnarray}
&& \overline{\chi_s^2} =\overline{ (  \langle u^2 \rangle  -
\langle u \rangle ^2 )^2 } =
\frac{1}{2}  \langle (u_a - u_b)^2 (u_c - u_d)^2  \rangle \nonumber  \\
&& =  \langle  u_a^2 u_b^2 - 2 u_a^2 u_b u_c + u_a u_b u_c u_d  \rangle   \\
&& =  \frac{T^2}{m^4} +  \frac{T^2}{m^8} R^{(4)}(0) \nonumber
\end{eqnarray}
The fourth cumulant $F_4$ cancels in this combination and $R''(0)$
also cancels in the disconnected part (see Appendix
\ref{sec:corr-funct}). In terms of the rescaled disorder the
variance of the susceptibility fluctuations becomes:
\begin{eqnarray}
&& \overline{\chi_s^2} - \overline{\chi_s}^2 = \frac{T^2}{m^8}
R^{(4)}(0) = \frac{T^2}{m^{4}} \frac{1}{4} \tilde{R}^{(4)}(0) .
\label{R40}
\end{eqnarray}
At this stage assuming a fixed point for $\tilde{R}^{(4)}(0)$
would imply conventional thermal scaling for the fluctuations
$\delta \chi_s = O(T)$. The TBLA allows to evade that, as
$\tilde{R}^{(4)}(0)$ blows up as $m \to 0$ with
$\tilde{R}^{(4)}(0) \simeq  \chi^2 r''''(0)/\tilde{T}$ with
$\tilde{T} = 2 T m^\theta$, which leads instead to much larger
fluctuations:
\begin{eqnarray}
&& \overline{\chi_s^2} - \overline{\chi_s}^2  \simeq \frac{1}{8}
(2-\theta) (-\tilde R^{*''}(0)) (T m^\theta) m^{- 4 \zeta} ,
\end{eqnarray}
using the exact result $r''''(0)=1$. The scaling coincide with the
result above, and the amplitude too, given the relation
(\ref{relation}) which exists at the fixed point
\begin{eqnarray}
&& \overline{\chi_s^2} - \overline{\chi_s}^2 \simeq  T \frac{2 -
\theta}{2} \overline{ \langle u^2 \rangle } m^{-2 \zeta}
\end{eqnarray}
It should be stressed that this relation originates from the exact
ERG identity:
\begin{eqnarray} \label{erg20}
&&  - m \partial m \tilde{R}''(0) = (2 -  \theta) \tilde{R}''(0) +
\tilde{T}_l \tilde{R}''''(0)
\end{eqnarray}
assuming that $\tilde{R}''(0)$ reaches a limit. Both droplets and
the TBLA are consistent with this exact relation. Note that
similar identities for the $2p$-th cumulant where shown from the
ERG (\ref{rel2p}) and result in relations similar to (\ref{erg20})
for higher order observables (see Appendix~\ref{sec:corr-funct}).

To check further consistency between the TBL and droplets, we now
compute the three well quantity $D_3$ defined in (\ref{d3}) above.
The calculation of the six point functions is performed in the
Appendix~\ref{sec:corr-funct}. We find:
\begin{eqnarray}
&&  D_3 =  \frac{1}{32} m^{- 6 \zeta}  \tilde T^3 (
6 \tilde R^{(4)}(0) + 2 \tilde T \tilde R^{(6)}(0) \nonumber \\
&& +  \tilde S^{(3)}_{222}(0,0,0)
+  3 \tilde R^{(4)}(0)^2 )
\end{eqnarray}
in rescaled variables. We note that from the TBLA $\tilde T \tilde
R^{(6)}(0) \sim 1/\tilde T^2$ as are all last three terms, so
naively $D_3 \sim \tilde T$. However this is not so, as we find
that these three terms actually cancel to leading order when using
the ERG equation. Indeed using the fourth derivative in zero of
the FRG equation (\ref{Wilson2cum}) for $\tilde{R}$, it simplifies
to:
\begin{eqnarray}
&& D_3 = \frac{1}{32} m^{- 6 \zeta}  \tilde T^3 ( 2 R^{(4)}(0) - m \partial_m  R^{(4)}(0) )
\\
&& \simeq  \frac{1}{32} (4 - \theta^2) (- \tilde R^{*\prime
\prime}(0) ) m^{- 6 \zeta} \tilde T^2
\end{eqnarray}
which is indeed $O(\tilde T^2)$. It is likely that obtaining the
correct prefactor here requires considering the next order
correction to the TBL. But clearly the TBL solution of the ERG
knows that $D_3$ is a three well quantity. Finally, the third
cumulant of the susceptibility is also considered in the Appendix
and shown to relate to the third cumulant TBL function $\tilde
s^{(3)}_{222}(0,0,0)$.

\subsection{comparison with the solution of the toy model}

It is interesting that one case of the model studied here, the
so-called toy model, has been exactly solved. In that case:
\begin{eqnarray}
&& u \sim m^{-4/3} \sim x^{4/3} \\
&& V(u) \sim u^{1/2} \sim u^{\theta/\zeta} \\
&& \zeta = 4/3 \quad , \quad \theta = 2/3
\end{eqnarray}
Correlations and probability distributions are known, or
accessible, and one can infer from them some characteristics of
the fixed point, e.g. the (universal) numerical value of $\tilde
R^{* \prime \prime}(0)$ (see below) etc..

We can in particular verify that the solution of the toy model
obtained in Ref. \cite{toy} obeys the exact relations from STS and
ERG obtained here. Let us start with the simple relation between
fourth and second moments. With the choice $K(u)=2 \sigma |u|$ at
large $u$ in (\ref{Ku}), with $\sigma=1$, one has at $T=0$ in the
limit $m \to 0$:
\begin{eqnarray}
&& \overline{ \langle u^2 \rangle } = -  \frac{R''(0)}{m^4} = c_2 m^{-8/3} \\
&& c_2 = - 2^{4/3} \int_{- \infty}^{+\infty} \frac{du}{2 \pi} \frac{Ai'(i u)^2}{ Ai(i u)^4} \\
&& =
 \frac{2^{4/3}}{3} \int_{- \infty}^{+\infty} \frac{du}{2 \pi} \frac{ - i u }{ Ai(i u)^2} =
1.05423856519
\end{eqnarray}
This implies:
\begin{eqnarray}
&& - \tilde R^{* \prime \prime}(0) = 4 c_2
\end{eqnarray}

On the other hand one has \cite{toy}:
\begin{eqnarray}
&& \overline{ (u -  \langle u \rangle  )^4 } = 3 \overline{ (  \langle u^2 \rangle  -  \langle u \rangle ^2 )^2 } \\
&& = \frac{T}{2} \mu_{4} m^{-14/3} \\
&& \mu_{4} = 2^{7/3} \int_{\infty}^{+\infty} \frac{du}{2 \pi} \frac{1}{ Ai(i u)^2}
( \frac{Ai'(i u)}{ Ai(i u)} )''''
\end{eqnarray}
The above relation (\ref{R40}) between this quantity and
$\tilde R^{(4)}(0)$ implies that one must have:
\begin{eqnarray}
&& \tilde R^{(4)}(0) = \frac{4}{3} \mu_4 \frac{1}{\tilde{T}}
\end{eqnarray}
where $\tilde{T} = 2 T m^\theta$. This is clearly
consistent with the TBL ansatz. Thus for the above prefactor to
be correct one needs the non trivial relation:
\begin{eqnarray}
&& \mu_{4} = 4 c_2
\end{eqnarray}
which we have confirmed numerically.

One can go further, since the full distribution $P(x_1)$ and the
droplet distribution $D(x_1,x_2)$ were obtained for the toy model.
First one notes that if $\hat P(x_1)$ and $\hat D(x_1,x_2)$
are solution of (\ref{eq:STSdrops}) and (\ref{eq:RGdrop})
then:
\begin{eqnarray}
&& P(x_1) = \rho \hat P(\rho x_1)  \\
&& D(x_1,x_2) =\rho^4  \hat D(\rho x_1,\rho x_2) \label{rescrho}
\end{eqnarray}
are also solutions (they correspond to different choices for
$\sigma$ and the rescaling factor of $u$). One member of this
family (corresponding to the choice $\rho=1$) is:
\begin{eqnarray}
&& P(x_1) = g(x_1) g(- x_1) \\
&& D(x_1,x_2) = \hat D(x_1,x_2) \theta(x_2 - x_1) \\
&& +
\hat D(x_2,x_1) \theta(x_1 - x_2) \\
&& \hat D(x_1,x_1+y) = 2 g(-x_1) d(y) g(x_1 +  y)
\label{shape}
\end{eqnarray}
with the specific forms:
\begin{eqnarray}
&& g(x) = \int_{-\infty}^{+\infty} \frac{d\lambda}{2 \pi}
e^{- i \lambda x} \frac{1}{Ai(i \lambda)} \\
&& d(y) = \int_{-\infty}^{+\infty} \frac{d\lambda}{2 \pi} e^{i \lambda y}
\frac{Ai'(i \lambda)}{Ai(i \lambda)} =
\sum_{s=1}^{\infty} e^{- |a_s| y}
\end{eqnarray}
in terms of the zeroes $a_s$ of the Airy function. Comparing with
\cite{toy} one further sees that the choice considered above
corresponds to $\rho=2^{-2/3}$.

We check in Appendix~\ref{sec:details-droplets} that these
functions indeed satisfy the relations (\ref{eq:STSdrops}) and
(\ref{eq:RGdrop}). An outstanding question is whether these
relations could allow some analytical progress in the other cases.

To conclude this Section we have established a precise mapping in
$d=0$ between droplet probabilities and some thermal boundary
layer quantities. Questions which deserve further investigation
are: (i) how much constraints the FRG-STS equations give on the
set of all thermal boundary layer functions (associated to all
disorder cumulants) (ii) whether $D(x_1,x_2)$ may contain all the
information encoded in this set (iii) what part of these functions
is universal. Answers to these questions in $d=0$ would certainly
be very helpful to understand the general case. Before discussing
$d>0$, we turn to the question of how the $T=0$ information,
namely $P(x_1)$, can be retrieved from the TBL study. This
necessitates considering the so-called matching problem.

\section{Matching}
\label{sec:matching}

In this section, we use the ERG method for $d=0$ developed in the
previous section to consider the matching problem in some detail.  As
discussed in the previous section, most physical quantities are formally
determined by derivatives of coupling functions in the effective action
evaluated at zero argument.  As such, they are defined deep within the
TBL.  Physically, however, one can divide these quantities into those
describing zero temperature (sample-averaged ground state) properties
and alternatively thermal ($T>0$) fluctuations.  While it is natural
that the latter are determined by the TBL, which, as we have discussed,
is an encapsulation of droplet physics, it is surprising for the former
zero temperature quantities to require a TBL analysis.  Thus one is led
to expect some sort of {\sl matching} of certain derivatives of coupling
functions at strictly zero argument (and $T\rightarrow 0^+$) to
corresponding limits as $u_{ab} \rightarrow 0^+$ of $T=0$ functions.
The subtlety in the latter definition is the order in which the limit of
coinciding points should be taken.  It will naturally emerge from a
careful study of the matching.

Unlike for $R(u)$, all higher cumulants are described by functions of
more than two separate points.  As the outer solution for such cumulants
is non-analytic whenever any two points are brought nearby, each such
non-analyticity must be resolved by some boundary layer
structure. Therefore for each cumulant we will introduce a set of
{\sl Partial Boundary Layers (PBLs)}, one for each distinct way of
bringing subset(s) of points together. We will adopt a transparent
notation to specify the distinct PBLs. Each PBL consists of a set of $q$
groups, all $n_i$ points within group $i$ having been brought together.
Thus for a given $k$-th cumulant there is one PBL for each
integer partition of $k$, $\sum_{i=1}^{q} n_i=k$. We label
the corresponding PBL functions by the integers $n_i$ in
decreasing order. For instance $s^{(3)}$ is the full
TBL introduced previously and $s^{(21)}$ is the only PBL associated
with the third cumulant, and we will denote simply the outer solution
$s^{(111)}=\tilde S^{(3)}$.

We will illustrate this procedure explicitly for the third and fourth
cumulants, making a full consistency check for the third and a partial
check on the fourth. In the end this construction enables and justifies
a expansion in powers of $R$ of the fixed point equation for the
function
$R(u)$, i.e. the beta function of the model. For
comparison with the $\epsilon$ expansion, we will keep
$\epsilon=4-d=4$ factors unsimplified.

\subsection{analysis of third cumulant}

The necessity of understanding matching in some detail is evident from
inspection of the equation (\ref{Wilson2cum}) for $\tilde{R}(u)$.  We
have already noted the feedback of the third cumulant in this equation,
but have not emphasized that even for $u$ of $O(1)$, this feedback term
$\tilde S^{(3)}_{110}(0,0,u)$ is required for two arguments {\sl
  coincident} with one another.  Since the outer solution is
non-analytic whenever pairs of arguments coincide, we should expect some
kind of boundary layer softening of this singularity to render $\tilde
S^{(3)}_{110}(0,0,u)$ well defined.  Because, however, $u$ remains of
$O(1)$, this requires a new {\sl Partial Boundary Layer (PBL)}
treatment.  The tantalizing question is how such a PBL analysis can
recover a sensible zero temperature limit, in which na\"ively $\tilde
S^{(3)}_{110}(0,0,u)$ above is replaced by some quantity defined
entirely outside all BLs.

We investigate the PBL for the third cumulant through its ERG
equation (\ref{Wilson3cum}).
The PBL will be taken to correspond to
$\tilde{u}_{12}=\epsilon\tilde\chi u_{12}/\tilde{T}_l=O(1)$,
$u_{13}=O(1)$, and $\tilde{T}_l\ll \epsilon^2$ as usual.  We would like
to constrain the PBL behavior by assuming that $\tilde S^{(3)}_{110}(0,0,u)$ is
$O(T^0)$ (and $O(\epsilon^3)$ hopefully in the $\epsilon$ expansion),
and moreover is quadratic in $u$ at small argument. It clearly should be,
so as to correct the fixed point value of $\tilde R''(0)$
compared to a lowest order truncation. A na\"ive scaling
ansatz would then be $\tilde S^{(3)}(u^{123}) \sim \tilde{T}_l^2 \chi
s^{(21)}(\tilde{u}_{12},u_{13})$.  However, it is readily seen that
this is inconsistent in the above equation, because then the thermal
term would be much larger than all others (i.e. $O(\tilde{T}_l)$ as
compared to $O(\tilde{T}_l^2)$).  Instead, the only way to consistently
obtain the proper behavior of $\tilde S^{(3)}_{110}(0,0,u)$ is to postulate a
{\sl purely quadratic} in $\tilde{u}_{12}$ term with the aforementioned
$\tilde{T}_l^2 \chi$ scaling, a fully complex PBL term appearing only at
$O(\tilde{T}_l^3 \chi)$, i.e.
\begin{eqnarray}
&& \hspace{-0.15in}  \tilde{S}^{(3)}(u_1,u_2,u_3)  \sim   \tilde{T}_l^2 \chi
  \tilde{u}_{12}^2 \phi(u_{13}) + \tilde{T}_l^3
  s^{(21)}(\tilde{u}_{12},u_{13}) \nonumber \\
&& + O(\tilde{T}_l^4) ,  \qquad
 \tilde{u}_{12} = O(1) \quad , \quad u_{13} = O(1)
. \label{eq:PBLscaling}
\end{eqnarray}
where $\phi$ and $\hat S$ are order $O(1)$ and even functions.
The function $\phi$ so defined is particularly significant because
it gives the feedback of the third cumulant into the second:
\begin{eqnarray}
&& \tilde S^{(3)}_{110}(0,0,u)=-2\chi^3\phi(u) \label{feedphi} .
\end{eqnarray}
Note that while $\phi$ does not depend on the choice of
parameterizing using $\tilde{u}_{12},u_{13}$ rather than
$\tilde{u}_{12},u_{32}$, the function $s^{(21)}(\tilde{u}_{12},u_{13})$
depends on this choice (as well as $\psi(u)$ defined below).

In term of the function $\phi(u)$ the outer equation for the second
cumulant thus reads:
\begin{eqnarray}
\label{eq:rout}
  && \partial_l \tilde{R}(u) = 0 =
  (\epsilon-4\zeta+\zeta u \partial_u) \tilde{R}(u) \\
&& +
  \frac{1}{2} \tilde{R}''(u)^2 - \tilde{R}''(0) \tilde{R}''(u) - 2
  \chi^3 \phi(u) .  \nonumber
\end{eqnarray}
Expanded to $O(u^2)$ it yields the exact relation:
\begin{eqnarray}
\label{r3inf}
  && {\sf R}'''(0^+)^2 = \chi^2 (1 + 2 \chi \phi''(0^+)) .\\&& \nonumber
\end{eqnarray}
Here and below we will omit the $^*$ to denote fixed point quantities.

If one considers the large $\tilde{u}_{12}$ limit, one may expect to
match to the outer solution.  In particular, taking the small $u_{12}$
limit of the outer solution, we expect (by a choice of gauge, see
Appendix \ref{sec:symm-tayl-expans}):
\begin{eqnarray}
&&  \tilde S^{(3)}(u_{123}) \sim \chi^3 u_{12}^2 \phi(u_{13}) + \chi^3 |u_{12}|^3
  \psi(u_{13}) + O(u_{12}^4)
\nonumber
\\ && u_{12} = O(1) \quad , \quad u_{13} = O(1) . \label{eq:outermatchpbl}
\end{eqnarray}
Matching the two forms is an extremely strong constraint, and completely
determines $\phi(u)$ from the outer solution (outside all BLs) -- for
this reason we have used the {\sl same} function $\phi(u)$ in
Eqs.~(\ref{eq:PBLscaling},\ref{eq:outermatchpbl}).  This fact explains
the conceptual dilemma mentioned above: although
$\phi(u)$ is a quantity defined inside the
PBL, it is wholly determined by the zero temperature solution.  Thus the
zero temperature limit of Eq.~(\ref{Wilson2cum}) is well-defined.  As a
check on this matching requirement, we will verify that the equation for
$\phi(u)$ obtained by analysis entirely with the PBL is equivalent to
the equation for $\phi(u)$ obtained from the small $u_{12}$ limit of the
outer solution.

Note that all terms in the series in Eq.~(\ref{eq:outermatchpbl}) are
powers of $|u_{12}|$, since any odd power of $u_{12}$ can be converted
to a higher even power by symmetrization.   Thinking
further about this, there should be a hierarchy of such terms, since the
$\phi$ term can only capture the $O(u_{12}^2)$ coefficient of the small
$u_{12}$ limit of the outer solution, the $s^{(21)}$ term only the
$O(|u_{12}|^3)$ etc. In general the coefficient of $O(|u_{12}|^n)$
in the outer solution is matched by the large $\tilde u_{12}$
( $\sim \tilde u_{12}^n$ ) of a generalization of $\hat S$ at
order $\tilde T_l^n$ in the PBL expansion (\ref{eq:PBLscaling}).
In particular one must have:
\begin{equation}
  \label{eq:psimatch}
  s^{(21)}(\tilde{u}_{12},u_{13}) \sim |\tilde{u}_{12}|^3
\psi(u_{13})  , \qquad \tilde u_{12} \to \infty
\end{equation}

In both regimes (inner PBL and outer small
$u_{12}$ limits), the higher and higher terms are progressively smaller,
so we may hope to truncate this successfully.  Unlike at $O(u_{12}^2)$
for $\phi(u)$, at $O(u_{12}^3)$, matching to the outer solution (i.e.
$\psi(u_{13})$) only determines the large $\tilde{u}_{12}$ limiting
behavior of $s^{(21)}(\tilde{u}_{12},u_{13})$, which has the full PBL
complexity otherwise.

We first consider the PBL constraints implied by the
$\tilde S^{(3)}$ equation.  One finds, applying the ansatz of
Eq.~(\ref{eq:PBLscaling}),  that all terms are $O(\tilde{T}^2)$ or
smaller (we do not, as appropriate to $d=0$, assume $\epsilon\ll 1$
for the moment).  Keeping only these terms, and for simplicity dropping
the feeding term from the fourth cumulant, one has
\begin{widetext}

\begin{eqnarray}
  && 0 = \partial_l \tilde S^{(3)}(u_{123}) = \tilde{T}_l^2  \tilde{u}_{12}^2 \chi\big[
  (2\epsilon-2-4\zeta)\phi(u_{13}) + \zeta
  u_{13}\phi'(u_{13})\nonumber \\
  && -\gamma
  \frac{{\sf R}''(u_{13})}{\chi}\left|\frac{{\sf
      }{\sf R}'''(u_{13})}{\chi}\right|^2 + \phi''(u_{13}) {\sf R}''(u_{13}) +
  3\phi'(u_{13}) {\sf R}'''(u_{13})\big]
  \nonumber \\
  & & +  \tilde{T}_l^2 {\sf R}''(u_{13})\big[r''(\tilde{u}_{12}) + \gamma
  (r''(\tilde{u}_{12}))^2 + s^{(3)}_{110}(0,0,\tilde{u}_{12})\big] \nonumber \\
  & & + \tilde{T}_l^2 \chi^2 \big[\left( s^{(21)}_{20}(\tilde{u}_{12},u_{13})
  -s^{(21)}_{20}(0,u_{13})\right) r''(\tilde{u}_{12}) +
s^{(21)}_{20}(\tilde{u}_{12},u_{13})\big] \nonumber \\
&&+ \tilde{T}_l^2  \chi^2 \big[ - \tilde{u}_{12}^2
 \psi^{(31) \prime}(u_{13})
 - \frac{1}{2} s^{(22)}_{020}(\tilde
u_{12},0,u_{13}) + s^{(31)}_{1100}(\tilde u_{112},u_{13})  \big] + O(\tilde T_l^3) . \label{eq:longpbl0}
\end{eqnarray}
\end{widetext}

To obtain this equation we have performed a expansion in small
$u_{12}$ discarding some ''gauge terms'' as detailed in the Appendix.
Note that the dangerous term, the contribution from $\phi$ to
$\tilde T_l \tilde S^{(3)}_{200}$ vanishes being a pure gauge.
For completeness we have included the feedback from the
PBL31 and the PBL22 of the fourth cumulant (last line) which will be discussed
in the next Section, but is not crucial for the following discussion.
In the third line of Eq.~\ref{eq:longpbl0}, we have grouped a contribution
arising from the $\tilde{T}_l[{\sf R}'']^2$, ${\sf R}'' \tilde S_{110}$, and
$[{\sf R}'']^3$ terms in Eq.~(\ref{Wilson3cum}). This is
simplified by eliminating
$s_{110}^{(3)}({{\tilde{u}}_1},{{\tilde{u}}_1},{{\tilde{u}}_2})$
using the full BL equation for the second cumulant:
\begin{equation}
  \label{eq:fulls3}
s_{110}^{(3)}({{\tilde{u}}_1},{{\tilde{u}}_1},{{\tilde{u}}_2}) =
\frac{1}{2} \tilde{u}_{12}^2 - \frac{1}{2} r''(\tilde{u}_{12})^2 -
r''(\tilde{u}_{12})
\end{equation}

The resulting equation can be studied at large $\tilde{u}_{12}$,
using the matching constraints Eq. (\ref{eq:psimatch})
and $r''(\tilde{u}_{12}) \sim
r'''_\infty |\tilde{u}_{12}|$ and the corresponding limit of the
fourth cumulant quantities (discussed below):
\begin{eqnarray}
 && \text{sym}_{12} s^{(31)}_{1100}(\tilde u_{112},u_{13}) \simeq - 2 \tilde
 u_{12}^2 g^{(31)}(u_{13}) \\
 && \text{sym}_{12} s^{(22)}_{020}(\tilde u_{12},0,u_{13}) \simeq 2 \tilde
 u_{12}^2 g^{(22)}(u_{13})
\end{eqnarray}
Collecting all $O(\tilde{u}_{12}^2)$
terms (the dominant ones in this limit), one finds
\begin{widetext}
\begin{eqnarray}
  \label{eq:PBL2}  \nonumber
 &&   (2\epsilon-2-4\zeta)\phi(u) + \zeta
  u\phi'(u) + \frac{{\sf R}''(u)}{\chi}\left(\frac{1}{2} + (\gamma - \frac{1}{2}) (r'''_\infty)^2
-\gamma
    \left|\frac{{\sf R}'''(u)}{\chi}\right|^2\right)
   \nonumber \\ && + \phi''(u){\sf R}''(u) + 3\phi'(u){\sf R}'''(u) +
   6\chi r'''_\infty \psi(u)  \\
&&   - \chi ( \psi^{(31) \prime}(u) + 2 g^{(31)}(u) + g^{(22)}(u)) = 0 \nonumber
\end{eqnarray}
This equation allows to compute $\phi(u)$ in an expansion in powers of
$R$ as we will explicitly show below.
\end{widetext}

Conceptually, one should regard $\phi(u)$ and $\psi(u)$ as determined by the
underlying zero temperature fixed point, since although they obtain in
the PBL, they are entirely constrained by matching. Thus $\phi(u)$ should be
obtainable by considering the outer equation for $\tilde S^{(3)}$:
\begin{widetext}
\begin{eqnarray}
&& 0 = (2 \epsilon-2-6\zeta+\zeta u_i\partial_{u_i})
  \tilde{S}^{(3)}(u_{123}) + 3 \, {\rm sym} \big( {\sf R}''(u_{12}) (
 - 2 \chi^3 \phi(u_{13}) -
    \tilde{S}^{(3)}_{110}(u_{123})) \\
&&  + \gamma {\sf R}''(u_{12}){\sf R}''(u_{13})^2  -
    \frac{\gamma}{3} {\sf R}''(u_{12}) {\sf R}''(u_{23})
  {\sf R}''(u_{13}) \big)
- 3 \chi^4 {\rm sym} \phi^{(211)}(u_{12},u_{13}) \nonumber
\end{eqnarray}
\end{widetext}
where the last term arises from the PBL211 of the fourth
cumulant. Eq.~(\ref{eq:PBL2}) should match the small $u_{12}$ limit of
the {\sl outer} equation. Inserting the expansion of
Eq.~(\ref{eq:outermatchpbl}),
together with fourth cumulant matching conditions (see below) and keeping
terms to $O(u_{12}^2)$.  One finds
\begin{widetext}
\begin{eqnarray}
  \label{eq:outerphieqn}
  && (2\epsilon-2-4\zeta)\phi(u) + \zeta
  u\phi'(u) + \gamma\frac{{\sf R}''(u)}{\chi}\left(\left|\frac{{\sf
          R}'''(0^+)}{\chi}\right|^2   -
    \left|\frac{{\sf R}'''(u)}{\chi}\right|^2\right) \nonumber \\
  && + (\phi''(u)-\phi''(0)){\sf R}''(u) + 3\phi'(u){\sf R}'''(u) +
   6 R'''(0^+) \psi(u) \\
&&   - \chi (\tilde  \psi^{(31) \prime}(u) + 2 \tilde g^{(31)}(u) +
\tilde g^{(22)}(u)) = 0 \nonumber
\end{eqnarray}
\end{widetext}
Note that these equations are similar but not identical, and in the
derivation of Eq.~(\ref{eq:outerphieqn}) we used neither the thermal
terms in Eq.~(\ref{Wilson3cum}) nor the BL equation for
$r(\tilde{u}_{12})$.  Nevertheless, they are equivalent as can be seen
by using (\ref{r3inf})  in Eq.~(\ref{eq:outerphieqn}) and the matching
requirement $R'''(0^+) = \chi r'''_\infty$, one directly obtains
Eq.~(\ref{eq:PBL2}). We have also used (\ref{reltilde1}, \ref{reltilde2})
relating the tilde quantities
to
untilde ones in the fourth cumulant matching. Note that $\phi(u)$ starts as $u^2$ (see
below). The corresponding term $\tilde{u}_{12} u_{13}^2$ is
allowed in the PBL, while in the full BL it is not (the third
cumulant starts as $\tilde{u}^6$).

The first true PBL quantity which is not fully determined by
matching is $s^{(21)}(\tilde{u}_{12},u_{13})$.
Using Eq.~(\ref{eq:outerphieqn}) to simplify
Eq.~(\ref{eq:longpbl0}), one finds
\begin{widetext}
\begin{eqnarray}
  && 0 =  \left( s^{(21)}_{20}(\tilde{u}_{12},u_{13})
  -s^{(21)}_{20}(0,u_{13})\right) r''(\tilde{u}_{12}) +
s^{(21)}_{20}(\tilde{u}_{12},u_{13})
- 6 r'''_\infty \tilde{u}_{12}^2 \psi(u_{13}) \\
&& \frac{{\sf R}''(u_{13}) }{\chi^2}
  (\gamma- \frac{1}{2})
  (r''(\tilde{u}_{12})^2 - (r'''_\infty)^2 \tilde{u}_{12}^2)  \nonumber \\
&& - \frac{1}{2} s^{(22)}_{020}(\tilde u_{12},0,u_{13}) +
\tilde{u}_{12}^2 g^{(22)}(u_{13})  + s^{(31)}_{1100}(\tilde
u_{112},u_{13}) + 2  \tilde{u}_{12}^2 g^{(31)}(u_{13}) \nonumber
\end{eqnarray}
\end{widetext}
supplemented by the {\sl boundary condition} ( \ref{eq:psimatch}).
We note that this equation can be formally but explicitly solved for
$s^{(21)}_{20}(\tilde{u},u)$. One finds
\begin{widetext}
\begin{eqnarray}
  \label{eq:Shatsol2}
&& s^{(21)}_{20}(\tilde u,u) - s^{(21)}_{20}(0 ,u)  =
\frac{1}{1+r''(\tilde{u})} \big( a(u)+b(\tilde u) + 6\tilde{u}^2
r'''_\infty \psi(u) - (\gamma - \frac{1}{2}) \frac{{\sf
R}''(u)}{\chi^2}
(r''(\tilde{u})^2 - (r'''_\infty)^2 \tilde{u}^2) \\
&& + \frac{1}{2} s^{(22)}_{020}(\tilde u,0,u) - \tilde{u}^2
g^{(22)}(u) - s^{(31)}_{1100}(0,0,-\tilde u, u) - 2  \tilde{u}^2
g^{(31)}(u) \big)
\end{eqnarray}
\end{widetext}
where $a$ and $b$ are arbitrary gauge functions. This
solubility is similar to the algebraic solution for
$r''(\tilde{u})$ possible in the full BL for the second cumulant.

We have thus solved the matching problem up to the third cumulant.
It is interesting to organise the
resulting implicit solution for $\phi(u)$ as an expansion in powers
of ${\sf R}$. To lowest order $O(R^3)$, Eq. (\ref{eq:outerphieqn}) becomes:
 \begin{eqnarray}
  \label{eq:outerphieqn0}
  && (2\epsilon-2-4\zeta)\phi(u) + \zeta
  u\phi'(u) \\
&& + \gamma\frac{{\sf R}''(u)}{\chi}\left(\left|\frac{{\sf
          R}'''(0^+)}{\chi}\right|^2   -
    \left|\frac{{\sf R}'''(u)}{\chi}\right|^2\right) =0 .
\end{eqnarray}
Solving this linear equation and inserting feedback (\ref{feedphi}) from
$\tilde S$ in the $\tilde R$ equation we obtain the ''beta-function''
to the same order, i.e. the condition
for the fixed point in the asymptotic
limit $\tilde T_l \to 0$:
\begin{widetext}
\begin{eqnarray}
  && \partial_l \tilde{R}(u) = 0 =
  (\epsilon-4 \zeta+\zeta u \partial_u) \tilde{R}(u) \label{exactbeta0} \\
&& +
\frac{1}{2} \tilde{R}''(u)^2 - \tilde{R}''(0) \tilde{R}''(u)
+ \cases{ \frac{- 2 \gamma}{\zeta}
\int_0^1 \frac{dx}{x} x^{\lambda} (\tilde{R}''(u x) - \tilde{R}''(0))
(\tilde{R}'''(u x)^2 - \tilde{R}'''(0^+)^2) + O(\tilde{R}^4) & $\lambda>0$ \cr
\frac{2 \gamma}{\zeta}
\int_{1}^\infty \frac{dx}{x} x^{\lambda} (\tilde{R}''(u x) - \tilde{R}''(0))
(\tilde{R}'''(u x)^2 - \tilde{R}'''(0^+)^2) + O(\tilde{R}^4) & $\lambda<0$ }
\nonumber \\
&& \lambda = \frac{2\epsilon-2-4\zeta}{\zeta}
\end{eqnarray}
\end{widetext}
Both formula match up to a constant when $\lambda=0$. Setting
$\epsilon=4$, $\gamma=3/4$, $\lambda=2-3 \alpha =
\frac{4(1-\theta)}{2 + \theta}$, the above result gives the {\it
exact beta function in $d=0$ } to order $O(\tilde{R}^3)$. Higher
powers can be obtained by studying (\ref{eq:PBL2}) or
(\ref{eq:outerphieqn}) to next order in the expansion in power of
$\tilde R$, and considering the feedback from the fourth cumulant
which is of the same order. Note that at this order, obtained here
rigorously, this result coincide with the naive zero temperature
procedure. This procedure consists in first solving the $S$
equation (\ref{Wilson3cum}) {\it for arbitrary } $\zeta$ by
dropping terms containing temperature and neglecting the $\tilde R
\tilde S$, and fourth cumulant term, then evaluating the feedback
term from $\tilde S$ in the $\tilde R$ equation (\ref{Wilson3cum})
which at this order has an unambiguous limit (see discussion near
Eq. \ref{lim}). It is unlikely at this stage that the absence of
ambiguity will exist at higher order. Note furthermore that in the
limit $\epsilon \to 0$ and $\zeta \to 0$ the convolution operator
in (\ref{exactbeta0}) becomes local, e.g. $\int_0^1 \frac{dx}{x}
x^{\lambda} f(x) \to f(1)/\lambda$, and we recover the result of
Ref. \cite{frg2loop}, (Eq. 3.43) with $X= 2 \gamma$. Thus the
Wilson value $\gamma=1$ seems to reproduce, up to a factor of $2$,
the result $X=1$ obtained in the $\epsilon$-expansion to two loop
in Ref. \cite{frg2loop}. This is not so bad, since this $d=0$
calculation does not involve any spatial integration. It raises
the hope that a lot of the structure of the putative
$\beta$-function in a true $\epsilon$-expansion may already be
contained in this $d=0$ model.

To close the study of the third cumulant we examine the
matching of the full BL to the partial one. This is complicated by
the fact that in the full BL regime the rescaling term is negligible,
while in fact it is not in the partial BL and the outer region.
Thus as one leaves the full BL region, presumably the
rescaling term becoming important tends to suppress $s$.
We find that to match we must have, coming from the full BL:
\begin{eqnarray}
&&  s^{(3)}(\tilde u_{123}) \sim_{\tilde u_{13} \to \infty} c_1 \chi \tilde u_{13}^2  \tilde u_{12}^2
\nonumber \\
&& + \chi \tilde u_{13} \rho(u_{12}) + \chi  |\tilde u_{13}| \hat \rho(u_{12}) + O(1)
\nonumber
\\ && \tilde u_{123} = O(1)  . \label{match2}
\end{eqnarray}
while the small argument behaviour from the partial BL side should be:
\begin{eqnarray}
 \hat S^{(3)}(\tilde u_{12},u_{13}) & \sim_{u_{13} \to 0} &  u_{13} \rho(\tilde u_{12})
+ |u_{13}| \hat \rho(\tilde u_{12}) +  O(u_{13}) \nonumber \\
 \phi(u_{13}) &\sim_{u_{13} \to 0} & c_1 u_{13}^2 + O(u_{13}^3)
\nonumber
\\ && \tilde u_{12} = O(1) \quad , \quad u_{13} = O(1) , \nonumber
\\ && \label{match2a}
\end{eqnarray}
with $c_1=\frac{1}{2} \phi''(0)$.

\subsection{analysis of fourth cumulant}

We now turn to the structure of matching for the fourth cumulant.
This is of interest for a variety of physical quantities. First,
the fourth cumulant of $u$ at zero temperature, (\ref{4conn}) can
be rewritten as:
\begin{equation}
  \label{eq:u4}
  \overline{u^4} - 3 (\overline{ u^2})^2  = \frac{1}{16} m^{- 4 \zeta} \chi^4 f_4
\end{equation}
and is the lowest order quantity expressing the deviation of the
distribution of $u$ from a Gaussian. Second, the beta function to
next order ($R^4$) requires the fourth cumulant feedback into the
third. More generally, the fourth cumulant is illustrative of the
nested structure of higher cumulant PBLs. Unlike for lower
cumulants there is more than one path of successive matchings
connecting the outer solution to the full TBL.

Using the notations defined at the beginning of this Section, we need to
discuss the PBLs : $s^{(4)}, s^{(31)}, s^{(22)}, s^{(211)},
s^{(1111)}=\tilde S^{(4)}$ for the full, various partial BL, and
finally, the outer solution.  Each of these has its zero T piece and
their finite T parts. Let us enumerate the various boundary layers. One has to first
non-trivial order in temperature, in each case:
\begin{widetext}
\begin{equation}
\begin{array}{lclcllr}
\tilde{S}^{(4)}(u_{1234}) &=& f_{4} \tilde T_l^4
\tilde u_1 \tilde u_2 \tilde u_3 \tilde u_4 & + & \tilde T_l^5 \chi^{-2}
 s^{(4)}(\tilde u_{1234})  , &
\tilde u_{1234}=O(1)& (\text{full BL}) \nonumber \\
 \tilde{S}^{(4)}(u_{1234}) &=&
\tilde{T}_l^3 \chi ~
  \text{sym}_{123} (\tilde{u}_{12} \tilde{u}_{23}^2)
\psi^{(31)}(u_{34}) & + & \tilde{T}_l^4
  s^{(31)}(\tilde{u}_{123};u_{34}) ,& \tilde{u}_{123},u_{34} = O(1)
& (\text{PBL 31}) \nonumber  \\
 \tilde{S}^{(4)}(u_{1234}) &=& && \tilde{T}_l^4
  s^{(22)}(\tilde{u}_{12};\tilde{u}_{34};u_{13})  ,& \tilde{u}_{12},\tilde{u}_{34},
u_{13} = O(1)
& (\text{PBL 22}) \nonumber  \\
\tilde{S}^{(4)}(u_{1234}) &=&   \tilde{T}_l^2 \chi^2
  \tilde{u}_{12}^2 \phi^{(211)}(u_{13},u_{14}) & +& \tilde{T}_l^3 \chi
  s^{(211)}(\tilde{u}_{12};u_{13},u_{14}) ,&
 \tilde{u}_{12},u_{13},u_{14},u_{34} = O(1)
& (\text{PBL 211}) \nonumber \\
&&&&&& \label{pbl4}
\end{array}
\end{equation}
\end{widetext}
where $\phi^{(211)}$, $s^{(211)}$, $\psi^{(31)}$ and $s^{(31)}$
are order $O(1)$ functions. The first terms in the full BL, PBL31
and PBL211 are zero temperature components and are anomalously
large (have a lower power of temperature) compared to their
associated $s$ functions. These terms are necessary for a zero
temperature limit to exist. Their form is constrained by the
requirements of permutation symmetry in each group of boundary
layer variables. They have the additional crucial property that
each contributes to its PBL equation at the same order in
temperature as the corresponding $s$ term, despite the lower power
of temperature in its definition. This is because when two
derivatives act on $\tilde u$ BL variables in these terms it
produces only gauge. The $\phi^{(211)}$ function is a simple
generalisation of the $\phi$ function introduced in PBL21 for the
third cumulant. In the PBL31 we have omitted the quadratic term
$(\tilde u_{12}^2+\tilde u_{13}^2+\tilde u_{23}^2)
\phi^{(31)}(u_{34})$ since the latter two terms in parenthesis are
pure gauge and the first can be rewritten up to gauge in the form
above using that $u_{13} \sim \tilde T_l$ (see Appendix
\ref{sec:symm-tayl-expans}). Finally, in PBL22 symmetry in both $1
\leftrightarrow 2$  and $3 \leftrightarrow 4$ forbids zero
temperature quadratic and cubic terms, while the naive quartic one
has been subsumed into $s^{(22)}$. Additional more detailed
properties of these functions are summarized in the
Appendix~\ref{sec:fourth-cumul-match}

\begin{figure}[h]
\centerline{\fig{2cm}{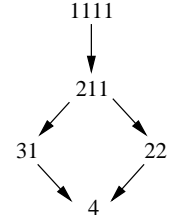}}
\caption{Nested partial boundary layer structure for the fourth
  cumulant.  The limit of two points in a given PBL coinciding takes one
  along the direction of the an arrow in the diagram.  \label{fig:nest}}
\end{figure}

Particular pairs (inner and outer) of these PBLs are connected by
taking a single argument (pair of points) in the outer PBL small
(points close) and reciprocally in the inner PBL large (points
far). This is illustrated in Fig.~\ref{fig:nest}.  The conditions
for such asymptotic matching sequentially going inwards from the
outer solution are considered in detail in
Appendix~\ref{sec:fourth-cumul-match}.  This determines the form
of various small and large argument limits of the PBL functions
defined above:
\begin{widetext}
\begin{equation}
\begin{array}{lclr}
\tilde S^{(4)}(u_{1234}) &=& \chi^4 u_{12}^2  \phi^{(211)}(u_{13},u_{14})
+ \chi^4 |u_{12}|^3 \tilde \psi^{(211)}(u_{13},u_{14})   & u_{12} \to 0 \nonumber \\
&& & \nonumber \\
s^{(211)}(\tilde{u}_{12};u_{13},u_{14}) &=& |\tilde u_{12}|^3 \tilde \psi^{(211)}(u_{13},u_{14})
 & \tilde{u}_{12} \to \infty \nonumber \\
s^{(211)}(\tilde{u}_{12};u_{13},u_{14}) &=& u_{13}
  \tilde\rho^{(31)}(\tilde{u}_{12},u_{14})+ |u_{13}|
  \tilde\sigma^{(31)}(\tilde{u}_{12},u_{14})  & u_{13} \to 0 \nonumber \\
s^{(211)}(\tilde{u}_{12};u_{13},u_{14}) &=& |u_{34}|
  \tilde{\sigma}^{(22)}(\tilde{u}_{12}, u_{13}) & u_{34} \to 0 \nonumber \\
&& & \nonumber \\
\phi^{(211)}(u_{13},u_{14}) &=&  \phi^{(211)}(0,u_{34}) - u_{13}
\tilde\psi^{(31)}(u_{34}) + u_{13}^2 \tilde{g}^{(31)}(u_{34}) +
  u_{13}^2 \text{sgn}
  (u_{13}) \tilde{h}^{(31)}(u_{34})   & u_{13} \to 0 \nonumber \\
\phi^{(211)}(u_{13},u_{14})  &=&
   u_{34}^2  \tilde{g}^{(22)}(u_{13}) & u_{34} \to 0 \nonumber \\
&& & \nonumber \\
 s^{(22)}(\tilde{u}_{12},\tilde{u}_{34};u_{13}) &=& \tilde{u}_{34}^2 \tilde{u}_{12}^2
  \tilde{g}^{(22)}(u_{13}) + |\tilde{u}_{34}|
  \tilde\sigma^{(22)}(\tilde{u}_{12};u_{13}) & \tilde u_{34} \to \infty,
\tilde u_{12}
\text{fixed} \nonumber \\
 s^{(22)}(\tilde u_{12},\tilde u_{34}, u_{13}) &=&
\tilde u_{12}^2 \tilde u_{34}^2 \frac{1}{4} s^{(22)}_{220}(0,0,0)
+ u_{13} \tilde \rho(\tilde u_{12},\tilde u_{34})
+ |u_{13}| \tilde \sigma(\tilde u_{12},\tilde u_{34})   & u_{13} \to 0 \\ && & \nonumber \\
 \psi^{(31)}(u_{34}) &=& u_{34} \psi^{(31) \prime }(0) +
\frac{1}{2} u_{34}^2 \text{sgn}(u_{34})
\psi^{(31) \prime \prime}(0^+)  & u_{34} \to 0 \nonumber \\ && & \nonumber \\
 s^{(31)}(\tilde{u}_{123};u_{34}) &=& \tilde u_{13}^2 \tilde{u}_{12}^2
  g^{(31)}(u_{34})
+ \tilde u_{13}^2 \text{sgn}(\tilde u_{13}) \tilde{u}_{12}^2
  h^{(31)}(u_{34}) +
\tilde u_{13} \tilde{u}_{12}^2
  \tilde{\rho}^{(31)}(u_{34}) + |\tilde
u_{13}| \tilde{u}_{12}^2
  \tilde{\sigma}^{(31)}(u_{34}) \hspace{-1.8cm}   & \tilde u_{13} \to \infty  \nonumber \\
s^{(31)}(\tilde u_{123} , u_{34}) &=&
\tilde \sigma^{(4)}(\tilde u_{123}) u_{34} + \tilde \rho^{(4)}(\tilde u_{123})
|u_{34}|  & u_{34} \to 0, \tilde u_{123}
\text{fixed} \nonumber \\ && & \nonumber \\
s^{(4)}(u_{1234})  &=&
\frac{1}{2}  \chi \tilde u_{34}^2 \text{sgn}(u_{34}) \tilde u_{12}^2 \psi^{(31) \prime \prime}(0^+)
+ \tilde u_{34} \chi \tilde \sigma^{(4)}(\tilde u_{123})
+ |\tilde u_{34}| \tilde \rho^{(4)}(\tilde u_{123}) & \tilde u_{34} \to \infty, \tilde u_{123}
\text{fixed} \nonumber \\
 s^{(4)}(u_{1234}) &=&  \tilde u_{13} \tilde \rho(\tilde u_{12},\tilde u_{34}) +
|u_{13}| \tilde \sigma(\tilde u_{12},\tilde u_{34}) & \tilde u_{13} \to \infty, \tilde u_{12},
\tilde u_{34}
\text{fixed} \nonumber \\
\end{array} \label{regimatch}
\end{equation}
\end{widetext}
The matching analysis results in some additional conditions not already
implicit above.  These are
\begin{eqnarray}
  \psi^{(31)}(u) & = & \tilde \psi^{(31)}(u)+ \phi_{01}^{(211)}(0,u),
  \label{eq:matchnew1} \\
  g^{(31)}(u_{34}) & = &  \tilde{g}^{(31)}(u_{34})- \frac{1}{2}\phi_{02}^{(211)}(0,u_{34}),
  \label{eq:matchnew2}
\end{eqnarray}

The various zero temperature terms can be determined sequentially by
matching from the outer solution.  In particular, the small argument
behavior of $S^{(4)}$ yields $\phi^{(211)}$, which in turn yields
$\psi^{(31)},g^{(31)}$ and $\tilde g^{(22)}$, which finally determine
$f_4$.  This Larkin term thus matches all the way through to give
\begin{eqnarray} \label{larkinf4}
&& f_4 = 2 \psi^{(31)  \prime}(0) = -2 \phi_{11}^{(211)}(0,0),\\
&& f_4 =  s^{(22)}_{220}(0,0,0) = 4 \tilde{g}^{(22)}(0) = -2
\phi_{11}^{(211)}(0,0).
\end{eqnarray}
As discussed in Appendix~\ref{sec:fourth-cumul-match}, these
derivatives are unambiguous, and determined by a definite
procedure discussed therein.

Unlike these zero temperature quantities, the PBL $s$ functions
themselves are not determined by matching alone.  Instead, they
obey PBL equations derived from the basic ERG fourth cumulant
equation (given in (\ref{ergS4})) and the PBL forms in
(\ref{pbl4}). The PBL211, PBL22, and PBL31 equations are given and
analyzed in Appendix~\ref{sec:fourth-cumul-match}.

\subsection{expansion in powers of $R$}
\label{expansion-power-R}

The PBL analysis of the fourth cumulant enables the calculation of the
beta function to $O(R^4)$.  To do so, we require $\phi(u)$ which feeds
in (\ref{eq:rout}).  To obtain it, we must solve (\ref{eq:outerphieqn})
perturbatively in $R$ (equivalently $\chi$).  We write:
\begin{eqnarray}
\phi(u)=\phi_0(u)+ \chi \phi_1(u) + \ldots
\end{eqnarray}
The zeroth order solution $\phi_0(u)$ obeys (\ref{eq:outerphieqn0}), the
solution of which is implicit in (\ref{exactbeta0}).
The next order satisfies
\begin{widetext}
\begin{eqnarray}
  && (d-3 \theta + 2 \zeta)\phi_1(u) + \zeta u \phi_1'(u) = \chi^{-1}\big[
(\phi_0''(0)- \phi_0''(u)) {\sf R}''(u) - 3\phi_0'(u)
 {\sf R}'''(u) -
   6  {\sf R}'''(0^+)  \psi_0(u) \big] + \eta_0(u),\label{eq:phi1}
\end{eqnarray}
\end{widetext}
where $\eta_0(u) =\psi_0^{(31) \prime}(u) + 2 g_0^{(31)}(u) +
g_0^{(22)}(u)$ represents the feeding from the fourth cumulant.

To determine $\psi_0(u)$ we use the expansion of the outer
equation for $S^{(3)}$ to cubic non-analytic order (see Appendix
~\ref{sec:symm-tayl-expans}) and keep only the feeding ($O(R^3)$)
term:
\begin{eqnarray}
  && (d-3 \theta + 3 \zeta)\psi_0(u) + \zeta u \psi_0'(u)
  \label{eq:psi0} \\
  && = -  \gamma \chi^{-3} \Big[\frac{1}{2}{\sf R}'''(0^+) \left|{\sf
R}'''(u)\right|^2 + {\sf R}''(u) {\sf R}'''(0^+){\sf
R}''''(0^+)\Big] \nonumber
\end{eqnarray}
Note that the r.h.s starts as $|u|$ at small $u$. The function
$\eta_0(u)$ is determined from the appropriate expansions of the
$\phi^{(211)}$ equation as detailed in
Appendix~\ref{sec:fourth-cumul-match}.  It satisfies
\begin{widetext}
\begin{eqnarray}
&& (d-4 \theta + 4 \zeta)\eta_0(u) + \zeta u \eta_0^{\prime}(u)
\nonumber \\
&&
  = -\gamma' \chi^{-4} \Big[\frac{7}{2} ( R'''(u)^2 - R'''(0^+)^2)^2 +
  12\,R''(u)\, ( R'''(u)^2 R''''(u) - R'''(0^+)^2\,R''''(0^+)) +
   R''(u)^2\, R''''(u)^2\Big] . \label{eq:eta0}
\end{eqnarray}
\end{widetext}
It is straightforward to solve the linear equations
(\ref{eq:psi0},\ref{eq:eta0}) for $\psi_0,\eta_0$, and substitute
them into (\ref{eq:phi1}) which in turn is easily solved.  For
arbitrary $\zeta$ the resulting beta function to this order now
involves double convolutions, generalizing the single one obtained
in ((\ref{exactbeta0}). Thus, for simplicity of presentation we
give the result explicitly for the particular case $\zeta=0$. This
choice is of formal interest for comparison with
$\epsilon$-expansion calculations. One obtains then
\begin{widetext}
\begin{eqnarray} \label{3loopbeta}
  && \partial_l \tilde{R}(u) = 0 =
  (\epsilon-4 \zeta+ \zeta u \partial_u) \tilde{R}(u)  \\
&& + \frac{1}{2} \tilde{R}''(u)^2 - \tilde{R}''(0) \tilde{R}''(u)
+ 2 \gamma \gamma_S (\tilde{R}''(u) - \tilde{R}''(0))
(\tilde{R}'''(u)^2 - \tilde{R}'''(0^+)^2) \nonumber \\
&& +  \gamma \gamma_S^2 \big[  {\sf R}''(u) [ 20 ( {\sf
R}'''(u)^2\,{\sf R}''''(u) - {\sf R}'''(0^+)^2\,{\sf R}''''(0^+) )
+ 2 {\sf R}''''(u) ({\sf R}'''(u)^2 - {\sf
R}'''(0^+)^2 ) \nonumber \\
  && + 6\,( {\sf R}'''(u)^2 - {\sf
R}'''(0^+)^2 )^2
  + 4\,{\sf R}''(u)^2\,( {\sf R}''''(u)^2 + {\sf R}'''(u)\,{\sf R}^{(5)}(u)) \big] \nonumber\\
  && + \gamma' \gamma_S \gamma_Q \Big[
  7 ( R'''(u)^2 - R'''(0^+)^2)^2 +
  24\,R''(u)\, ( R'''(u)^2 R''''(u) - R'''(0^+)^2\,R''''(0^+)) +
  2 \, R''(u)^2\, R''''(u)^2 \Big] \,.\nonumber
\end{eqnarray}
\end{widetext}
where here $\gamma_S=1/(2-2 \epsilon)$ and $\gamma_Q=1/(4 - 3
\epsilon)$ (respectively $\gamma_{3,0}$ and $\gamma_{4,0}$ in the
notations introduced below).

This result is also of interest if one performs the $R$-expansion
in another consistent way, closer to the $\epsilon$ expansion, by
expanding simultaneously in $\zeta$ taking $\zeta=O(R)$. In that
case, and to this order, there is only one additional $O(R^3)$
term appearing in (\ref{3loopbeta}), $2 \chi^3 \zeta (2 \phi_0(u)
+ u \phi_0'(u))$ with $\phi_0$ given by (\ref{eq:outerphieqn0})
setting $\zeta=0$.

A remarkable property of the above $\beta$-function is that, apart
from an immaterial constant, its Taylor expansion in $|u|$ does
not contain a $O(|u|)$ but starts as $u^2 + |u|^3 + ..$. It thus
satisfies the physical requirement of the absence of ''supercusp''
(as discussed in \cite{frg2loop,frgdep2loop} in the dynamical
version this can be traced to a requirement about potentiality of
the model). It implies some non trivial identities between the
zero temperature functions introduced here. Examinations of the
various terms on the right hand side of (\ref{eq:outerphieqn})
shows that a sufficient condition for the coefficient of $|u|$ to
vanish is:
\begin{eqnarray}
&& \phi''(0^+) + 2 \psi'(0^+) = 0 \\
&& \eta'(0^+) = 0
\end{eqnarray}
One easily checks that these identities are indeed verified to
lowest order in $R$, i.e. by the functions $\phi_0$, $\psi_0$ and
$\eta_0$ ( note that $\eta_0$ is a linear combination of functions
which do start as $|u|$ so cancellations occur). A quite non
trivial property is that there hold actually {\it to all orders}.
From the {\it exact } equation (\ref{eq:rout}), it is again a
sufficient condition for the absence of supercusp. Although we
hold no general proof we have checked that it holds on some higher
order terms. It seems to work term by term in the FRG equation
with cancellations reminiscent of those occurring in properties
such as dimensional reduction.

We give here the correction to the second derivative, the $u^2$
term in the $\beta$-function, for any $\zeta$. One notes indeed
that one can obtain relations between derivatives valid for
arbitrary $\zeta$. We define $\gamma_{m,n}=-1/x_{m,n}$ where the
$x_{m,n}=d - m \theta + n \zeta = - (2m - 4 - (m-1) \epsilon + (2m
- n) \zeta)$ are the naive bare eigenvalues (rescaling factor) of
a $n$-th derivative of a $m$-th cumulant at zero. These generalize
the one defined above in (\ref{eq:BL2d0}), $x_{m,m}=x_m$.
Computing $\phi''(0)$ with the above method, one finds:
\begin{widetext}
\begin{eqnarray}
&& 0= (\epsilon - 2 \zeta) {\sf R}''(0) + {\sf R}'''(0^+)^2 + 8
\gamma \gamma_{3,4} {\sf R}'''(0^+)^2 {\sf R}''''(0^+) + 48
\gamma_{3,4} (\gamma_{3,5} \gamma + \gamma_{4,2} \gamma') {{\sf
R}'''(0)}^3\,{\sf R}'''''(0) \\
&& + 12 \gamma_{3,4} ((2 \gamma_{3,4} + 4 \gamma_{3,5} ) \gamma +
13 \gamma_{4,2} \gamma') {\sf R}'''(0^+)^2\, {\sf R}''''(0^+)^2 +
O(R^5)
\end{eqnarray}
\end{widetext}
this can be equivalently obtained by taking two derivative at zero
of the beta function for arbitrary $\zeta$, not shown here.

We now compute the Larkin term to leading order in $R$. It can be
expressed by taking the derivative at zero of the equation for
$\psi^{(31)}$ which, to lowest order reads (see Appendix
\ref{sec:fourth-cumul-match}):
\begin{eqnarray}
  && (d-4 \theta + 3 \zeta)\psi^{(31)}_0(u) + \zeta u \psi^{(31) \prime}_0(u) \nonumber \\
  && =
  3 \gamma' \chi^{-4} {\sf R}''(u) {\sf R}'''(u)^3 \label{eq:psi310}
\end{eqnarray}
Using Eq. (\ref{larkinf4}) yields, for arbitrary $\zeta$ one
finds:
\begin{eqnarray}
  && f_4 = \frac{6}{d-4 \theta + 4 \zeta} \gamma' \chi^{-4} {\sf R}'''(0^+)^4 +
  O(R^5) \label{eq:f4lowest}
\end{eqnarray}
that gives:
\begin{eqnarray}
  && \kappa_4 = \frac{\overline{u^4}}{(\overline{u^2})^2} - 3 =
  \frac{6 (\epsilon - 2 \zeta)^2}{\epsilon - 2 \theta} \gamma' +
  O(R) \label{eq:f4lowesta}
\end{eqnarray}
given that the exact value in $d=0$ for the case $\zeta=4/3$,
$\theta=2/3$ is \cite{toy} $\kappa = -0.16173..$ it clearly
appears that the $R$ expansion should also be performed as an
expansion in small $\epsilon$ and $\zeta$. For the (relevant)
random field case $\zeta=\epsilon/3$ one finds:
\begin{eqnarray}
  && \kappa_4 = - \frac{1}{12} \epsilon^2 + O(\epsilon^3) \label{eq:f4lowestb}
\end{eqnarray}
thus values of order the exact result in $d=0$ are already reached
for $\epsilon \approx 1$.

To obtain a reasonable estimate directly in $d=0$ (within a fixed
dimension expansion) one should clearly pursue the expansion to
large order, i.e compute the $f_4(R)$ ''function'' up to values of
$R$ (i.e. its anomalous derivatives) where it changes sign. $R$ is
presumably not small in that regime. In fact we now the exact
value of $\tilde R''(0)= 4 C_2 \approx 4.2$ at the fixed point.

Let us now reflect on what has been accomplished. We have shown
explicitly on the third cumulant, and to some degree of detail on
the fourth one that in $d=0$ matching through the various PBL was
possible from the outer solution all the way down to the full BL
where the true derivatives are defined. Since there is little
doubt that this extends to any cumulant in $d=0$ a consequence is
that one should be able to compute directly, e.g. the beta
function, from the outer equation, performing a non-analytic
expansion of $R(u)$ and taking the proper derivatives at the limit
of coinciding points, in a manner free of ambiguities.

This is what we have performed, using mathematica, up to four
loops (order $R^5$). The three loop result (order $R^4$)
reproduces exactly (\ref{3loopbeta}) above and the four loop
result is detailed in the Appendix~\ref{sec:higher-orders-}

Let us summarize the procedure here. We use schematic notations,
parenthesis means a symmetrized expression, square bracket means
taking the limit of two coinciding points and symmetrizing the
result. We indicate schematically the derivatives. Here subscripts
do not indicate derivatives but the order in the expansion. The
iteration reads as follows. The lowest (two loop, $R^3$) order
reads as:
\begin{eqnarray}
&& S_0^{(3)}= \gamma_S \gamma (R'' R'' R'') \\
&& \beta_0 = [S_0^{(3) \prime \prime}]
\end{eqnarray}
the three loop order reads:
\begin{eqnarray}
&& S_0^{(4)}= \gamma_Q \gamma' (R'' R'' R'' R'') \\
&& S_1^{(3)}= \gamma_S ( (R'' S_0^{(3) \prime \prime}) + [S_0^{(4)\prime \prime}] ) \\
&& \beta_1 = [S_1^{(3) \prime \prime}]
\end{eqnarray}
We have explicitly checked that the evaluation of the limits at
coinciding arguments $[..]$ never contain sign ambiguities. This
is the case also for the direct evaluation of
\begin{eqnarray}
&&  \lim_{u_i \to 0} (\prod_{i=1}^{4} \partial_{u_i})
S^{(4)}_{0}(u_{1234}) = - 6 \gamma_Q \gamma' \tilde R'''(0^+)^4 \nonumber \\
&&
\end{eqnarray}
Upon taking the four derivatives, a choice of ordering of the
arguments can be made. The limit can then be taken, and, since the
expression is permutation invariant in $1,2,3,4$, is found to be
independent of the choice. Thus the function can be defined by
continuity at zero. That this is also equal to the true derivative
at zero, which requires some knowledge of the full TBL is of
course the non trivial property. We have demonstrated in some
detail here how it arises.

\newpage

\section{Discussion and prospects}
\label{sec:conclusion}

We have shown that the naive thermal boundary layer structure
found in previous one loop studies of the FRG is really only the
tip of the iceberg. In fact the TBL is, as we have shown, a truly
non-perturbative feature of the field theory which extends to all
cumulants. For the special case $d=0$ and $N=1$ we have explored
the structure in some detail and unveiled the connection of the
TBL to droplet physics. Also for this case, we have also been able
to describe in detail the matching between the TBL and the zero
temperature outer solution. This matching {\it solves} the
ambiguity problem and explains how correlation functions
calculated in the field theory have their expected zero
temperature limits. It enables a systematic evaluation of the
$T=0$ beta function in powers of the renormalized disorder $R(u)$,
given explicitly here up to four loops.

The detailed understanding of the thermal boundary layer described
here in $d=0$ is clearly a necessary first step towards progress
in higher dimensions. The ultimate goal of such a program is a
first principles derivation of the $\epsilon$-expansion, of
matching, and a test of the droplet picture (or its extensions) in
$d>0$. The ERG described in Section~\ref{sec:exact-rg-general}
appears as a promising technique to approach the problem. We now
indicate some first steps in this program.

We employ the multilocal expansion of the full effective action
functional
discussed already in (\ref{eq:multilocalexp}):
\begin{eqnarray}
&&  {\cal V}[u] = \sum_{j=1}^\infty \int_{r_1\cdots r_j}
V_l^{(j)}(\vec u_{r_1} \cdots \vec u_{r_j} ; r_1 \cdots r_j)
\end{eqnarray}
in terms of translationally-invariant {\sl functions} (``j-local
terms'') of $j$ $u_i$ and $r_i$ variables, with $V_l^{(1)}(\vec u)
= V(\vec u)$ in previous notations (and here $l=\ln (1/m)$). The
ERG equation (\ref{ergeq}) can itself be expanded in locality
\cite{ergchauve,erg} yielding a set of FRG flow equations for the
functions $V_l^{(j)}$ and more conveniently for the rescaled
functions $V_l^{(j)}(\vec u_1 \cdots \vec u_j ; r_{1\cdots j}) =
m^{j d} \tilde V_l^{(j)}(\vec u_1 m^\zeta \cdots \vec u_j m^\zeta
; x_{1..j})$ with $x=m r$. By similar considerations as in Section
we obtain a generalization of the TBLA
\begin{widetext}
\begin{eqnarray}
\tilde V_l^{(j)}(\vec u_1 \cdots \vec u_j ; x_{1 \cdots j}) & = &
\sum_{p \geq 1} f_{j,2 p}(x_{1 \cdots j}) \sum_{a_1,..a_{2 p}}
\tilde{u}_1^{a_1} .. \tilde{u}_{2 p}^{a_{2 p}} +
\frac{\tilde{T}_l}{\epsilon^2} v^{(j)}(\vec{\tilde u}_1 \cdots
\vec{\tilde u}_j ; x_{1 \cdots j}),
\end{eqnarray}
\end{widetext}
with $ \vec{\tilde u} = \epsilon \tilde{\chi} \vec{u}/\tilde{T}
_l$ and $\tilde T = m^{\theta} T$. The $ f_{j,2p}(x_{1 \cdots j})
$ are the multilocal extension of the random force terms and are
proportional to the $j$ point, $k$-th cumulant of the random force
(with $f_{1,2} = - \tilde{R}''(0)/(\tilde\chi\epsilon)^2$). The
TBLA is applied by first separating the ERG equations into coupled
sets by a cumulant expansion (i.e. by number of replicas), and
then applying the ansatz.  Each $q^{th}$ cumulant equation is
satisfied by the ansatz to the leading $O(\tilde T_l^q)$ order for
$\vec{\tilde{u}}\sim O(1)$. As in $d=0$ and in the one loop Wilson
calculation, in this balance the random force terms contribute
only through rescaling terms while the $v^{(j)}$ functions survive
in the remaining terms, the flow ($-m\partial_m$) term being
subdominant.  Thus one obtains a set of coupled fixed point
equations for a set of appropriately defined $j$-local $k^{th}$
cumulant boundary layer functions $s^{(j,k)}$.  These equations
are regular, thereby solving the $1/\tilde T_l$ divergence problem
mentionned in the Introduction.

The apparently self-consistency of this generalized TBLA is
encouraging.  Furthermore, it embodies the lowest-order droplet
scaling as we now describe.    Indeed, since the ERG gives the exact
effective action $\Gamma[u]$, the above scaling form is a very strong
statement with implications for all correlations.  As in $d=0$,
these are obtained exactly by their tree level calculation
from $\Gamma[u]$.  For instance, for the center of mass fluctuations ($u_0
= L^{-d} \int_r u_r$),
\begin{eqnarray}
  &&  \overline{(\langle u_0^2\rangle - \langle u_0\rangle^2)^n}
    \\
    && \sim    \frac{T m^\theta}{(mL)^{d(2n-1)} } m^{-2n\zeta}
    \sum_{n_v}
    \left.\partial^{2n+2n_i}_{\tilde u}[v^{(0)}(\tilde u)]^{n_v}\right|_{\tilde u =0}
    ,\nonumber
\end{eqnarray}
where the sum is over the number $n_v$ of vertices,
$\left.\partial^{2n+2n_i}_{\tilde u}[v^{(0)}(\tilde u)]^{n_v}\right|_{\tilde u =0}$
indicates schematically a combination of derivatives of the local
replica TBL function $v^{(0)}(\vec{u})$, with $2(n+n_i)^{\rm th}$ total
derivatives spread over $n_v^{\rm th}$ factors.  The number $n_i$ is the
number of internal lines, and for tree diagrams $n_i=n_v-1$.
Remarkably this precisely
reproduces the droplet behavior of thermal fluctuations,
(\ref{eq:twowell}), with an amplitude determined from the local TBL
functions.  Of course, the full TBLA for the effective action determines
{\sl all} correlation functions at low temperature, including those at
non-zero wavevector, which will have scaling dependent upon $q$ in the
manner expected from droplets.

While these successes are encouraging, the further steps of
matching and connecting to the outer solution (hopefully
describable in an $\epsilon$-expansion) remain a formidable task
ahead.  In Appendix~\ref{sec:diff-extend-erg}, we sketch some of
the problems encountered in approaching these goals.  Another
intriguing question is the extension of the present methods to
$N>1$ in $d=0$.  There too, a preliminary study indicates that the
$N=1$ mechanism for the resolution of zero temperature ambiguities
is in some way modified. A better understanding of the mechanism
of the (super-cusp) cancellations, which appear to occur term by
term for $N=1$, is necessary. Of course, despite the progress made
in this paper, we should not be surprised that some formidable
challenges remain in the glass problem. We look forward to
successful application and extension of the ideas elaborated
herein.

\begin{acknowledgements}
  L.B. was supported by NSF grant DMR-9985255, and the Packard
  foundation.  Both L.B. and P.L.D. were supported by the NSF-CNRS
  program through NSF grant INT-0089835, and CNRS Projet 10674.
\end{acknowledgements}

\newpage

\appendix

\begin{widetext}

\section{general structure of the FRG equation}
\label{sec:gener-struct-frg}

In this Appendix we study the equation (\ref{eq:Wilsoncp}), in
terms of the rescaled function $\tilde{V}(\vec u)$ defined in
(\ref{eq:Vexpand2}), namely:
\begin{eqnarray}
&& \partial_l \tilde{V}(\vec u) = {\cal O} \tilde{V}(\vec u) +
\sum_{p \geq 1} c_p \tilde T_l^p Tr( (\tilde{V}'')^p )
\end{eqnarray}
where ${\cal O}$ is the rescaling operator, for an arbitrary set of coefficients $c_p$.
We expand the trace and project onto $k$ replica terms.
We want to evaluate:
\begin{eqnarray}
&& Tr (V'')^p
\end{eqnarray}
we decompose the matrix $V''=A+B$ as:
\begin{eqnarray}
&& V''_{ab} = \delta_{ab} V''_{aa} + (1 - \delta_{ab}) V''_{ab}
\end{eqnarray}
We can use the matrix identity:
\begin{eqnarray}
&& Tr (A+B)^p = Tr (A^p) + \sum_{v=1}^p ~~~~~~~ \sum_{m,m_1,..m_v \geq 0 ; v+m+\sum m_i=p}
~~~ Tr (A^m B A^{m_1} B A^{m_2} .. B A^{m_v}) \\
&& =  Tr (A^p) + \sum_{v=1}^p ~~~~~~~ \sum_{m_1,..m_v \geq 0 ; v+\sum m_i=p}
~~~ (m_v+1) Tr (B A^{m_1} B A^{m_2} .. B A^{m_v})
\end{eqnarray}
as can be easily checked by multiplying with $z^p$ and summing over $p$. Thus we obtain
the convenient representation:
\begin{eqnarray}
&&  Tr (V'')^p = \sum_a (V''_{aa})^m \\
&& + \sum_{v=1}^p ~~~~~~~ \sum_{m_1,..m_v \geq 0 ; v+\sum m_i=p}
(m_v + 1) \sum_{a_1,..a_v} (V''_{a_v, a_1} (V''_{a_1 a_1})^{m_1}
V''_{a_1, a_2} (V''_{a_2 a_2})^{m_2} ..  V''_{a_{v-1}, a_v}
(V''_{a_v a_v})^{m_v} \nonumber
\end{eqnarray}
Using the definition of $V$ in terms of the cumulants, we can now
obtain the full form of the flow of the rescaled $q$ replica term.

To express it we define the following sets of constraints, $C_0(q)$ as:
\begin{eqnarray}
&& 1 \leq m \leq q -1 \quad, \{ 0  \leq p_i  \}_{i=1,..m} \quad, \sum_{i=1}^m p_i = q-1 -m
\end{eqnarray}
the set $C_1(q)$ as:
\begin{eqnarray}
&& 1 \leq v  \leq q \quad , \{ 0  \leq  m_i \leq q - v \}_{i=1,..v}  \quad ,
 1 \leq p = v + \sum_{i=1}^v m_i \leq q
\end{eqnarray}
and $C_2(q,v,m,\{ m_i \},p)$ as:
\begin{eqnarray}
&&
\{ p_{i0},p_{i1},..p_{i m_i} \}_{i=1,..v} \quad , p_{ij} \geq 0 , \quad
\sum_{i=1}^v ( p_{i0} + p_{i1} + ..+ p_{i m_i} ) = q - p
\end{eqnarray}

The FRG equations for the rescaled cumulants (in the Wilson
approach) read:
\begin{eqnarray}
&& \partial \tilde{S}^{q}(u_\alpha) = (d - q \theta + \zeta u_i \partial u_i) \tilde{S}^{q}(u_\alpha)
\\
&& + \tilde T_l \sum_{m,\{ p_i \}_{i=1,..m} \in C_0(q)}  \frac{c_m
q!}{(1+p_1)! .. (1+p_m)!} sym [ \tilde
S^{2+p_{1}}_{200..0}(u_1,u_{\alpha_{1}})\tilde
S^{2+p_{2}}_{200..0}(u_1,u_{\alpha_{2}}) ..
\tilde S^{2+p_{m}}_{200..0}(u_1,u_{\alpha_{m}}) ] \nonumber \\
&& + \sum_{ v,\{ m_i \},p \in C_1(q)} ~~~ \sum_{ \{
p_{i0},p_{i1},..p_{i m_i} \}_{i=1,..v} \in C_2(q,v,\{ m_i \},p)}
c_p q!  (m_v+1) \prod_{i=1}^v \frac{1}{p_{i0}! (1+p_{i 1})! ..
(1+p_{i m_i})! } \nonumber
\\
&& sym [ \tilde S^{2+p_{10}}_{110..0}(u_v,u_1,u_{\alpha_{10}}) ~~
\left(
\prod_{k_1=1}^{m_1} \tilde S^{2+p_{1 k_1}}_{200..0}(u_1,u_{\alpha_{1 k_1}}) \right) \nonumber \\
&& \times \tilde S^{2+p_{20}}_{110..0}(u_1,u_2,u_{\alpha_{20}}) ~~
\left( \prod_{k_2=1}^{m_2} \tilde S^{2+p_{2
k_2}}_{200..0}(u_2,u_{\alpha_{2 k_2}}) \right)  .. \tilde
S^{2+p_{v0}}_{110..0}(u_{v-1},u_v,u_{\alpha_{v0}}) \left(
\prod_{k_v=1}^{m_v} \tilde S^{2+p_{v k_v}}_{200..0}(u_v,
u_{\alpha_{v k_v}}) \right) ] \nonumber
\end{eqnarray}
where the first $sym$ is symmetrization with respect to the union of sets
$u_1,..u_m,u_{\alpha_{1}},..u_{\alpha_{m}}$, with $card(u_{\alpha_{i}}) = p_{i} + 1$
and the second
$sym$ is symmetrization with respect to the union of sets
$u_1,..u_v, u_{\alpha_{10}},..u_{\alpha_{1 m_1}},..
u_{\alpha_{v 0}},..u_{\alpha_{v m_v}}$, with $card(u_{\alpha_{i0}}) = p_{i0}$ and
$card(u_{\alpha_{i k_i}}) = p_{i k_i} + 1$ for $1 \leq k_i \leq m_i$.

Let us now derive the flow equations for the second, third and
fourth cumulant. The definitions of the rescaled quantities are
the Wilson ones, Eq. (\ref{resc}), only at the end we give the
equations for the ERG defnitions (\ref{eq:Srescale}).

For $q=2$, the possibilities for $C_0$, $C_1$, $C_2$ are:
\begin{eqnarray}
&& m=1, p_1=0 \\
&& v=1 , m_1=0 , p=1 ,  p_{10}=1 \nonumber \\
&& v=1 , m_1=1 , p=2 , p_{10}=p_{11}=0 \nonumber \\
&& v=2 , m_1=m_2=0 , p=2 , p_{10}=p_{20}=0 \nonumber
\end{eqnarray}
which yield the following terms, in the same order:
\begin{eqnarray}
&& \partial_l \tilde{R}(u) = (\epsilon - 4 \zeta + \zeta u \partial_u) \tilde{R}(u)
+ 2 c_1 \tilde{T} _l \tilde{R}''(u) + 2 c_1 \tilde{S}^{(3)}_{110}(0,0,u) - 4 c_2 \tilde{R}''(0) \tilde{R}''(u)
+ 2 c_2 \tilde{R}''(u)^2
\end{eqnarray}

For $q=3$, one gets:
\begin{eqnarray}
&& m=1, p_1=1 \\
&& m=2 , p_1=p_2=0 \nonumber \\
&& v=1 , m_1=0 , p=1 ,  p_{10}=2 \nonumber \\
&& v=1 , m_1=1 , p=2 , p_{10}=1 , p_{11}=0 \nonumber \\
&& v=1 , m_1=1 , p=2 , p_{10}=0 , p_{11}=1 \nonumber \\
&& v=1 , m_1=2 , p=3 , p_{10}=p_{11}=p_{12}=0 \nonumber \\
&& v=2 , m_1=m_2=0 , p=2 , p_{10}+p_{20}=1 \nonumber \\
&& v=2 , m_1=1, m_2=0 , p=3 , p_{10}=p_{11}=p_{20}=0 \nonumber \\
&& v=2 , m_1=0, m_2=1 , p=3 , p_{10}=p_{20}=p_{21}=0  \nonumber \\
&& v=3 , m_1=m_2=m_3=0 , p=3 , p_{10}=p_{20}=p_{30}=0 \nonumber
\end{eqnarray}
These yields, in the order in which they appear (denoting $u_{ij}\equiv u_i - u_j$):
\begin{eqnarray}
&& \partial_l \tilde{S}^{(3)}(u_1,u_2,u_3) = (-2 + 2 \epsilon - 6 \zeta + \zeta u_i \partial_{u_i})
\tilde{S}^{(3)}(u_1,u_2,u_3) + \,{\rm sym} [
3 c_1 \tilde{T} _l \tilde{S}^{(3)}_{200}(u_1,u_2,u_3)
+ 6 c_2 \tilde{T} _l \tilde{R}''(u_{12}) \tilde{R}''(u_{13}) \nonumber \\
&& + 3 c_1 \tilde{S}^{(4)}_{1100}(u_1,u_1,u_2,u_3)
+ 12 c_2 \tilde{S}^{(3)}_{110}(u_1,u_1,u_2) \tilde{R}''(u_{13})
- 6 c_2 \tilde{R}''(0) \tilde{S}^{(3)}_{200}(u_1,u_2,u_3) \nonumber \\
 &&  -
18 c_3 \tilde{R}''(0) \tilde{R}''(u_{12}) \tilde{R}''(u_{13})
- 12 c_2 \tilde{S}^{(3)}_{110}(u_1,u_2,u_3) \tilde{R}''(u_{12})
+ 6 c_3 \tilde{R}''(u_{21}) \tilde{R}''(u_{13}) \tilde{R}''(u_{12}) \nonumber \\
 &&
+ 12 c_3 \tilde{R}''(u_{21}) \tilde{R}''(u_{12}) \tilde{R}''(u_{23})
- 6 c_3 \tilde{R}''(u_{31}) \tilde{R}''(u_{12}) \tilde{R}''(u_{23}) ]
\end{eqnarray}
This is rearranged by redefining ${\sf R}''(u) = \tilde{R}''(u) -
\tilde{R}''(0)$, using the gauge invariance (terms which depend on
less than $q$ replicas can be erased in a $q$-rep term) as well as
identities from translational invariance (STS) such as $\text{sym}
\tilde{S}^{(3)}_{200}(u_1,u_2,u_3) = - 2 \text{sym}
\tilde{S}^{(3)}_{110}(u_1,u_2,u_3)$. Finally setting $c_p=1/(2 p)$
one obtains the one loop Wilson FRG equations for $q=2,3$.

The same is applied to $q=4$. One checks that the change
$\tilde{R}''(u) \to \tilde{R}''(0) + {\sf R}''(u)$ produces
terms proportional to $\tilde{R}''(0)$ which cancel or are only
gauge terms. It thus simplifies into:
\begin{eqnarray} \label{ergS4}
&& \partial_l \tilde{S}^{(4)}(u_{1234}) = (d - 4 \theta + \zeta u_i \partial_{u_i}) \tilde{S}^{(4)}(u_{1234})
 + \,{\rm sym} [ 2 \tilde{S}^{(5)}_{11000}(u_{11234})
+ 2 \tilde{T} _l \tilde{S}^{(4)}_{2000}(u_{1234}) \\
&& +
3 \tilde{T} _l \tilde{S}^{(3)}_{200}(u_{123}) {\sf R}''(u_{14}) +
4 \gamma \tilde{T} _l {\sf R}''(u_{12}) {\sf R}''(u_{13}) {\sf R}''(u_{14}) \nonumber \\
&&
+ 6  {\sf R}''(u_{12}) (\tilde{S}^{(4)}_{1100}(u_{1134}) -
\tilde{S}^{(4)}_{1100}(u_{1234}) ) + 6 ( \tilde{S}^{(3)}_{110}(u_{112}) \tilde{S}^{(3)}_{200}(u_{134})
+ \tilde{S}^{(3)}_{110}(u_{123}) \tilde{S}^{(3)}_{110}(u_{124}) )
\nonumber \\
&&
+ 12 \gamma (
\tilde{S}^{(3)}_{110}(u_{112}) {\sf R}''(u_{13}) {\sf R}''(u_{14})
+ \tilde{S}^{(3)}_{110}(u_{123}) {\sf R}''(u_{14})
(- 2 {\sf R}''(u_{12}) + {\sf R}''(u_{24}) )
+ \frac{1}{2} \tilde{S}^{(3)}_{200}(u_{123}) {\sf R}''(u_{14})^2 )
\nonumber \\
&&
+ 6 \gamma' ( {\sf R}''(u_{12})^2 {\sf R}''(u_{13})
(2 {\sf R}''(u_{14}) + {\sf R}''(u_{24}) )
- 2 {\sf R}''(u_{31}) {\sf R}''(u_{14}) {\sf R}''(u_{12}) {\sf R}''(u_{23})
+ \frac{1}{2} {\sf R}''(u_{41})
{\sf R}''(u_{12}) {\sf R}''(u_{23}) {\sf R}''(u_{34}) ) ]
\nonumber
\end{eqnarray}
with $\gamma=\gamma'=1$ for Wilson and $\gamma=3/4$, $\gamma'=1/2$ for ERG in $d=0$.

\newpage

\section{structure of cumulants}
\label{sec:structure-cumulants}

We discuss the structure for $n=0$. The disorder cumulant
$S^{(k)}(u_1,..u_k)$ is a sum of (symmetrized) monomials of even
degree $N$, $u_1^{p_1} u_2^{p_2} .. u_k^{p_k}$ with $1 \leq p_k
\leq p_{k-1} \leq p_1$ and $\sum_{i=1}^k p_i = N$. Symmetrization
is implicit below so we do not always write the $sym$ symbols. The
number of possible such terms of same degree $N$ is $A(N,k)$.
These however must come in a smaller number of independent linear
combinations (of same $N$) satisfying translational invariance. We
denote it $B(N,k)$. Shifting $u_i \to u_i + \lambda$ each monomial
yields:
\begin{eqnarray} \label{translat}
&& u_1^{p_1} u_2^{p_2} .. u_k^{p_k} \to u_1^{p_1} u_2^{p_2} .. u_k^{p_k} +
\lambda \sum_{i=1}^k \frac{p_i}{u_i} u_1^{p_1} u_2^{p_2} .. u_k^{p_k} + O(\lambda^2)
\end{eqnarray}
imposing that combinations give zero - up to terms which can be gauge with respect
to $k$-th cumulant, allows to find them. It is sufficient to impose
the condition on the $O(\lambda)$ term, i.e. focus on infinitesimal translations.
Also each replica must appear at least once (otherwise it is a gauge term).
Let us determine the lowest possible value of $N$ for a given $k$ and
the structure of the lowest terms. We must distinguish $k$ even or odd.

For $k$ even the lowest possible term is simple and unique and has
$N=k$ (i.e. $A(k,k)=B(N,k)=1$):
\begin{eqnarray}
&& S^{(k)}(u_1,..u_k) = S^{(k)}_{11..1}(0,..0) u_1 u_2 ....u_k
\end{eqnarray}
which is by itself symmetric and translationally invariant since
the translation (\ref{translat}) produces only $k-1$-rep terms
(pure gauge). One can determine easily the next term $N=k+2$. Let
us do it first on the fourth cumulant. There are two choices,
which give under translation (terms proportional to $\lambda$):
\begin{eqnarray}
&& u_1^3 u_2 u_3 u_4 \quad \to \quad 3 u_1^2 u_2 u_3 u_4 \\
&& u_1^2 u_2^2 u_3 u_4 \quad \to \quad 4  u_1^2 u_2 u_3 u_4
\end{eqnarray}
using symmetrization, so the translationally invariant combination
is ${\rm sym} [ \frac{1}{3} u_1^3 u_2 u_3 u_4 - \frac{1}{4} u_1^2
u_2^2 u_3 u_4 ] $. The result for the $k$-th cumulant is the
obvious generalization:
\begin{eqnarray}
&& S^{(k)}(u_1,..u_k) = \frac{1}{2} k S^{(k)}_{31..1}(0,..0)~{\rm
sym} ( \frac{1}{3} u_1^3 u_2 u_3 .. u_k - \frac{1}{4} u_1^2 u_2^2
u_3 .. u_k )
\end{eqnarray}.
There is thus a single term for $N=k+2$, i.e. $A(k+2,k)=2$,
$B(k+2,k)=1$. In particular one can parameterize the fourth
cumulant up to $O(u^8)$ as:
\begin{eqnarray}
&& S^{(4)}(u_{1234}) = q_4 u_1 u_2 u_3 u_4 - \frac{1}{4} q_6 u_1
u_2 u_3 u_4  (2 u_1^2 + 2 u_2^2 + 2 u_3^2 + 2 u_4^2 - u_1 u_2 -
u_1 u_3 - u_1 u_4 - u_2 u_3 - u_2 u_4 - u_3 u_4 ) \nonumber
\end{eqnarray}
with $q_4=S^{(4)}_{1111}(0,0,0,0)$ and
$q_6=S^{(4)}_{2211}(0,0,0,0)$.

For odd $k$ it is more subtle. The naive choice $N=k+1$ yields again a single term:
\begin{eqnarray}
&& S^{(k)}(u_1,..u_k) = \frac{1}{2} S^{(k)}_{21..1}(0,..0) ~{\rm
sym} u_1^2 u_2 ....u_k
\end{eqnarray}
but, it is in fact forbidden. Indeed, the translation produces $
\lambda u_1 u_2 ....u_k$,  a variation which cannot be cancelled,
since there is only one term. One then has to try $N=k+3$. To find
the possible combinations let us discuss first the three replica
term $k=3$. There are three monomials at order $u^6$:
\begin{eqnarray}
&& u_1^4 u_2 u_3  \quad \to \quad 4  u_1^3 u_2 u_3 \\
&& u_1^3 u_2^2 u_3 \quad \to \quad 3  u_1^2 u_2^2 u_3 + 2 u_1^3 u_2 u_3 \\
&& u_1^2 u_2^2 u_3^2 \quad \to \quad 6  u_1^2 u_2^2 u_3
\end{eqnarray}
since there are two constraints, there is only one possible
combination $u_1^4 u_2 u_3 - 2 u_1^3 u_2^2 u_3 + u_1^2 u_2^2
u_3^2$. Again, this extends easily to any $k$ odd, the lowest $N$
is thus $N=k+3$ with a single term:
\begin{eqnarray}
&& S^{(k)}(u_1,..u_k) = \frac{1}{4!} k S^{(k)}_{41..1}(0,..0)
~{\rm sym} (u_1^4 u_2 u_3 - 2 u_1^3 u_2^2 u_3 + u_1^2 u_2^2 u_3^2)
u_4 .. u_k
\end{eqnarray}

It is interesting to note that for the third cumulant complex
analysis can be used. Indeed $S^{(3)}(u_{123})$ can be expressed
as a function of $z=u_1+j u_2+ j^2 u_3$ and its complex conjugate
$\bar z$ ($j$ is the third root of unity satisfying $j^3=1$). That
automatically enforces translational invariance. Since the
permutation group is generated by $z \to j z$ and $z \to \bar z$
it is easy to express symmetrization. One drawback of this
representation however is that gauge freedom is not immediately
apparent, so we have not explored further this method.

\section{correlation functions}
\label{sec:corr-funct}

We give here some details of the calculation of correlations from
the effective action. We compute all four and six point
correlations in terms of the disorder cumulants. In each case we
also give the STS and ERG identities which relate these
correlations.

The calculation was outlined in Section
\ref{sec:exact-rg-general}. Correlations are obtained from tree
diagrams using the exact two point function and the proper
vertices $\Gamma^{(4)}$, $\Gamma^{(6)}$, .. with $4$, $6$ external
legs etc. These are then separated in number of replicas. The
exact two point function is given in (\ref{twopoint}).

\subsection{four point functions}

The calculation of the four point function requires only one
$\Gamma^{(4)}$ vertex. Since $S^{(3)}$ starts as $u^6$ only $R$
and $S^{(4)}$ can give a contribution. One simplification in the
calculation is that dressing the external lines with a $R''(0)$
gives zero, so any propagator from an external point to a vertex
can be replaced by the free propagator $T \delta_{ab}/m^2$. This
is because if one contracts first these terms one has, for
instance:
\begin{eqnarray}
&& R''(0) \sum_f \partial_{u_f} \sum_{egh} S(u_e,u_g,u_h) = 0
\end{eqnarray}
from translational invariance. Next, a limited range of
$q$-replica interactions can contribute to a given average. At
linear order in $\Gamma$ one finds that for a given correlation
with $p$ distinct replicas, interactions with $q<p$ cannot
contribute (since one can equivalently use ''excluded vertices''
where replica are constrained to be distinct). For a given
correlation which is a monomial of order $m$, interactions with $q
> m$ cannot contribute (using un-excluded vertices, there remain free replica
sums). This is why an observable such as $ \langle u_a u_b u_c u_d
\rangle $ with four distinct replicas gives zero when contracted
to linear order with $R$ and $S^{(3)}$, and cannot involve fifth
and higher cumulants either.

The result for the connected averages (graphs where all proper
vertices are connected) for the five possible monomials at order
$u^4$ is given in the text in (\ref{conn4}). In each (except the
last) a term proportional to $S^{112}(0,0,0) + S^{121}(0,0,0) +
S^{211}(0,0,0)$ naively arises but vanishes by translational
invariance. Here we give the disconnected parts:
\begin{eqnarray} \label{disc4}
&&  \langle  u_a^4  \rangle _{disc} = 3  \langle  u_a^2  \rangle ^2 \\
&&  \langle  u_a^3 u_b  \rangle _{disc} = 3  \langle  u_a^2  \rangle   \langle  u_a u_b  \rangle  \\
&&  \langle  u_a^2 u_b^2  \rangle _{disc} =  \langle u_a^2 \rangle ^2 + 2  \langle  u_a u_b \rangle ^2 \\
&&  \langle  u_a^2 u_b u_c  \rangle _{disc} =    \langle u_a^2 \rangle   \langle  u_a u_b \rangle
+ 2  \langle  u_a u_b \rangle ^2  \\
&&  \langle  u_a u_b u_c u_d  \rangle _{disc} = 3  \langle  u_a u_b \rangle ^2
\end{eqnarray}
These can be used to check that the part involving $R''(0)$
cancels in the calculation of the second moment of the sample
dependent susceptibility $\chi =  \langle u^2 \rangle  - \langle u
\rangle ^2$. In the disconnected part:
\begin{eqnarray}
&&  \langle  u_a^2 u_b^2 - 2 u_a^2 u_b u_c + u_a u_b u_c u_d
\rangle _{disc} = ( \langle u_a^2 \rangle - \langle u_a u_b
\rangle )^2 = T^2 \frac{1}{m^4}
\end{eqnarray}

Let us now give the STS and ERG identities from (\ref{stsid0}) and
(\ref{ergid0}) . We find {\it five} relations between the two
two-point functions and the five four-point functions. The ERG
relations read:
\begin{eqnarray}
&& 2 \zeta T \langle  u_a u_b  \rangle  \simeq -  T m \partial_m
\langle u_a u_b  \rangle  = 2 m^2 ( \langle  u_a^3 u_b   \rangle -
\langle u_a^2 u_b u_c  \rangle ) \nonumber
 \\
&& 2 \zeta T \langle  u_a^2  \rangle  \simeq - T m \partial_m
\langle u_a^2  \rangle  = m^2 ( \langle u_a^4   \rangle  - \langle
u_a^2 u_b^2  \rangle  )
\end{eqnarray}
where we recall $a,b,c$ are all distinct replicas, and one must
distinguish several cases in the sum over $f$ in (\ref{ergid0}).
Choosing respectively $O=u_a^3$, $O=u_a^2 u_b$ and $O=u_a u_b
u_c$, one finds three STS relations:
\begin{eqnarray}
&& 3 \frac{T}{m^2}  \langle u_a^2 \rangle  =   \langle  u_a^4  \rangle  -  \langle  u_a^3 u_b  \rangle  \\
&& \frac{T}{m^2} ( 2  \langle u_a u_b \rangle  +  \langle u_a^2
\rangle  ) = \langle  u_a^3 u_b  \rangle  +  \langle  u_a^2 u_b^2
\rangle  - 2  \langle  u_a^2 u_b u_c \rangle
\\
&&  3 \frac{T}{m^2}  \langle u_a u_b \rangle  = 3 (  \langle u_a^2
u_b u_c \rangle  -   \langle  u_a u_b u_c u_d \rangle  ) \nonumber
\end{eqnarray}
These five equations are not independent, otherwise {\it all} four
point correlations would be determined from the two point ones.
The difference of the two ERG equations is equivalent to the
difference between the two first STS ones. Thus only four are
independent and we have five fourth order correlations. Generally
at order $n$ one will have $N_n$ $n$-th order correlations,
$N_{n-2}$ RG equations and $N_{n-1}$ STS equations. $N_n$ can be
obtained from the boson partition sum $\prod_{k=1}^\infty
\frac{1}{1 - z^k} = \sum_n N_n z^n$ (with $N_4=5$, $N_6=11$,
$N_8=22$, $N_10=42$, ..).

Note that the relation (\ref{stsu4}) is obtained by forming the
linear combination of the three STS relations (with coefficients
$1,-3,2$ for the three successive lines, respectively).

If we now compute all above averages using the droplet theory, one
finds that the above five equations reduce to only two:
\begin{eqnarray}
&& 3  \langle  x_1^2  \rangle _P =  \langle  x_1^4  \rangle _D -   \langle  x_1^3 x_2  \rangle _D \\
&& 2 \zeta  \langle  x_1^2  \rangle _P =  \langle  x_1^4  \rangle
_D -   \langle  x_1^2 x_2^2  \rangle _D
\end{eqnarray}
where $\langle O(x_1,x_2) \rangle_D=\int dx_1 dx_2 O(x_1,x_2)
D(x_1,x_2)$. The three STS equations yield a single equivalent one
and the two RG equations a single one. Since there are three
fourth order unknown we are left with e.g $ \langle x_1^4 \rangle
_D$ undetermined.

\subsection{six point functions}

The calculation of the six point connected function requires
either one $\Gamma^{(6)}$ vertex or two $\Gamma^{(4)}$ vertices.
It can require a priori up to the sixth cumulant. However, terms
of order $(S^{(6)})^2$ and higher contain too many replica sums to
match the incoming ones. In both vertices $S^{(5)}$ can be dropped
as it starts as $u^8$. The $\Gamma^{(4)}$ vertices involve four
derivatives in zero and thus can only involve $R$ or $S^{(4)}$.
Thus to obtain e.g. $ \langle u_a^6 \rangle _c$, we need to
contract $u_a^6$ with
\begin{eqnarray}
&& \sum_{ef} \frac{1}{2 T^2} R_{ef} + \sum_{efg} \frac{1}{6 T^3}
S^{(3)}_{efg} + \sum_{efgh} \frac{1}{24 T^4} S^{(4)}_{efgh} +
\sum_{efghij} \frac{1}{6! T^6} S^{(6)}_{efghij} \\
&& + \sum_{efg} \frac{1}{2 (2 T^2)^2} (4 T) R'_{ef} R'_{eg} +
\sum_{efghi} \frac{1}{2 (2 T^2) (24 T^4)} (16 T) R'_{ef}
(S^{(4)})^{(1,0,0,0)}_{eghi}
\end{eqnarray}
where in this equation only derivatives are indicated by
superscript and schematic notations are used. We find the
following result for the eleven distinct connected sixth point
functions:
\begin{eqnarray}
&& \langle u_a^6  \rangle_c = 15 q_4 T - 15 q_6 T^2 - 15 T^3
R^{(4)}(0) - T^4 R^{(6)}(0) + m_{6}
\\
&& \langle u_a^5 u_b \rangle_c = 10 q_4 T - 15 q_6 T^2 - 10 T^3
R^{(4)}(0) - T^4 R^{(6)}(0) + m_{6}
\\
&& \langle u_a^4 u_b^2 \rangle_c = 7 q_4 T - 3 q_6 T^2 + 2 s_6 T^3
+ 5 T^3 R^{(4)}(0) + T^4 R^{(6)}(0) + 12 T^3 R^{(4)}(0)^2 + m_{6}
\\
&& \langle u_a^3 u_b^3 \rangle_c = 6 q_4 T + 3 q_6 T^2 - 3 s_6 T^3
- 6 T^3 R^{(4)}(0) - T^4 R^{(6)}(0) - 18 T^3 R^{(4)}(0)^2 + m_{6}
\\
&& \langle u_a^4 u_b u_c \rangle_c =  6 q_4 T - 9 q_6 T^2 + s_6
T^3 + 6 T^3 R^{(4)}(0)^2 + m_{6}
\\
&& \langle u_a^3 u_b^2 u_c \rangle_c = 4 q_4 T - \frac{1}{2} s_6
T^3 - T^3 R^{(4)}(0) - 3 T^3 R^{(4)}(0)^2 + m_{6}
 \\
&& \langle u_a^3 u_b u_c u_d \rangle_c = 3 q_4 T - 3 q_6 T^2  +
m_{6}
 \\
&& \langle u_a^2 u_b^2 u_c^2  \rangle_c = 3 q_4 T + 3 q_6 T^2  +
s_6 T^3 + 3 T^3 R^{(4)}(0)  + 6 T^3 R^{(4)}(0)^2 + m_{6}
\\
&& \langle u_a^2 u_b^2 u_c u_d \rangle_c = 2 q_4 T +  q_6 T^2 +
m_{6}
\\
&& \langle u_a^2 u_b u_c u_d u_e \rangle_c = q_4 T + m_{6}
\\
&& \langle u_a u_b u_c u_d u_e u_f \rangle_c  = m_{6}
\end{eqnarray}
Here for simplicity the mass $m$ has been set to unity, but is
easy to restore. We have denoted $s_6=S^{(3)}_{222}(0,0,0)$,
$q_4=S^{(4)}_{1111}(0,0,0,0)$, $q_6=S^{(4)}_{2211}(0,0,0,0)$ and
$m_6=S^{(6)}_{111111}(0,0,0,0,0,0)$.

We now list the ERG and STS identities to sixth order. Choosing
the observable $O$ in (\ref{stsid0}) successively as
$O=u_a^5$,$O=u_a^4 u_b$,$O=u_a^3 u_b^2$,$O=u_a^3 u_b u_c$,$O=u_a^2
u_b^2 u_c$,$O=u_a^2 u_b u_c u_d$, $O=u_a u_b u_c u_d u_e$ one
finds seven STS identities:

\begin{eqnarray}
&& \frac{T}{m^2} 5  \langle u_a^4 \rangle  =   \langle u_a^6 \rangle  -  \langle u_a^5 u_b\rangle  \\
&& \frac{T}{m^2} ( 4  \langle u_a^3 u_b\rangle  +  \langle u_a^4 \rangle  ) =
\langle u_a^5 u_b \rangle  +  \langle  u_a^4 u_b^2  \rangle  - 2  \langle  u_a^4 u_b u_c \rangle  \\
&& \frac{T}{m^2} ( 3  \langle u_a^2 u_b^2 \rangle  + 2  \langle u_a^3 u_b\rangle  ) =  \langle u_a^4 u_b^2  \rangle  +  \langle  u_a^3 u_b^3  \rangle  - 2  \langle  u_a^3 u_b^2 u_c \rangle  \\
&& \frac{T}{m^2} (3  \langle u_a^2 u_b u_c \rangle  + 2  \langle u_a^3 u_b \rangle  )
 =  \langle  u_a^4 u_b u_c \rangle   + 2  \langle  u_a^3 u_b^2 u_c \rangle  -
 3  \langle  u_a^3 u_b u_c u_d \rangle  \\
&& \frac{T}{m^2} (4  \langle u_a^2 u_b u_c \rangle  +  \langle u_a^2 u_b^2 \rangle  ) =
2  \langle  u_a^3 u_b^2 u_c \rangle   +  \langle  u_a^2 u_b^2 c^2 \rangle
-  3  \langle  u_a^2 u_b^2 u_c u_d \rangle  \\
&& \frac{T}{m^2} (2  \langle u_a u_b u_c u_d \rangle  + 3  \langle u_a^2 u_b u_c \rangle  ) =
\langle  u_a^3 u_b u_c u_d \rangle   + 3  \langle  u_a^2 u_b^2 u_c u_d \rangle
 -  4  \langle  u_a^2 u_b u_c u_d u_e \rangle  \\
&& \frac{T}{m^2} 5  \langle u_a u_b u_c u_d \rangle   =  5 \langle
u_a^2 u_b u_c u_d u_e \rangle   - 5  \langle  u_a u_b u_c u_d u_e
u_f \rangle
\end{eqnarray}

There are also five ERG relations:
\begin{eqnarray}
&& - m \partial_m T  \langle  u_a^4  \rangle  = m^2 (  \langle  u_a^6   \rangle  -  \langle  u_a^4 u_b^2  \rangle   ) \\
&& - m \partial_m T  \langle  u_a^3 u_b  \rangle  = m^2 (  \langle  u_a^5 u_b  \rangle  +  \langle  u_a^3 u_b^3  \rangle  - 2  \langle  u_a^3 u_b^2 u_c  \rangle   ) \\
&& - m \partial_m T  \langle  u_a^2 u_b u_c  \rangle  = m^2 (  \langle  u_a^4 u_b u_c  \rangle
+ 2  \langle  u_a^3 u_b^2 u_c  \rangle  - 3  \langle  u_a^2 u_b^2 u_c u_d \rangle   ) \\
&& - m \partial_m T  \langle  u_a^2 u_b^2  \rangle  = m^2 ( 2  \langle  u_a^4 u_b^2  \rangle
- 2  \langle  u_a^2 u_b^2 u_c^2 \rangle   ) \\
&& - m \partial_m T  \langle  u_a u_b u_c u_d  \rangle  = m^2 ( 4
\langle u_a^3 u_b u_c u_d  \rangle  - 4  \langle  u_a^2 u_b u_c
u_d u_e \rangle )
\end{eqnarray}

If one now estimates the above correlations using the droplet
theory, these twelve relations (seven STS and five ERG), to lowest
order in $T m^\theta$, reduce again to two ! They read:
\begin{eqnarray}
&& 5 T m^\theta  \langle  x_1^4  \rangle _P =  \langle  x_1^6  \rangle _D -   \langle  x_1^5 x_2  \rangle _D \\
&& T m^\theta ( 4  \zeta - m \partial_m    \langle  x_1^4  \rangle
_P =
  \langle  x_1^6  \rangle _D -   \langle  x_1^4 x_2^2  \rangle _D
\end{eqnarray}
for STS and ERG respectively.

Finally, let us indicate the calculation of $D_3$. One has:
\begin{eqnarray}
&& D_3 =
 \langle 2 a^3 u_b^2 c - a^2 u_b^2 c^2 - a^4 b c + a^4 u_b^2 - a^3 u_b^3
 \rangle_c \\
 && = 6 \frac{T^3}{m^{10}} R^{(4)}(0) + 2 \frac{T^4}{m^{12}} R^{(6)}(0)
+ 2 s_6 \frac{T^3}{m^{12}} + 12  \frac{T^3}{m^{14}} R^{(4)}(0)^2
\nonumber
\end{eqnarray}
with $s_6=S^{(3)}_{222}(0,0,0)$. Upon rescaling this yields the
expression in the text.

Finally, the third cumulant of the susceptibility reads:
\begin{eqnarray}
&&  \overline{ (\chi_s - \overline{\chi_s})^3 } = \overline{ (
\langle u^2 \rangle  -  \langle u \rangle ^2)^3 } \\
&& = \frac{1}{8}  \langle (u_a - u_b)^2 (u_c - u_d)^2 (u_e -
u_f)^2 \rangle \\
&& =
\langle  u_a^2 u_b^2 u_c^2 - 3 u_a^2 u_b^2 u_c d + 3 u_a^2 u_b u_c u_d u_e
-  u_a u_b u_c u_d u_e u_f  \rangle _c \\
&& =  s_6 \frac{T^3}{m^{12}} + 6 \frac{T^3}{m^{14}} R^{(4)}(0)^2
\end{eqnarray}
It is of order $T m^\theta$ as here no cancellation occurs. It
directly gives a physical interpretation to the sixth cumulant TBL
function (since we have already related $R^{(4)}(0)$ to the
variance of $\chi_s$).

\section{Details on droplet calculations}
\label{sec:details-droplets}

\subsection{some thermal moments}

Let us compute some thermal moments predicted by the two well
droplets. We use the rescaled variables $u=x m^{- \zeta}$.
Denoting $y=x_1-x_2$, one has:
\begin{eqnarray}
&&  \langle  (x -  \langle x \rangle )^m   \rangle  = (p (1-p)^m +
(-1)^m (1-p) p^m) y^m
\end{eqnarray}
for any power $m$. One can define the general correlations:
\begin{eqnarray}
&& D_{m_1,m_2..m_q} = \overline{ \prod_{i=1}^q  \langle  (x -
\langle x \rangle )^{m_i}   \rangle  }
\end{eqnarray}
with $m_1 \geq m_2 \geq .. m_q \geq 2$ and $\sum_i m_i$ is even.
Defining:
\begin{eqnarray}
&& D(y) = \int_{-\infty}^{\infty} dx_1 D(x_1,x_1+y) \label{defD}
\end{eqnarray}
One then obtains:
\begin{eqnarray}
&& D_{m_1,m_2..m_q} = C_{m_1,m_2..m_q} T \int y^{m_1 +.. + m_q} D(y) \\
&& C_{m_1,m_2..m_q} =
\frac{1}{2} \int_0^{+\infty} \frac{dz}{z}
\prod_{i=1}^q \frac{(z + (-z)^{m_i})}{(1+ z)^{m_i+1}}
\end{eqnarray}
The symmetry $z \to 1/z$ can be used. One finds for example for
$m$ even:
\begin{eqnarray}
&& C_{m} = \frac{1}{m} \\
&& C_{m,m} = \frac{1}{2 m + 4 m^2} + \frac{4^{-m} \sqrt{\pi}
\Gamma[1+m] }{ 2 \Gamma[\frac{3}{2} +m] }
\\
&& C_{m,m,m} = \frac{2}{6 m + 27 m^2 + 27 m^3} + \frac{3
\Gamma[2+m] \Gamma[1 + 2 m] }{ \Gamma[3 + 3 m] }
\end{eqnarray}

There are three such quantities for the sixth cumulants:

\begin{eqnarray}
&& D_{6} = \overline{  \langle  (x -  \langle x \rangle )^6
\rangle }
= T \frac{1}{6} \int_{-\infty}^{\infty} y^{6} D(y) \\
&& D_{33} = \overline{  \langle  (x -  \langle x \rangle )^3
\rangle ^2 }
= T \frac{1}{60} \int_{-\infty}^{\infty} y^{6} D(y) \\
&& D_{222} = \overline{  \langle  (x -  \langle x \rangle )^2
\rangle ^3 } = T  \frac{1}{60}  \int_{-\infty}^{\infty} y^{6} D(y)
\end{eqnarray}

\subsection{Check that the solution of toy model satisfies the
ERG-STS equations} \label{sec:check-that-solution}

We will need:

\begin{eqnarray}
&& g'(x) = \int_{-\infty}^{+\infty} \frac{d\lambda}{2 \pi}
e^{- i \lambda x} \frac{- i \lambda}{Ai(i \lambda)} \\
&& x g(x) = \int_{-\infty}^{+\infty} \frac{d\lambda}{2 \pi}
e^{- i \lambda x} \frac{- Ai'(i \lambda)}{Ai(i \lambda)^2} \\
&& x^2 g(x) = \int_{-\infty}^{+\infty} \frac{d\lambda}{2 \pi}
e^{- i \lambda x}
( \frac{- i \lambda}{Ai(i \lambda)} +
2 \frac{Ai'(i \lambda)^2}{Ai(i \lambda)^3} ) \\
&& g(x) + x g'(x) = \int_{-\infty}^{+\infty} \frac{d\lambda}{2 \pi}
e^{- i \lambda x} \frac{i \lambda Ai'(i \lambda)}{Ai(i \lambda)^2} \\
&& g(x) + x^3 g(x) = \int_{-\infty}^{+\infty} \frac{d\lambda}{2 \pi}
e^{- i \lambda x} ( 5 \frac{i \lambda Ai'(i \lambda)}{Ai(i \lambda)^2} -
6 \frac{Ai'(i \lambda)^3}{Ai(i \lambda)^4} )
\end{eqnarray}
obtained taking successive derivatives. The STS equation reads,
plugging in (\ref{shape}):

\begin{eqnarray}
&& - P'(x_1) = - g'(x_1) g(- x_1) + g'(- x_1) g(x_1) = 2
\int_0^{+ \infty} dy y d(y) [ - g(- x_1) g(x_1 + y) + g(x_1) g(- x_1 + y) ]
\end{eqnarray}

It can be satisfied if:

\begin{eqnarray}
&& 2 \int_0^{+ \infty} dy y d(y) g(x_1 + y) =  g'(x_1) + h(x_1^2)
g(x_1)
\end{eqnarray}

One has:

\begin{eqnarray}
&& \int_0^{+ \infty} dy y d(y) g(x_1 + y) = \int_{- \infty}^{+
\infty} \frac{d \lambda_1}{2 \pi} \frac{Ai'(i \lambda_1)}{Ai(i
\lambda_1)} \int_{-\infty}^{+\infty} \frac{d \lambda_2}{2 \pi}
\frac{1}{Ai(i \lambda_2)} \int_0^{+ \infty} dy y
e^{i \lambda_1 y } e^{ - i \lambda_2 (x_1 + y) } \\
&& = \int_{-\infty}^{+\infty} \frac{d \lambda_2}{2 \pi}
\frac{1}{Ai(i \lambda_2)} e^{ - i \lambda_2 x_1} \int_{- i
\infty}^{+ i \infty} \frac{d z_1}{2 i \pi}
\frac{Ai'(z_1)}{Ai(z_1)}
\frac{1}{(z_1 - i \lambda_2)^2} \\
&& = \int_{-\infty}^{+\infty} \frac{d \lambda_2}{2 \pi} e^{ - i
\lambda_2 x_1} \frac{1}{Ai(i \lambda_2)}
(\frac{Ai'}{Ai})'(i \lambda_2) \\
&& = \int_{-\infty}^{+\infty} \frac{d \lambda_2}{2 \pi} e^{ - i
\lambda_2 x_1} (
\frac{i \lambda_2}{Ai(i \lambda_2)} - \frac{Ai'(i \lambda_2)^2}{Ai(i \lambda_2)^3} ) \\
&& =  \frac{1}{2} g'(x_1) + \frac{1}{2} x_1^2 g(x_1)
\end{eqnarray}
It works, with $h(x_1^2)=x_1^2$. Similarly the FRG equation reads:
\begin{eqnarray}
&&
\zeta ( P(x_1) + x_1 P'(x_1) ) = 2
\int_0^{+ \infty} dy d(y) [ (2 x_1 y + y^2) g(-x_1) g(x_1 + y)
+ (- 2 x_1 y + y^2) g(x_1) g(-x_1 + y) ] \\
&& = 2 x_1 P'(x_1) + 2
\int_0^{+ \infty} dy d(y) y^2 [ g(- x_1) g(x_1 + y) + g(x_1) g(- x_1 + y) ]
\end{eqnarray}
if we use the STS equation. Thus we need to check that:
\begin{eqnarray}
&&
\zeta P(x_1) - (2 - \zeta ) x_1 P'(x_1)  = 2
\int_0^{+ \infty} dy d(y) y^2 [ g(- x_1) g(x_1 + y) + g(x_1) g(- x_1 + y) ]
\end{eqnarray}
Now we have again:
\begin{eqnarray}
&& \int_0^{+ \infty} dy y^2 d(y) g(x_1 + y) = \int_{- \infty}^{+
\infty} \frac{d \lambda_1}{2 \pi} \frac{Ai'(i \lambda_1)}{Ai(i
\lambda_1)} \int_{-\infty}^{+\infty} \frac{d \lambda_2}{2 \pi}
\frac{1}{Ai(i \lambda_2)} \int_0^{+ \infty} dy y^2
e^{i \lambda_1 y } e^{ - i \lambda_2 (x_1 + y) } \\
&& =
\int_{-\infty}^{+\infty} \frac{d \lambda_2}{2 \pi} \frac{1}{Ai(i \lambda_2)} e^{ - i \lambda_2 x_1}
\int_{- i \infty}^{+ i \infty} \frac{d z_1}{2 i \pi} \frac{Ai'(z_1)}{Ai(z_1)}
\frac{2}{(- z_1 + i \lambda_2)^3} \\
&& = \int_{-\infty}^{+\infty} \frac{d \lambda_2}{2 \pi} e^{ - i \lambda_2 x_1}
\frac{1}{Ai(i \lambda_2)}
(\frac{Ai'}{Ai})''(i \lambda_2)  \\
&& = \int_{-\infty}^{+\infty} \frac{d \lambda_2}{2 \pi} e^{ - i \lambda_2 x_1}
( \frac{1}{Ai(i \lambda_2)} - 2
\frac{i \lambda_2 Ai'(i \lambda_2)}{Ai(i \lambda_2)^2} +
2 \frac{Ai'(i \lambda_2)^3}{Ai(i \lambda_2)^4} ) \\
&& = g(x_1) - \frac{1}{3} (g(x_1) + x_1 g'(x_1))
- \frac{1}{3} (g(x_1) + x_1^3 g(x_1)) = \frac{1}{3} g(x_1) - \frac{1}{3} x_1 g'(x_1) -
\frac{1}{3} x_1^3 g(x_1)
\end{eqnarray}
which we have also checked numerically. One has thus:
\begin{eqnarray}
&& 2 \int_0^{+ \infty} dy d(y) y^2 [ g(x_1) g(- x_1 - y) + g(-
x_1) g(x_1 - y) ] = \frac{4}{3} g(x_1) g(-x_1) - \frac{2}{3} x_1
g'(x_1) g(-x_1) + \frac{2}{3} x_1 g'(- x_1) g(x_1) \nonumber \\
&& = \zeta P(x_1) - (2 - \zeta ) x_1 P'(x_1)
\end{eqnarray}
and the ERG equation is also satisfied.

Let us close by noting that under the rescaling (\ref{rescrho})
the function defined in (\ref{defD}) transforms as:
\begin{eqnarray}
&& D(y) = \rho^3 \hat D(\rho y)
\end{eqnarray}
with:
\begin{eqnarray}
&& \hat D(y) = 2 d(y) \int_{-\infty}^{+\infty} dx_1 g(-x_1)
g(x_1+y) = 2 d(y) \int_{-\infty}^{+\infty} \frac{d \lambda}{2 \pi}
\frac{e^{- i \lambda y}}{Ai(i \lambda)^2}
\end{eqnarray}
One can check that the cumulants computed in \cite{toy} are
recovered with the choice $\rho=2^{-2/3}$.

\section{Symmetrized Taylor expansion of cumulants}
\label{sec:symm-tayl-expans}

The matching calculations in the main text extensively involve Taylor
expansions of cumulants. This procedure exhibits some subtleties
since the cumulants themselves are defined only up to
the gauge transformation:
\begin{eqnarray}
S^{(k)}(u_1, \cdots , u_k) \rightarrow S^{(k)}(u_1, \cdots , u_k) +
sym_k ( f(u_1, \cdots , u_{k-1}) ).
\end{eqnarray}
with arbitrary $f$. Thus in general since the specific function $S^{(k)}$ is not defined
without specifying a gauge the results of the Taylor
expansion may themselves not be gauge invariant.

In particular when two arguments $u_1$ and $u_2$
of $S^{(k)}(u_1, \cdots , u_k)$
are brought close together it is desirable to
avoid odd terms in $u_1- u_2$. One may for instance expand
around the midpoint $(u_1 + u_2)/2$, or perform
the half sum of two expansions around each point respectively
using the permutation symmetry of $S^{(k)}$. For instance:
\begin{eqnarray}
&& S^{(3)}(u_1,u_2,u_3) = S^{(3)}(\frac{u_1 + u_2}{2}, \frac{u_1 + u_2}{2},u_3)
+ \frac{1}{4} (u_1- u_2)^2 \left[ S^{(3)}_{200}(u_1,u_1,u_3) - S^{(3)}_{110}(u_1,u_1,u_3)
\right]
+ O((u_1- u_2)^3)\nonumber \\
&&  S^{(3)}(u_1,u_2,u_3) = \frac{1}{2}
\left[ S^{(3)}(u_1,u_1,u_3) + S^{(3)}(u_2,u_2,u_3) \right]
-  \frac{1}{2} (u_1- u_2)^2 S^{(3)}_{110}(u_1,u_1,u_3)  + O((u_1- u_2)^3) .
\end{eqnarray}
Note that the zero-th order term in the first case is {\it not}
gauge-invariant, while it is in the second. We will thus use the
second method to perform short distance expansions.
Note that taking the derivative $\partial_1 \partial_2 \partial_3$
on both sides recovers a gauge invariant quantities identical
in both cases.

Let us illustrate the procedure to higher orders. Let us consider
for simplicity an analytic and symmetric function $f(x,y)=f(y,x)$.
The result is:
\begin{eqnarray}
f[u_{13},u_{23}] = f[u_{13},u_{13}] - \frac{1}{2} u_{12}^2
f_{11}[u_{13},u_{13}] + \frac{1}{4!} u_{12}^4 (2
f_{13}[u_{13},u_{13}] + 3 f_{22}[u_{13},u_{13}]) + O(u_{12}^6)
\end{eqnarray}
This is performed iteratively. When the lowest order term is even
one extracts it from the sum and keeps iterating the rest. When it
is odd, one first symmetrizes the full expression in $1,2$ then
writes $u_{23}=u_{13}-u_{12}$ and finally Taylor expand again in
$u_{12}$. What is actually done is to use the freedom:
\begin{eqnarray}
f[u_{13},u_{23}] \to f[u_{13},u_{23}] + g[u_{13},u_{23}]
\end{eqnarray}
where $g$ is antisymmetric in its two arguments. Then $g$ is
determined order by order so that all odd terms are made to
vanish. One can formalize this procedure as follows. If one
assumes the result to be:
\begin{eqnarray}
f[x,x-y] + g[x,x-y] = \sum_n y^{2 n} \phi_{2 n}(x)
\end{eqnarray}
then one has
\begin{eqnarray}
&& f(x,y) = f(y,x) = \frac{1}{2} \sum_{n} (x-y)^{2 n} ( \phi_{2
n}(x) + \phi_{2 n}(y)) \label{newdec} \\
&& g(x,y) = - g(y,x) = \frac{1}{2} \sum_{n} (x-y)^{2 n} ( \phi_{2
n}(x) - \phi_{2 n}(y)) \nonumber
\end{eqnarray}
One can show that any symmetric and globally even function can
indeed be decomposed as (\ref{newdec}).

Using the gauge freedom, one can also make to vanish any
particular Taylor coefficient. Note the following property of the
Taylor expansion upon a gauge transformation. Consider:
\begin{eqnarray}
&& S^{(3)}(u_{123})  \to S^{(3)}(u_{123}) + h(u_{23})
\end{eqnarray}
Consider the expansion:
\begin{eqnarray}
&& S^{(3)}(u_{123}) = \sum_n u_{12}^n \phi_n(u_{13})
\end{eqnarray}
writing $h(u_{23})=h(u_{13}-u_{12})$ and expanding in $u_{12}$ one
sees that the gauge tranformation becomes:
\begin{eqnarray}
&& \phi_n(u_{13}) \to  \phi_n(u_{13}) + (-1)^n \frac{1}{n!}
h^{(n)}(u_{13})
\end{eqnarray}
Note that each term in that expansion is not gauge invariant, it
is only when the series is resummed to all orders that gauge
invariance becomes manifest. This helps to clarify the proper
definition of the function $\phi(u)$ in the PBL21. First let us
note that the feedback from third to second cumulant involves the
combination $2 \phi_2(u) + \phi_1'(u)$ which is indeed invariant
upon the above gauge transformation, as physically expected.
Although one could have in general a non vanishing $\phi_1(u)$ it
can be set to zero by the above gauge transformation. The small
argument expansion and the definition of $\phi(u)$ given in the
text thus corresponds to a choice of gauge, and thus with this
definition $\phi(u)$ is gauge invariant (up to a constant).

Let us illustrate the symmetrized expansion in the calculation of
the feeding terms for the third cumulant functions $\phi(u)$ and
$\psi(u)$. Consider the $R^3$ feeding term of the outer $S$
equation. It can be rewritten, up to gauge:
\begin{eqnarray}
&&
    \frac{\gamma}{2} {\sf R}''(u_{12}) ({\sf R}''(u_{13}) - {\sf R}''(u_{23}))^2 +
    \frac{\gamma}{2} {\sf R}''(u_{12})^2  ({\sf R}''(u_{13}) + {\sf
    R}''(u_{23}))+ \frac{\gamma}{6} ({\sf R}''(u_{13}) + {\sf R}''(u_{23}))^3
\end{eqnarray}
The first term gives $ \frac{\gamma}{2} {\sf R}'''(0^+) {\sf
R}'''(u_{13})^2 |u_{12}|^3 + O(u_{12}^4)$, the second gives:
\begin{eqnarray}
&& \frac{\gamma}{2} {\sf R}'''(0^+)^2 u_{12}^2 ( 2 {\sf
R}''(u_{13}) - {\sf R}'''(u_{13}) u_{12} ) + \gamma {\sf
R}'''(0^+) {\sf R}''''(0^+) {\sf R}''(u_{13}) |u_{12}|^3 +
O(u_{12}^4)
\end{eqnarray}
The cubic analytic odd term is eliminated when performing the
symmetrized expansion explained in the Appendix, which we also use
to perform the expansion of the last term. The latter is analytic
and gives $- u_{12}^2 \gamma {\sf R}''(u_{13}) {\sf
R}'''(u_{13})^2 + O(u_{12}^4)$. Putting everything together we
find
\begin{eqnarray}
\gamma [ ({\sf R}'''(0^+)^2 - {\sf R}'''(u_{13})^2) {\sf
R}''(u_{13}) ]  u_{12}^2 + \gamma  [ \frac{1}{2} {\sf R}'''(0^+)
{\sf R}'''(u_{13})^2  +  {\sf R}'''(0^+) {\sf R}''''(0^+) {\sf
R}''(u_{13}) ] |u_{12}|^3 + O(u_{12}^4)
\end{eqnarray}
yielding the feeding terms given in the text.

\section{Fourth cumulant matching}
\label{sec:fourth-cumul-match}

In this Appendix we give details about the matching procedure
explained in the text. We first derive general relations between
asymptotic behaviors of the various functions which parameterize
successive partial boundary layers, based on consistency
requirements that the TBL scaling holds. Then we derive exact
fixed point equations for these functions, for which, in some
cases formal solutions can be given. The general matching
relations can then in turn be checked.

\subsection{general analysis of the matching}

The possible forms of the full BL, PBL211, PBL31, PBL22 and outer
solution, and their functions, are defined in (\ref{regimatch}).
Let us first give some useful properties of these functions. We
can assume permutation symmetry in the $\tilde u$ variables inside
each TBL function, e.g. that $s^{(31)}(\tilde{u}_{123};u_{34})$ is
symmetric in $\tilde{u}_{123}$ independently of the variable
$u_{34}$ since it is true up to corrections that are higher order
in $\tilde T_l$. The other symmetries are the following:
$\psi^{(31)}$ is odd,
$s^{(22)}(\tilde{u}_{12};\tilde{u}_{34};u_{13})=s^{(22)}(\tilde{u}_{34};
\tilde{u}_{12};u_{13})=s^{(22)}(\tilde{u}_{12};
\tilde{u}_{34};-u_{13})$, $\phi^{(211)}$ and $s^{(211)}$ are
symmetric in the two last arguments (exchange of $3$ and $4$).
Note a quartic polynomial can be incorporated in $s^{(31)}$ and
thus one need to consider only degree two and three polynomial.
Degree one polynomial produce gauge terms. Other possible cubic
combinations vanish, $\text{sym}_{123} (\tilde{u}_{12}
\tilde{u}_{23} \tilde{u}_{21}) =0$, $\text{sym}_{123}
(\tilde{u}_{12}^3)=0$. All quadratic combinations are equivalent
up to gauge terms. Finally, note that only $\phi^{(211)}_{11}$ and
$\psi^{(31) \prime}$ are gauge invariant.

One might alternatively choose to express the form in the PBL31 in
terms of functions of $u_{14}$ rather than $u_{34}$, i.e.
  \begin{eqnarray}
    \label{eq:u14pbl211}
    && \tilde{S}^{(4)}(u_{1234}) =
\tilde{T}_l^3 \chi ~
  \text{sym}_{123} (\tilde{u}_{12} \tilde{u}_{23}^2)
\hat \psi^{(31)}(u_{14}) + \tilde{T}_l^4
  \hat s^{(31)}(\tilde{u}_{123};u_{14}) \quad , \quad \tilde{u}_{123},u_{34} = O(1)
\quad (\text{PBL31}) .
  \end{eqnarray}
Clearly since $u_3$ and $u_1$ are completely equivalent in the
PBL31, $\hat{\psi}^{(31)}=\psi^{(31)}, \hat{s}^{(31)}=s^{(31)}$.
However, the expression in terms of $u_{14}$ is more awkward when
taking the necessary limit of large
$\tilde{u}_{13},\tilde{u}_{23}$, since this form singles out $u_1$
over $u_2$, artificially breaking the symmetry $u_1\leftrightarrow
u_2$, which is preserved by this limit.

First consider matching of the outer solution with the PBL211. One
can expand the outer solution at small $u_{12}$. We can always
choose this expansion to be even in $u_{12}$ (see
Appendix~\ref{sec:symm-tayl-expans}) and the zero-th order term
vanishes being pure gauge. Hence we expect:
\begin{eqnarray}
&&  \tilde S^{(4)}(u_{1234}) \sim \chi^4 u_{12}^2 \tilde \phi^{(211)}(u_{13},u_{14})
+ \chi^4 |u_{12}|^3
\tilde \psi^{(211)}(u_{13},u_{14}) + O(u_{12}^4)
\nonumber
\\ && u_{12},u_{13},u_{14},u_{34}  = O(1) . \label{eq:outermatchpbl4}
\end{eqnarray}
As already discussed for the third cumulant, for the TBL scaling
to hold, the first non-analyticity must arise at order
$|u_{12}|^3$. This should be compared to the large
$\tilde{u}_{12}$ limit of PBL211. This implies:
\begin{eqnarray}
&& s^{(211)}(\tilde{u}_{12};u_{13},u_{14}) \sim_{\tilde{u}_{12} \to
  \infty} |\tilde u_{12}|^3 \psi^{(211)}(u_{13},u_{14}),
\end{eqnarray}
which defines $\psi^{(211)}$.  To match, one has then
$\tilde\phi^{(211)} = \phi^{(211)}, \tilde\psi^{(211)}=\psi^{(211)}$.

Next consider matching PBL211 to PBL31.  From PBL211, we must take the
limit of small $u_{13}$:
\begin{eqnarray}
  && \phi^{(211)}(u_{13},u_{14}) =  \phi^{(211)}(0,u_{34}) - u_{13}
  \tilde\psi^{(31)}(u_{34}) + u_{13}^2 \tilde{g}^{(31)}(u_{34}) +
  u_{13}^2 \text{sgn}
  (u_{13}) \tilde{h}^{(31)}(u_{34}) +  O(u_{13}^3), \label{eq:phi211smallu13}\\
  && s^{(211)}(\tilde{u}_{12};u_{13},u_{14}) = s^{(211)}(\tilde{u}_{12};0,u_{34}) +
u_{13}
  \tilde\rho^{(31)}(\tilde{u}_{12},u_{34})+ |u_{13}|
  \tilde\sigma^{(31)}(\tilde{u}_{12},u_{34}) +O(u_{13}^2).
\end{eqnarray}
We have omitted a na\"ively possible $|u_{13}|$ term in the
expansion of $ \phi^{(211)}(u_{13},u_{14})$ which again is
forbidden by matching and TBL scaling. As discussed in Section
\ref{sec:matching} such a term would give rise to a supercusp.

It is useful to note at this stage that $\phi^{(211)}(u_{13},u_{14})$
has some gauge redundancy.  In particular, one may tranform
$\phi^{(211)}(u_{13},u_{14}) \rightarrow \phi^{(211)}(u_{13},u_{14}) +
f(u_{13}) + \tilde{f}(u_{14})$  changing $S^{(4)}$ by gauge terms.  The
implications for the expansion coefficients in
(\ref{eq:phi211smallu13}) are straightforward to determine.  In
particular, the transformation by $f(u_{13})$ leads to independent
constant shifts in $\tilde{\psi}^{(31)},
\tilde{g}^{(31)},\tilde{h}^{(31)}$ and all
higher functions.  The transformation by
$\tilde{f}(u_{14})$ leads to
\begin{eqnarray}
  \label{eq:gauge211}
  \phi^{(211)}(0,u_{34}) & \rightarrow & \phi^{(211)}(0,u_{34})  +
  \tilde{f}(u_{34}), \\
  \tilde{\psi}^{(31)}(u_{34}) & \rightarrow &
  \tilde{\psi}^{(31)}(u_{34}) - \tilde{f}'(u_{34}), \\
  \tilde{g}^{(31)}(u_{34}) & \rightarrow &
  \tilde{g}^{(31)}(u_{34}) +\frac{1}{2} \tilde{f}''(u_{34}).
\end{eqnarray}
Non-analytic coefficients such as $\tilde{h}^{(31)}(u_{34})$ are
invariant under the latter transformation.

We should compare to the large
$\tilde{u}_{13}$ limit of PBL31.  This requires some care due to the
cubic invariant.  It is useful to rewrite
\begin{eqnarray}
  \label{eq:cubictrivial}
  \text{sym}_{123} (\tilde{u}_{12} \tilde{u}_{23}^2) =
  \frac{1}{6}(\tilde{u}_{13}^3 + \tilde{u}_{23}^3) +
  \frac{1}{2}\tilde{u}_{12}^3 - \tilde{u}_{13}\tilde{u}_{12}^2.
\end{eqnarray}
When multiplied by $\psi^{(31)}(u_{34})$, the first two terms are gauge, and
the second term is both antisymmetric (hence higher order in $T$) and
does not grow at large $\tilde{u}_{13}$, hence is subdominant to the
dominant final term $\sim\tilde{u}_{13}$.  Hence
\begin{eqnarray}
&&
\tilde{T}_l^3 \chi ~ \text{sym}_{123} (\tilde{u}_{12} \tilde{u}_{23}^2)
\psi^{(31)}(u_{34}) =
-   \tilde{T}_l^2 \chi^2 \tilde{u}_{12}^2 u_{13} \psi^{(31)}(u_{34})
+O(\tilde{T}_l^4)
\end{eqnarray}
in this limit. Next consider the large  $\tilde{u}_{13}$ behaviour of
$ s^{(31)}$:
\begin{eqnarray}
&& s^{(31)}(\tilde{u}_{123};u_{34}) = \tilde u_{13}^2
g^{(31)}(\tilde{u}_{12};u_{34}) + \tilde u_{13}^2
\text{sgn}(\tilde u_{13}) h^{(31)}(\tilde{u}_{12};u_{34}) + \tilde
u_{13} \rho^{(31)}(\tilde{u}_{12};u_{34}) + |\tilde
u_{13}|\sigma^{(31)}(\tilde{u}_{12};u_{34}) + O(1). \nonumber
\end{eqnarray}
It cannot grow faster because it must match PBL 211 which is no larger
than $\tilde T^2$.  Comparing the two limits, we find:
\begin{eqnarray}
  \label{eq:211to31}
  \psi^{(31)}(u) & = & \tilde \psi^{(31)}(u)+ \phi_{01}^{(211)}(0,u),
  \label{matchnew1} \\
  g^{(31)}(\tilde{u}_{12},u_{34}) & = & \tilde{u}_{12}^2 g^{(31)}(u_{34}) =
  \tilde{u}_{12}^2
  (\tilde{g}^{(31)}(u_{34})- \frac{1}{2}\phi_{02}^{(211)}(0,u_{34})),
  \label{matchnew2} \\
  h^{(31)}(\tilde{u}_{12},u_{34}) & = & \tilde{u}_{12}^2
  \tilde{h}^{(31)}(u_{34}), \\
  \rho^{(31)}(\tilde{u}_{12},u_{34}) & = & \tilde{u}_{12}^2
  \tilde{\rho}^{(31)}(u_{34}),
  \\
  \sigma^{(31)}(\tilde{u}_{12},u_{34}) & = & \tilde{u}_{12}^2
  \tilde{\sigma}^{(31)}(u_{34}).
\end{eqnarray}
To obtain the dependence on $\phi^{(211)}$ in the first two lines
above, one must take some care with the $O(u_{13}^0)$ term in
(\ref{eq:phi211smallu13}):
\begin{eqnarray}
  \label{eq:crazymanips}
  \tilde{u}_{12}^2 \phi^{(211)}(0,u_{34}) & = &   \tilde{u}_{12}^2
  \phi^{(211)}(0,u_{14}-u_{13}) \nonumber \\
  & = &  \tilde{u}_{12}^2 \phi^{(211)}(0,u_{14}) - u_{13} \tilde{u}_{12}^2
  \phi_{01}^{(211)}(0,u_{14}) +  \frac{1}{2}u_{13}^2 \tilde{u}_{12}^2
  \phi_{02}^{(211)}(0,u_{14}) + \cdots
\end{eqnarray}
The first term above is gauge and can be dropped.  After dropping
this, we can express the result back in terms of $u_{13}$ and $u_{34}$
by writing $u_{14}=u_{34}+u_{13}$ and expanding in $u_{13}$.  Because
a gauge term has been discarded at the intermediate stage, one does
not arrive back at the starting expression!  Indeed,
\begin{eqnarray}
  \label{eq:crazymanips2}
  \tilde{u}_{12}^2 \phi^{(211)}(0,u_{34}) & = &   - u_{13} \tilde{u}_{12}^2
  \phi_{01}^{(211)}(0,u_{34}) -  \frac{1}{2}u_{13}^2 \tilde{u}_{12}^2
  \phi_{02}^{(211)}(0,u_{34}) + \cdots
\end{eqnarray}
This leads directly to (\ref{matchnew1},\ref{matchnew2}). Such
symmetrization procedure is also discussed in Appendix
~\ref{sec:symm-tayl-expans}. It is useful for calculations to note
the alternative expansion:
\begin{equation}
   \phi^{(211)}(u_{13},u_{14}) =  \phi^{(211)}(0,u_{14}) - u_{13}
  \psi^{(31)}(u_{14}) + u_{13}^2 (g^{(31)}(u_{14}) + \psi^{(31)\prime}(u_{14}))  +
  u_{13}^2 \text{sgn}
  (u_{13}) \tilde{h}^{(31)}(u_{14}) +  O(u_{13}^3)
\end{equation}
which defines the partial derivatives. Finally note that
$\psi^{(31)}$ (up to a constant) and $g^{(31)}$ are gauge
invariant.

Next we match PBL211 to PBL22.  This requires the (different from
above) small $u_{34}$ limit of the PBL211:
\begin{eqnarray}
  && \phi^{(211)}(u_{13},u_{14}) = |u_{34}|\tilde{\psi}^{(22)}(u_{13}) +
  u_{34}^2 \tilde{g}^{(22)}(u_{13})+O(|u_{34}|^3), \\
  && s^{(211)}(\tilde{u}_{12};u_{13},u_{14}) = |u_{34}|
  \tilde{\sigma}^{(22)}(\tilde{u}_{12}, u_{13})+O(|u_{34}|^2).
\end{eqnarray}
In the large $\tilde{u}_{34}$ limit of PBL22, one has generally
\begin{eqnarray}
  \label{eq:PBL22large}
  && s^{(22)}(\tilde{u}_{12},\tilde{u}_{34};u_{13})= \tilde{u}_{34}^2
  g^{(22)}(\tilde{u}_{12};u_{13})+ |\tilde{u}_{34}|
  \sigma^{(22)}(\tilde{u}_{12};u_{13}),
\end{eqnarray}
up to gauge terms independent of $\tilde{u}_{34}$, and we used the
fact that $s^{(22)}$ is even in $\tilde{u}_{34}$.  Matching gives then
$\sigma^{(22)}=\tilde\sigma^{(22)}$, and
\begin{eqnarray}
  \label{eq:psizero}
  g^{(22)}(\tilde{u}_{12},u_{13}) & = & \tilde{u}_{12}^2
  \tilde{g}^{(22)}(u_{13}), \\
  \tilde\psi^{(22)}& =&0.
\end{eqnarray}

Next we match PBL31 to BL4. The small $u_{34}$ expansion of
PBL31 can be written as:
\begin{eqnarray}
&& \psi^{(31)}(u_{34}) = u_{34} \psi^{(31) \prime }(0) +
\frac{1}{2} u_{34}^2 \text{sgn}(u_{34})
\psi^{(31) \prime \prime}(0^+) + .. \\
&& s^{(31)}(\tilde u_{123} , u_{34}) =
\tilde \sigma^{(4)}(\tilde u_{123}) u_{34} + \tilde \rho^{(4)}(\tilde u_{123})
|u_{34}| + O(u_{34}^2)
\end{eqnarray}
where $\tilde \sigma^{(4)}$ is odd and $\tilde \rho^{(4)}$ is even. We have used that
$\psi$ is odd. On the other
hand the large argument limit of PB4, taking point $4$ away from
$1,2,3$ (i.e. large $\tilde u_{34}$), reads:
\begin{eqnarray}
&& s^{(4)}(u_{1234}) = (\tilde u_{34})^2 g^{(4)}(\tilde u_{123})
+ (\tilde u_{34})^2 \text{sgn}(u_{34}) h^{(4)}(\tilde u_{123})
+ \tilde u_{34} \sigma^{(4)}(\tilde u_{123})
+ |\tilde u_{34}| \rho^{(4)}(\tilde u_{123})
+ cst \nonumber \\
&&
\end{eqnarray}
where the matching implies:
\begin{eqnarray}
&& \sigma^{(4)} = \chi \tilde \sigma^{(4)} \quad , \quad
\rho^{(4)} = \chi \tilde \rho^{(4)} \\
&& h^{(4)}(\tilde u_{123}) = \chi \frac{1}{2} \tilde u_{12}^2 \psi^{(31) \prime \prime}(0^+) \\
&& g^{(4)}(\tilde u_{123}) = 0
\end{eqnarray}
and $g^{(4)}(\tilde u_{123}) = 0$ because $\psi^{(31)}$ is odd. Note that by adding gauge
terms the quadratic functions $h^{(4)}$ and $f^{(4)}$ can be made symmetric in
$123$. Finally, the matching of the Larkin term in BL4 and the
small argument behaviour of the PBL31 yields:
\begin{eqnarray}
&& f_4 = 2 \psi^{(31)  \prime}(0)
\end{eqnarray}
(as we find below, these derivatives exist and are non-ambiguous).

Finally, we match PBL22 to BL4. The small $u_{13}$ expansion of
PBL22 can be written as:
\begin{eqnarray}
&& s^{(22)}(\tilde u_{12},\tilde u_{34}, u_{13}) =
\tilde u_{12}^2 \tilde u_{34}^2 \frac{1}{4} s^{(22)}_{220}(0,0,0)
+ u_{13} \tilde \rho(\tilde u_{12},\tilde u_{34})
+ |u_{13}| \tilde \sigma(\tilde u_{12},\tilde u_{34})
\end{eqnarray}
On the other hand we must consider the limit of PB4
when taking {\it both} points $34$ far from $12$,
i.e. the large $u_{13}$ limit with $\tilde u_{12}$
and $\tilde u_{34}$ fixed:
\begin{eqnarray}
&& s^{(4)}(u_{1234}) = \tilde u_{13} \rho(\tilde u_{12},\tilde u_{34}) +
|u_{13}| \sigma(\tilde u_{12},\tilde u_{34}) + cst
\end{eqnarray}
One sees however that one must have $\rho=0$ because $\rho(-u,v)=\rho(u,v)$,
$\rho(u,-v)=\rho(u,v)$ thus $\rho(-u,-v)=\rho(u,v)$, but $\rho$
must be odd. Thus the matching implies $\rho = 0$ and
\begin{eqnarray}
&& \sigma(\tilde u_{12},\tilde u_{34}) = \chi \tilde \sigma(\tilde u_{12},\tilde u_{34})
\end{eqnarray}

Finally we can combine the various matching stages to obtain
expressions for the Larkin term $f_4$.  This is an important test
since one may approach the full BL from either the PBL22 or PBL31.
These two approaches give
\begin{eqnarray}
&& f_4 = 2 \psi^{(31)  \prime}(0), \label{f431}\\
&& f_4 =  s^{(22)}_{220}(0,0,0) = 4 \tilde{g}^{(22)}(0) .\label{f422}
\end{eqnarray}
Consider first the approach from PBL211.  We require
$\tilde\psi^{(31)}$ which is obtained using
$\phi^{(211)}(u_{13},u_{14}) = \phi^{(211)}(u_{13},u_{34}+u_{13})$:
\begin{eqnarray}
  \label{eq:psit31exp}
  \tilde\psi^{(31)}(u_{34}) & = & - \phi_{01}^{(211)}(0,u_{34})-
  \phi_{10}^{(211)}(0,u_{34}).
\end{eqnarray}
Thus using (\ref{matchnew1})
\begin{eqnarray}
  \label{eq:psi31exp}
 \psi^{(31)}(u_{34}) & = & -  \phi_{10}^{(211)}(0,u_{34}).
\end{eqnarray}
We also require $\tilde{g}^{(22)}(u_{13})$.  From the PBL211, taking
small $u_{34}$, one writes
\begin{eqnarray}
  \label{eq:morecrazy}
  \phi^{(211)}(u_{13},u_{13}+u_{34}) & = & \phi^{(211)}(u_{13},u_{13})
  + \phi_{01}^{(211)}(u_{13},u_{13})u_{34} +
  \frac{1}{2}\phi_{02}^{(211)}(u_{13},u_{13})u^2_{34} +  \cdots
\end{eqnarray}
The first term is gauge and can be dropped.  The second is usefully
rewritten by symmetrizing in $34$ (allowed in PBL211).
\begin{eqnarray}
  \phi^{(211)}(u_{13},u_{14}) & = &
  \frac{1}{2}(\phi_{01}^{(211)}(u_{13},u_{13})-
  \phi_{01}^{(211)}(u_{14},u_{14})) u_{34} +
  \frac{1}{2}\phi_{02}^{(211)}(u_{13},u_{13})u^2_{34} +  \cdots
  \nonumber \\
  & = & -
  \frac{1}{2}\phi_{11}^{(211)}(u_{13},u_{13})u^2_{34}+\cdots
\end{eqnarray}
We therefore read off
\begin{eqnarray}
  \label{eq:gtres}
  \tilde{g}^{(22)}(u_{13}) & = &
  -\frac{1}{2}\phi_{11}^{(211)}(u_{13},u_{13}).
\end{eqnarray}
Note that first derivatives at the point $(u,u)$ such as
$\phi_{10}^{(211)}(u_{13},u_{13})$ is changed upon adding a gauge
term $f$, and can thus be set to zero upon choice of gauge.

Finally (\ref{eq:psi31exp}) and (\ref{eq:gtres}) can be applied to
equations (\ref{f431}) and (\ref{f422}) respectively to obtain the
Larkin term.  Both happily agree, and give
\begin{eqnarray}
  \label{eq:Larkinans}
  f_4 & = & -2 \phi_{11}^{(211)}(0,0),
\end{eqnarray}
the derivative being unambiguous thanks to the TBLA. Note that
$\phi^{(211)}$ has unambiguous first derivatives at the points
$(u,0)$ and $(u,u)$. It has also unambiguous second crossed
derivatives $11$ at these points $(u,u)$. There are ambiguities
however in the second derivatives $20$ at point $(u,0)$ (because
of the $\tilde h$ term) and one should be careful there.

Let us note the useful properties:
\begin{eqnarray}
 && s^{(31)}_{1100}(\tilde u_{112},u_{13}) \simeq - 2 \tilde
 u_{12}^2 g^{(31)}(u_{13}) \\
 && s^{(22)}_{020}(\tilde u_{12},0,u_{13}) \simeq 2 \tilde
 u_{12}^2 g^{(22)}(u_{13})
\end{eqnarray}
\begin{eqnarray}
&& - \phi^{(211)}_{11}(0,u) + \frac{1}{2} \phi^{(211)}_{11}(u,u) -
\phi^{(211)}_{20}(0,u) =   - ( \tilde \psi^{(31) \prime}(u) + 2
\tilde g^{(31)}(u) + \tilde g^{(22)}(u))
\end{eqnarray}
which is consistent with the equation from the PBL21 above since:
\begin{eqnarray}
&& \tilde \psi^{(31) \prime}(u) + 2 \tilde g^{(31)}(u) =
\psi^{(31) \prime}(u) + 2 g^{(31)}(u) \label{reltilde1} \\
&& \tilde g^{(22)}(u) = g^{(22)}(u) \label{reltilde2}
\end{eqnarray}

\subsection{Outer equation}

Having made precise the various limiting behaviors of the PBL
functions we now derive the FRG equations that they obey.

We start by writing the fourth cumulant equation (\ref{ergS4}) in
the outer region where all $u_{ij} \sim O(1)$. Neglecting feedback
from fifth cumulant it reads:
\begin{eqnarray} \label{outerS40}
&& \partial_l \tilde{S}^{(4)}(u_{1234}) = 0 = (d - 4 \theta +
\zeta u_i \partial_{u_i}) \tilde{S}^{(4)}(u_{1234})
 + \,{\rm sym} [ 6  {\sf R}''(u_{12}) (- 2 \chi^4 \phi^{(211)}(u_{13},u_{14}) -
\tilde{S}^{(4)}_{1100}(u_{1234}) ) \\
&& + 6 ( - 2 \chi^3 \phi(u_{12}) \tilde{S}^{(3)}_{200}(u_{134}) +
\tilde{S}^{(3)}_{110}(u_{123}) \tilde{S}^{(3)}_{110}(u_{124}) )
\nonumber
\\
&& + 12 \gamma ( - 2 \chi^3 \phi(u_{12}) {\sf R}''(u_{13}) {\sf
R}''(u_{14}) + \tilde{S}^{(3)}_{110}(u_{123}) \tilde {\sf
R}''(u_{14}) (- 2 {\sf R}''(u_{12}) + \tilde {\sf R}''(u_{24}) ) +
\frac{1}{2} \tilde{S}^{(3)}_{200}(u_{123}) \tilde {\sf
R}''(u_{14})^2 ) \nonumber
\\
&& + 6 \gamma' ( {\sf R}''(u_{12})^2 \tilde {\sf
R}''(u_{13}) (2 {\sf R}''(u_{14}) + \tilde {\sf
R}''(u_{24}) ) - 2 {\sf R}''(u_{31}) \tilde {\sf
R}''(u_{14}) {\sf R}''(u_{12}) {\sf R}''(u_{23}) +
\frac{1}{2} {\sf R}''(u_{41}) {\sf R}''(u_{12})
{\sf R}''(u_{23}) {\sf R}''(u_{34}) ) ] \nonumber
\end{eqnarray}

We are now going to study the small $u_{12}$ limit of this
equation. Before doing so, it is instructive to give the detailed
result for the various small argument expansions defined in the
previous Section of the $R^4$ feeding term. These will feed in the
various RG equations for their corresponding functions. We denote
these feeding terms by the subscript $R^4$ but one should not
forget that they are only feeding terms, not the true function to
this order. Symmetrized Taylor expansion to order $u_{12}^2$
yields:
\begin{eqnarray}
\label{phi211n} && \left.\partial_l S^{(4)}\right|_{R^4
\phi^{(211)}} = \phi^{(211)}_{R^4}(u_{13},u_{14}) = {\rm sym}_{34}
\big[ 3\,{\sf R}''({u_{13}})\,{\sf R}''({u_{14}})\,{{\sf
R}^{(3)}(0^+)}^2 - 2 {\sf
  R}''({u_{13}})\,{\sf R}''({u_{14}})\,{{\sf R}^{(3)}({u_{13}})}^2 \\
  && -
  {{\sf R}''({u_{14}})}^2\,{{\sf R}^{(3)}({u_{13}})}^2
  - 4\,{\sf R}''({u_{13}})\,{\sf R}''({u_{34}})\,{{\sf
R}^{(3)}({u_{13}})}^2 +
  2{\sf R}''({u_{14}})\,{\sf R}''({u_{34}})\,{{\sf R}^{(3)}({u_{13}})}^2
  - {{\sf R}''({u_{34}})}^2\,{{\sf R}^{(3)}({u_{13}})}^2 \nonumber \\
&&  -
  {{\sf R}''({u_{13}})}^2\,{\sf R}^{(3)}({u_{13}})\,{\sf R}^{(3)}({u_{14}}) -
  {\sf R}''({u_{13}})\,{\sf R}''({u_{14}})\,{\sf R}^{(3)}({u_{13}})\,{\sf R}^{(3)}({u_{14}}) \nonumber
  \\ && +
  2{\sf R}''({u_{13}})\,{\sf R}''({u_{34}})\,{\sf R}^{(3)}({u_{13}})\,{\sf R}^{(3)}({u_{14}}) -
  \frac{1}{2}{{\sf R}''({u_{34}})}^2\,{\sf R}^{(3)}({u_{13}})\,{\sf
      R}^{(3)}({u_{14}})\big] \nonumber
\end{eqnarray}

We now expand this in $u_{13}$ at fixed $u_{34}$.  One finds
\begin{eqnarray*}
&&   \phi^{(211)}_{R^4}(0^+,u_{34}) = - \frac{5}{2} {\sf
      R}''({u_{34}})^2\,{{\sf R}^{(3)}({u_{34}})}^2  \\
&& \tilde\psi^{(31)}_{R^4}(u_{34}) = 2\,{\sf R}''({u_{34}})\,{{\sf
    R}^{(3)}({u_{34}})}^3 + 5\,{{\sf R}''({u_{34}})}^2\,{\sf
  R}^{(3)}({u_{34}})\,{\sf R}^{(4)}({u_{34}}) \\
&& \psi^{(31)}_{R^4}(u_{34}) = -3\,{\sf R}''({u_{34}})\,{{\sf
    R}^{(3)}({u_{34}})}^3 \\
&& \tilde{g}^{(31)}_{R^4}(u_{34}) = -2\,{{\sf R}^{(3)}(0^+)}^2\,{{\sf R}^{(3)}({u_{34}})}^2 - 6\,{\sf R}''({u_{34}})\,{{\sf R}^{(3)}(0^+)}^2\,{\sf R}^{(4)}(0^+) -
  5\,{\sf R}''({u_{34}})\,{{\sf R}^{(3)}({u_{34}})}^2\,{\sf
    R}^{(4)}({u_{34}})\\ && - \frac{5}{2} [ {\sf R}''({u_{34}})^2\,{{\sf
        R}^{(4)}({u_{34}})}^2  +
  {\sf R}''({u_{34}})^2\,{\sf R}^{(3)}({u_{34}})\,{\sf
      R}^{(5)}({u_{34}}) ] \\
&& g^{(31)}_{R^4}(\tilde{u}_{12},u_{34}) =\tilde{u}_{12}^2\big[
-2\,{{\sf R}^{(3)}(0^+)}^2\, {{\sf R}^{(3)}({u_{34}})}^2 +
\frac{5}{2} {\sf R}^{(3)}({u_{34}})^4 \\
&& - 6\,{\sf R}''({u_{34}})\,{{\sf R}^{(3)}(0^+)}^2\,{\sf
R}^{(4)}(0) +
  \frac{15}{2} {\sf R}''({u_{34}})\,{{\sf R}^{(3)}({u_{34}})}^2\,{\sf
      R}^{(4)}({u_{34}}) \big] \\
&& \tilde{h}^{(31)}_{R^4}(u_{34}) = \frac{3}{2}\,{{\sf R}^{(3)}(0^+)}^3\,{\sf R}^{(3)}({u_{34}})
- \frac{3}{2}\,{\sf R}^{(3)}(0^+)\,{{\sf R}^{(3)}({u_{34}})}^3
\end{eqnarray*}

We next expand instead the $\phi^{(211)}$ feeding term in $u_{34}$ at
fixed $u_{13}$, discarding the zeroth order gauge piece (and
symmetrizing this by averaging with the same expansion at fixed
$u_{14}$).  This confirms that $\tilde\psi^{(22)}(u_{13})=0$ and gives
\begin{eqnarray*}
&& \tilde{g}^{(22)}_{R^4}(u_{13})=  -3\,{{\sf
R}^{(3)}(0^+)}^2\,{{\sf R}^{(3)}({u_{13}})}^2 + \frac{3}{2} {\sf
R}^{(3)}({u_{13}})^4 +
  6\,{\sf R}''({u_{13}})\,{{\sf R}^{(3)}({u_{13}})}^2\,{\sf R}^{(4)}({u_{13}})
  + {{\sf R}''({u_{13}})}^2\,{{\sf R}^{(4)}({u_{13}})}^2.
\end{eqnarray*}

The small argument behaviours are:

\begin{eqnarray*}
&&   \phi^{(211)}_{R^4}(0^+,u_{34}) = - \frac{5}{2} {\sf
R}'''(0^+)^4 u^2
- \frac{15}{2}  {\sf R}'''(0^+)^3 {\sf R}''''(0^+) |u|^3 + O(u^4) \\
&& \tilde\psi^{(31)}_{R^4}(u) = 2 {\sf R}'''(0^+)^4 u
+ 12 {\sf R}'''(0^+)^3 {\sf R}''''(0^+) u^2 \text{sgn}(u) + O(u^3) \\
&& \psi^{(31)}_{R^4}(u) = - 3 {\sf R}'''(0^+)^4 u
- \frac{21}{2} {\sf R}'''(0^+)^3 {\sf R}''''(0^+) u^2 \text{sgn}(u) + O(u^3) \\
&& \tilde{g}^{(31)}_{R^4}(u) = - 2 {\sf R}'''(0^+)^4
- 15  {\sf R}'''(0^+)^3{\sf R}''''(0^+) |u| + O(u^2) \\
&& g^{(31)}_{R^4}(u) = \frac{1}{2} {\sf R}'''(0^+)^4
+ \frac{15}{2}  {\sf R}'''(0^+)^3 {\sf R}''''(0^+) |u| + O(u^2) \\
&& \tilde{h}^{(31)}_{R^4}(u) = - 3
{\sf R}'''(0^+)^3 {\sf R}''''(0^+) u + O(u^2 \text{sgn}(u)) \\
&& \tilde{g}^{(22)}_{R^4}(u)= - \frac{3}{2} {\sf R}'''(0^+)^4 + 6
{\sf R}'''(0^+)^3 {\sf R}''''(0^+) |u| + O(u^2)
\end{eqnarray*}

Similarly, the Taylor expansion to order $|u_{12}|^3$ yields the
feeding $R^4$ term for $\tilde \psi^{(211)}(u_{13},u_{14})$. It
reads:
\begin{eqnarray*}
&&   \psi^{(211)}_{R^4}(u_{13},u_{14}) = 2 {\rm sym} \big[ {\sf
R}''({u_{14}})\,{\sf R}'''({u_{13}})^2 \,{{\sf R}^{(3)}(0^+)} +
\frac{1}{2}{{\sf R}''({u_{34}})}^2\,{\sf R}^{(3)}(0^+)\,{\sf
      R}^{(3)}({u_{13}})^2 \\
      && +
      {{\sf R}''({u_{13}})} \,{\sf R}^{(3)}({0^+}) {\sf R}^{(3)}({u_{13}})\,{\sf
      R}^{(3)}({u_{14}}) - \frac{1}{2}
{{\sf R}''({u_{34}})}\,{\sf R}^{(3)}(0^+)\,{\sf
      R}^{(3)}({u_{13}}) {\sf
      R}^{(3)}({u_{14}}) + \frac{3}{2} {{\sf R}'''(0^+)}\,{\sf R}^{(4)}(0^+)\,{\sf
      R}''({u_{13}}) {\sf
      R}''({u_{14}}) \big]
\end{eqnarray*}
with the following small argument behaviour:
\begin{eqnarray}
&&   \psi^{(211)}_{R^4}(u_{13},u_{14}) =
\psi^{(211)}_{R^4}(0,u_{14}) + u_{13} ( \frac{1}{2} {{\sf
R}'''(0^+)}^3 \, {\sf
      R}'''({u_{14}}) - \frac{1}{2} {{\sf
R}'''(0^+)} \, {\sf
      R}'''({u_{14}})^3 ) \nonumber \\
      && + 2 |u_{13}| {{\sf
R}'''(0^+)}^2 ( {\sf R}'''({u_{14}})^2 + 3 {\sf R}''({u_{14}})
{\sf R}''''({0^+}) ) \label{psi211small}
\end{eqnarray}

We can now perform Taylor expansion to $O(u_{12}^2)$ of the
complete outer equation (\ref{outerS40}). One finds:

\begin{eqnarray}
&& 0 = \chi^4 u_{12}^2 [ (d- 4 \theta + 2 \zeta + \zeta u_{13}
\partial_{u_{13}} + \zeta u_{14}
\partial_{u_{14}} )  \phi^{(211)}(u_{13},u_{14}) \\
&& + 6 {\sf R}'''[0^+] \tilde \psi^{(211)}(u_{13},u_{14}) + 2
\text{sym}_{34} \big[ - ( {{\psi}^{(31)\prime}}({u_{14}}) + 2
g^{(31)}({u_{14}}) ) \,{\sf R}''({u_{13}}) +
  {{\psi}^{(31)}}({u_{14}})\,{\sf R}^{(3)}({u_{13}})
- g^{(22)}(u_{13}) {\sf R}''({u_{34}}) \nonumber \\
&& +
  2\,{\sf R}^{(3)}({u_{14}})\,  \phi_{01}^{(211)}
({u_{13}},{u_{14}}) +
  {\sf R}''({u_{14}})\,  \phi_{02}^{(211)}
({u_{13}},{u_{14}}) -
  {\sf R}^{(3)}({u_{13}})\,  \phi_{10}^{(211)}
(-{u_{13}},{u_{34}}) \nonumber \\
&&  +
  {\sf R}''({u_{13}})\,  \phi_{11}^{(211)}
({u_{13}},{u_{14}}) - \frac{1}{2}
  {\sf R}''({u_{34}})\,  \phi_{11}^{(211)}
({u_{13}},{u_{14}})  \big] \\
&& + 2 \chi^2 u_{12}^2 \text{sym}_{34} \big[
 \chi^4 ( 18 \psi(u_{13}) \psi(u_{14})
+ 3 \phi '({u_{13}})\,\phi '({u_{14}}) - \phi ({u_{14}})\,\phi
''({u_{13}}) -
 \phi ({u_{34}})\,\phi ''({u_{13}}))  \nonumber \\
&& - \chi \,\phi ''({u_{14}})\,S_{101}^{(3)}({u_1},{u_3},{u_4}) -
\frac{1}{2} \chi \,\phi ''(0)\,S_{200}^{(3)}({u_1},{u_3},{u_4})
+
  \chi \,\phi '({u_{14}})\,S_{102}^{(3)}({u_1},{u_3},{u_4}) \\
  && -
  \frac{{S_{111}^{(3)}({u_1},{u_3},{u_4})}^2}{4\,{\chi }^2} -
  2\,\chi \,\phi '({u_{14}})\,S_{201}^{(3)}({u_1},{u_3},{u_4})
  \big]
  \\
  &&
   + 2  \gamma u_{12}^2 \chi^3 \text{sym}_{34} \big[ 12
   \psi(u_{14}) {\sf R}''(u_{13}) {\sf R}'''(0^+) +
{\sf R}''({u_{13}})\,{\sf R}''({u_{14}})\,( \phi ''({u_{13}})-
\phi ''(0)) \\
&& + \frac{1}{2} {{\sf R}''({u_{14}})}^2\,\phi ''({u_{13}}) +
  {\sf R}''({u_{13}})\,{\sf R}''({u_{34}})\,\phi ''({u_{13}}) - {\sf
R}''({u_{14}})\,{\sf R}''({u_{34}})\,\phi ''({u_{13}}) \nonumber \\
&& + \frac{1}{2}
 {{\sf R}''({u_{34}})}^2\,\phi ''({u_{13}}) +
  2\,\phi '({u_{14}})\,{\sf R}''({u_{13}})\,{\sf R}^{(3)}({u_{13}}) +
  2\,\phi '({u_{13}})\,{\sf R}''({u_{14}})\,{\sf R}^{(3)}({u_{13}}) +
  2\,\phi '({u_{14}})\,{\sf R}''({u_{14}})\,{\sf R}^{(3)}({u_{13}}) \nonumber \\
&& +
  4\,\phi '({u_{13}})\,{\sf R}''({u_{34}})\,{\sf R}^{(3)}({u_{13}}) -
  2\,\phi '({u_{14}})\,{\sf R}''({u_{34}})\,{\sf R}^{(3)}({u_{13}}) + \phi
({u_{14}})\,{{\sf R}^{(3)}({u_{13}})}^2 +
  \phi ({u_{34}})\,{{\sf R}^{(3)}({u_{13}})}^2 ) \big] \nonumber \\
&& +  2 \gamma  u_{12}^2  \text{sym}_{34} \big[ \frac{1}{2} {\sf
R}'''(0^+)^2 S_{200}^{(3)}({u_1},{u_3},{u_4}) + {{\sf
R}^{(3)}({u_{14}})}^2\,S_{101}^{(3)}({u_1},{u_3},{u_4}) -
  {\sf R}''({u_{14}})\,{\sf
R}^{(3)}({u_{14}})\,S_{102}^{(3)}({u_1},{u_3},{u_4}) \nonumber \\
&& -
  {\sf R}''({u_{14}})\,{\sf
R}^{(3)}({u_{13}})\,S_{111}^{(3)}({u_1},{u_3},{u_4}) +
  {\sf R}''({u_{34}})\,{\sf
R}^{(3)}({u_{13}})\,S_{111}^{(3)}({u_1},{u_3},{u_4})  +
  {\sf R}''({u_{14}})\,{\sf
R}^{(3)}({u_{14}})\,S_{201}^{(3)}({u_1},{u_3},{u_4})   \big]
\nonumber \\
&& + \gamma' u_{12}^2 \phi^{(211)}_{R^4}(u_{13},u_{14})
\end{eqnarray}
where $\phi^{(211)}_{R^4}(u_{13},u_{14})$ as defined in
(\ref{phi211n} ) above and we have used extensively all the
already derived relations between partial derivatives of
$\phi^{(211)}$. Note that we have performed an expansion symmetric
in $1$,$2$ as explained in the Appendix. Thus terms such as
$u_{12}^2 \text{sgn} u_{12} \tilde h(u_{14}) {\sf R}''(u_{13})$
which do appear, e.g., in expanding $\phi^{(211)}(u_{21},u_{24})
{\sf R}''(u_{23})$ are in fact of higher order $\sim |u_{12}|^3$.

Next, the resulting equation for $\phi^{(211)}(u_{13},u_{14})$ can
be further expanded for small $u_{13}$ (one uses $\psi(0)=0$,
$\phi(0)=0$), to obtain an equation for $\psi^{(31)}$. In
principle we should be able to obtain it by taking the first
derivative with respect to $u_{13}$ at fixed $u_{14}$. It indeed
works for the $R^4$ feeding term, but for the other terms it
produces a non analytic piece $|u_{13}|$, i.e. the above equation
expands as (schematically):
\begin{eqnarray}
0 = u_{12}^2 ( u_{13} H(u_{14}) + |u_{13}| \tilde H(u_{14}) +
O(u_{13}^2) )
\end{eqnarray}
Let us give each piece separately. The $H$ equation gives:
\begin{eqnarray}
&& 0= (d- 4 \theta + 3 \zeta  + \zeta u \partial_{u} )
\psi^{(31)}(u) - 6 {\sf R}'''(0^+) \tilde \psi^{(211)}_{10}(0,u) \\
&& + {\sf R}'''(u) ( - g^{(22)}(0) - 3 g^{(22)}(u) -
\psi^{(31)\prime}(0) + 3 \psi^{(31)\prime}(u)  )  + {\sf
R}''(u) \psi^{(31) \prime \prime}(u)  - 2 {\sf R}'''(0^+) \tilde h^{(31)}(u) \nonumber  \\
&& + \chi^2 ( - 15 \phi''(0) \phi'(u) - 9 \phi''(u)
\phi'(u) ) \\
&& + \gamma \chi^{-1} (  3 \phi'(u) {\sf R}'''(0^+)^2 + 6
\phi''(u) {\sf R}''(u) {\sf R}'''(u) + 9 \phi'(u)
{\sf R}'''(u)^2 ) \\
&& - 3 \gamma' \chi^{-4}  {\sf R}''(u) {\sf R}'''(u)^3
\end{eqnarray}
note that we write the term $\tilde \psi^{(211)}_{10}(0,u)$ in
both equations since it has an analytic piece and a non-analytic
one (subscript NA) as one sees from (\ref{psi211small}) above. The
$\tilde H$ equation gives:

\begin{eqnarray}
&& 0=  {\sf R}''(u) ( g^{(22)\prime}(0^+) + 2 g^{(31)\prime}(0^+)
+ \psi^{(31)\prime \prime}(0^+)  )
 - {\sf R}'''(0^+) ( 6 g^{(31)}(u) + 5 \psi^{(31) \prime}(u) ) \nonumber  \\
&& - 6 {\sf R}'''(0^+) \tilde \psi^{(211)}_{10}(0,u)_{NA} - 18 \chi^2 ( \phi''(0) + 2 \psi'(0^+) ) \psi(u)  \\
&& - 6 \gamma \chi^{-1} [ \phi''(0) + 2 \psi'(0^+) ] {\sf R}''(u)
{\sf R}'''(0^+)
\end{eqnarray}
This should vanish identically, in the same way that a $u_{12}^2
|u_{13}|$ term does not appear in the third cumulant, which would
produce $\phi(u) \sim |u|$ and the infamous supercusp!

One easily checks that to lowest order in $R$ the various
combinations in this equation do actually vanish. First the
combination $\phi''(0) + 2 \psi'(0^+)$ vanishes. Next to lowest
order, one has (from above):
\begin{eqnarray}
&& - {\sf R}'''(0^+) ( 6 g^{(31)}(u) + 5 \psi^{(31) \prime}(u) ) =
12 {\sf R}'''(0^+)^3 ( {\sf R}'''(u)^2 + 3 {\sf R}''(u) {\sf
R}''''(0^+)^2 ) \\
&& - 6 {\sf R}'''(0^+) \tilde \psi^{(211)}_{10}(0,u)_{NA} = - 12
{\sf R}'''(0^+)^3 ( {\sf R}'''(u)^2 + 3 {\sf R}''(u) {\sf
R}''''(0^+)^2 )
\end{eqnarray}
so that the non-analytic piece of these two terms in the $R
S^{(4)}$ term also cancel. Finally we note that the feeding terms
to order $O(R^4)$ displayed above:
\begin{eqnarray}
&& g^{(22)\prime}_{R^4} (0^+) + 2 g^{(31)\prime}_{R^4}(0^+) +
\psi^{(31)\prime \prime}_{R^4}(0^+) = 0
\end{eqnarray}
also exactly cancel.

When the feedback from fifth cumulant is also included, these
cancellations should keep occurring order by order. We have
checked on the equation for $\psi$ that $\phi''(0) + 2 \psi'(0^+)$
also vanishes to next order from the $R S$ term in the equation
for $S^3$. There seems to be additional exact relations to higher
orders, we find:
\begin{eqnarray}
&& \phi''(0) + 2 \psi'(0^+) = 0 \\
&& 24 \rho'(0^+) + 14 \psi''(0^+) + 4 \phi'''(0^+) = 0
\end{eqnarray}
this seems to work in some algebraic way, term by term,
reminiscent of dimensional reduction type cancellations, according
to a mechanism which remains to be understood to all orders.

\subsection{Equation for PBL 211}

We now proceed as in the study of the third cumulant matching in
the text, where we derived two equations obeyed by $\phi(u)$ (then
shown to be consistent). We have derived above the equation for
$\phi^{(211)}(u_{13}, u_{14})$ as a small argument limit of the
outer solution. We now derive the equation for
$\phi^{(211)}(u_{13}, u_{14})$ directly in the PBL211, by
examining (\ref{ergS4}) in the regime where $\tilde u_{12}=O(1)$,
all other $u_{ij}$ being of order one. We use the various PBL
functions defined in \ref{regimatch}. Grouping all terms of lowest
order $O(\tilde T_l^2)$ we obtain:
\begin{eqnarray}
&& 0 = \chi^2 \tilde u_{12}^2 [ d+2 \zeta -4 \theta
+ \zeta (u_{13} \partial_{u_{13}} + u_{14} \partial_{u_{14}}) ] \phi^{(211)}(u_{13}, u_{14})
 +  \chi^3 s^{(211)}_{200}( \tilde u_{12}, u_{13}, u_{14}) \\
&& \\
&& + 2  \chi^2 \tilde u_{12}^2 \text{sym}_{34} \big[ -
{{\psi}^{(31)\prime}}({u_{14}})\,{\sf R}''({u_{13}})  +
  {{\psi}^{(31)}}({u_{14}})\,{\sf R}^{(3)}({u_{13}})   \nonumber \\
&& +
  2\,{\sf R}^{(3)}({u_{14}})\,  \phi_{01}^{(211)}
({u_{13}},{u_{14}}) +
  {\sf R}''({u_{14}})\,  \phi_{02}^{(211)}
({u_{13}},{u_{14}}) -
  {\sf R}^{(3)}({u_{13}})\,  \phi_{10}^{(211)}
(-{u_{13}},-{u_{34}}) \nonumber \\
&&  +
  {\sf R}''({u_{13}})\,  \phi_{11}^{(211)}
({u_{13}},{u_{14}}) - \frac{1}{2}
  {\sf R}''({u_{34}})\,  \phi_{11}^{(211)}
({u_{13}},{u_{14}})  \big] \nonumber \nonumber \\
&&  +  \chi^2  \text{sym}_{34} \big[  -
 \chi  \, r''({{\tilde{u}}_{12}})\,s_{200}^{(211)}(0,{u_{13}},{u_{14}}) +
\chi \,
r''({{\tilde{u}}_{12}})\,s_{200}^{(211)}({{\tilde{u}}_{12}},{u_{13}},{u_{14}})
\nonumber \\
&& -
  {\sf R}''({u_{34}})\,s_{200}^{(22)}(0,{{\tilde{u}}_{12}},-{u_{13}}) +
  {\sf
R}''({u_{14}})\,s_{1100}^{(31)}({{\tilde{u}}_1},{{\tilde{u}}_1},{{\tilde{u}}_2},{u_{13}})
+ {\sf
R}''({u_{14}})\,s_{1100}^{(31)}({{\tilde{u}}_2},{{\tilde{u}}_2},{{\tilde{u}}_1},{u_{13}})
 \big] \nonumber \\
&& +
\chi^2 \text{sym}_{34} {\sf R}''(u_{14}) s^{(21)}_{20}(\tilde u_{12},u_{13})
+ \frac{1}{2} r''(\tilde u_{12}) S^{(3)}_{200}(u_{134}) \nonumber \\
&&  \nonumber \\
&& +
2 \tilde{u}_{12}^2  \text{sym}_{34} \big[ {\chi }^4 (
3 \phi '({u_{13}})\,\phi '({u_{14}}) - \phi
({u_{14}})\,\phi ''({u_{13}}) -
 \phi ({u_{34}})\,\phi ''({u_{13}}))  \nonumber \\
&& - \chi \,\phi
''({u_{14}})\,S_{101}^{(3)}({u_1},{u_3},{u_4}) +
  \chi \,\phi '({u_{14}})\,S_{102}^{(3)}({u_1},{u_3},{u_4}) -
  \frac{{S_{111}^{(3)}({u_1},{u_3},{u_4})}^2}{4\,{\chi }^2} -
  2\,\chi \,\phi '({u_{14}})\,S_{201}^{(3)}({u_1},{u_3},{u_4})  \big] \nonumber \\
&&
 +  \text{sym}_{34} \big[
  - 2
  {\chi
}^4\,s_{20}^{(21)}(0,{u_{13}})\,s_{20}^{(21)}({{\tilde{u}}_{12}},{u_{14}})
+
  {\chi
}^4\,s_{20}^{(21)}({{\tilde{u}}_{12}},{u_{13}})\,s_{20}^{(21)}({{\tilde{u}}_{12}},{u_{14}})
+
  s_{011}^{(3)}({{\tilde{u}}_1},{{\tilde{u}}_2},{{\tilde{u}}_2})\,S_{200}^{(3)}
({u_1},{u_3},{u_4})
 \big]
\nonumber \\
&&  \nonumber \\
&&  + 2  \gamma \tilde{u}_{12}^2 \chi \text{sym}_{34} \big[
{\sf R}''({u_{13}})\,{\sf R}''({u_{14}})\,\phi ''({u_{13}}) + \frac{1}{2} {{\sf
R}''({u_{14}})}^2\,\phi ''({u_{13}}) +
  {\sf R}''({u_{13}})\,{\sf R}''({u_{34}})\,\phi ''({u_{13}}) - {\sf
R}''({u_{14}})\,{\sf R}''({u_{34}})\,\phi ''({u_{13}}) \nonumber \\
&& + \frac{1}{2}
 {{\sf R}''({u_{34}})}^2\,\phi ''({u_{13}}) +
  2\,\phi '({u_{14}})\,{\sf R}''({u_{13}})\,{\sf R}^{(3)}({u_{13}}) +
  2\,\phi '({u_{13}})\,{\sf R}''({u_{14}})\,{\sf R}^{(3)}({u_{13}}) +
  2\,\phi '({u_{14}})\,{\sf R}''({u_{14}})\,{\sf R}^{(3)}({u_{13}}) \nonumber \\
&& +
  4\,\phi '({u_{13}})\,{\sf R}''({u_{34}})\,{\sf R}^{(3)}({u_{13}}) -
  2\,\phi '({u_{14}})\,{\sf R}''({u_{34}})\,{\sf R}^{(3)}({u_{13}}) + \phi
({u_{14}})\,{{\sf R}^{(3)}({u_{13}})}^2 +
  \phi ({u_{34}})\,{{\sf R}^{(3)}({u_{13}})}^2 ) \big] \nonumber \\
&& +  2 \gamma  \tilde{u}_{12}^2 \chi^{-2} \text{sym}_{34} \big[
{{\sf R}^{(3)}({u_{14}})}^2\,S_{101}^{(3)}({u_1},{u_3},{u_4}) -
  {\sf R}''({u_{14}})\,{\sf
R}^{(3)}({u_{14}})\,S_{102}^{(3)}({u_1},{u_3},{u_4}) \nonumber \\
&& -
  {\sf R}''({u_{14}})\,{\sf
R}^{(3)}({u_{13}})\,S_{111}^{(3)}({u_1},{u_3},{u_4}) +
  {\sf R}''({u_{34}})\,{\sf
R}^{(3)}({u_{13}})\,S_{111}^{(3)}({u_1},{u_3},{u_4})  +
  {\sf R}''({u_{14}})\,{\sf
R}^{(3)}({u_{14}})\,S_{201}^{(3)}({u_1},{u_3},{u_4})   \big]
\nonumber \\
&& +
2 \gamma  \text{sym}_{34} \big[ -2\,{\chi }^2\,r''({{\tilde{u}}_{12}})\,{\sf
R}''({u_{14}})\,s_{20}^{(21)}(0,{u_{13}}) +
  2\,{\chi }^2\,r''({{\tilde{u}}_{12}})\,{\sf
R}''({u_{14}})\,s_{20}^{(21)}({{\tilde{u}}_{12}},{u_{13}}) \nonumber \\
&& +
  {\sf R}''({u_{13}})\,{\sf
R}''({u_{14}})\,s_{110}^{(3)}({{\tilde{u}}_1},{{\tilde{u}}_1},{{\tilde{u}}_2})
+ \frac{1}{2}
  {r''({{\tilde{u}}_{12}})}^2\,S_{200}^{(3)}({u_1},{u_3},{u_4}) \big]
\nonumber \\
&&  \nonumber \\
&& + (2\,\gamma \,r''({{\tilde{u}}_{12}}) + 3 \gamma' r''(\tilde
u_{12})^2) {\sf R}''({u_{13}}) {\sf R}''({u_{14}}) + \gamma'
\chi^{-2} \tilde u_{12}^2 ( \phi^{(211)}_{R^4}(u_{13},u_{14}) -
3\,{\sf R}''({u_{13}})\,{\sf R}''({u_{14}})\,{{\sf
R}^{(3)}(0^+)}^2 )
 \nonumber
\end{eqnarray}
we recall $\gamma=3/4$ and $\gamma'=1/2$ for ERG.

Taking the large $\tilde u_{12}^2$ limit one checks that one
recovers the previous small $u_{12}^2$ limit of the outer equation
for $S^4$. The non trivial terms which grow as $\tilde u_{12}^2$
give respectively:
\begin{eqnarray}
&& \tilde T_l^2 \chi^3 \text{sym}_{34} r''(\tilde u_{12})
s^{(211)}_{200}(\tilde u_{12},u_{13},u_{14}) \to 6 \chi^4 u_{12}^2
{\sf
R}'''(0^+) \psi^{(211)}(u_{13},u_{14}) \\
&& - \tilde T_l^2 \chi^2 \text{sym}_{34} {\sf R}''(\tilde u_{34})
s^{(22)}_{200}(0,\tilde u_{12},-u_{13}) \to - 2 \chi^4 u_{12}^2
\text{sym}_{34} {\sf R}''(\tilde u_{34}) g^{(22)}(u_{13}) \\
&& 2 \tilde T_l^2 \chi^2 \text{sym}_{34} {\sf R}''(\tilde u_{14})
s^{(31)}_{1100}(\tilde u_{112},u_{13}) \to - 4  \chi^4 u_{12}^2
\text{sym}_{34} {\sf R}''(\tilde u_{14}) g^{(31)}(u_{13}) \\
\end{eqnarray}
Using the full BL of $R$, $s^{(3)}_{011}(\tilde u_{122}) =
\frac{1}{2} \tilde u_{12}^2 - \frac{1}{2} r''(\tilde u_{12})^2 -
r''(\tilde u_{12}) \simeq \frac{1}{2} \tilde u_{12}^2 (1-
(r'''_\infty)^2)$ and that $(r'''_\infty)^2 = 1 + 2 \chi
\phi''(0)$ yields:
\begin{eqnarray}
&& \tilde T_l^2 \text{sym}_{34} s^{(3)}_{011}(\tilde u_{122})
S^{(3)}_{200}(u_{134}) \to - \chi^3 u_{12}^2 \phi''(0)
S^{(3)}_{200}(u_{134})
\end{eqnarray}
and similarly:
\begin{eqnarray}
&& 2 \tilde T_l^2 \text{sym}_{34} \big[
  2\,{\chi }^2\,r''({{\tilde{u}}_{12}})\,{\sf
R}''({u_{14}})\,s_{20}^{(21)}({{\tilde{u}}_{12}},{u_{13}}) +
  {\sf R}''({u_{13}})\,{\sf
R}''({u_{14}})\,s_{110}^{(3)}({{\tilde{u}}_1},{{\tilde{u}}_1},{{\tilde{u}}_2})
+ \frac{1}{2}
  {r''({{\tilde{u}}_{12}})}^2\,S_{200}^{(3)}({u_1},{u_3},{u_4}) \big]
\nonumber \\
&& \simeq u_{12}^2 \text{sym}_{34} \big[ 24 \chi^3 \psi(u_{14})
{\sf R}''({u_{13}}) {\sf R}'''(0^+) - 2 \chi^3 \phi''(0)  {\sf
R}''({u_{13}})  {\sf R}''({u_{14}}) + {\sf R}'''(0^+)^2
S_{200}^{(3)}({u_1},{u_3},{u_4}) \big]
\end{eqnarray}
the non trivial term in the $R^4$ feeding term is:
\begin{eqnarray}
&& 3 \gamma' \tilde T_l^2 r''({{\tilde{u}}_{12}})^2 {\sf
R}''({u_{13}})  {\sf R}''({u_{14}}) \simeq 3 \gamma'  u_{12}^2
{\sf R}'''(0^+)^2 {\sf R}''({u_{13}})  {\sf R}''({u_{14}})
\end{eqnarray}

Let us write the resulting equation obtained from PBL211 when the
leading order equation is used. One gets:

\begin{eqnarray}
&& 0 =  \chi^3 s^{(211)}_{200}( \tilde u_{12}, u_{13}, u_{14}) +
\chi^3
 r''({{\tilde{u}}_{12}})\,( s_{200}^{(211)}({{\tilde{u}}_{12}},{u_{13}},{u_{14}})
 - s_{200}^{(211)}(0,{u_{13}},{u_{14}}) ) - 6 \chi^3 \tilde u_{12}^2
r'''_\infty \psi^{(211)}(u_{13},u_{14})
\nonumber \\
&&\chi^2  \text{sym}_{34} \big[  -
  {\sf  R}''({u_{34}})\,( s_{200}^{(22)}(0,{{\tilde{u}}_{12}},-{u_{13}})
- 2 \tilde u_{12}^2 g^{(22)}({u_{13}}) ) \big]
\nonumber \\
&& + 2 \chi^2  \text{sym}_{34} \big[
  {\sf R}''({u_{14}})\, (s_{1100}^{(31)}(\tilde{u}_{112},{u_{13}})
  + 2 \tilde u_{12}^2 g^{(31)}({u_{13}}) )
 \big] \nonumber \\
&&
 + {\chi}^4
(s_{20}^{(21)}({{\tilde{u}}_{12}},{u_{14}}) -
s_{20}^{(21)}(0,{u_{14}}))(s_{20}^{(21)}({{\tilde{u}}_{12}},{u_{13}})
- s_{20}^{(21)}(0,{u_{13}}) ) - 36 {\chi}^4 \tilde u_{12}^2
\psi(u_{13}) \psi(u_{14})
\nonumber \\
&& + 4 \gamma {\chi }^2 \text{sym}_{34} \big[ {\sf
R}''({u_{14}})\,[ r''({{\tilde{u}}_{12}})\, (
s_{20}^{(21)}({{\tilde{u}}_{12}},{u_{13}}) -
s_{20}^{(21)}(0,{u_{13}}) ) + \frac{1}{4 \gamma}
s_{20}^{(21)}({{\tilde{u}}_{12}},{u_{13}}) - 6 r'''_\infty
{\tilde{u}}_{12}^2 \psi({u}_{13}) ] \big]
 \nonumber \\
&& + [  \frac{1}{2} r''(\tilde u_{12}) + s_{110}^{(3)}(\tilde
u_{112}) - \frac{1-(r'''_\infty)^2}{2} \tilde u_{12}^2 + \gamma
(r''(\tilde u_{12})^2 - (r'''_\infty)^2 \tilde u_{12}^2 ) ]
S_{200}^{(3)}({u_1},{u_3},{u_4})
\nonumber \\
&&  + [ 2 \gamma (s_{110}^{(3)}(\tilde u_{112}) -
\frac{1-(r'''_\infty)^2}{2} \tilde u_{12}^2 ) + 2\,\gamma
\,r''({{\tilde{u}}_{12}})\, + 3 \gamma' (r''(\tilde u_{12})^2 -
(r'''_\infty)^2 \tilde u_{12}^2 )] {\sf R}''({u_{13}}) {\sf
R}''({u_{14}})
\end{eqnarray}
It can be simplified slightly by using the full BL of $R$. Note
that it is a linear equation for $s^{(211)}$, the homogeneous part
has the same structure as found in the third cumulant matching in
the text, for $s^{(21)}$ above and $s^{(3)}$. Thus it can be
similarly solved, formally, in terms of the (more complicated)
homogeneous part.

\subsection{PBL 31}

Considering the ERG equation (\ref{ergS4}) for $S^{(4)}$ in the
regime where $u_1$, $u_2$ and $u_3$ are close together, and
inserting the various PBL forms from (\ref{regimatch}) as
appropriate, we find at leading order $O(\tilde T_l^3)$:

The equation is of order $O(\tilde T_l^3)$. One finds:
\begin{eqnarray}
&& 0 = - \chi \text{sym}_{123} [ \tilde u_{12}^2 \tilde u_{13}]
( d+3 \zeta -4 \theta
+ \zeta (u_{34} \partial_{u_{34}} ) \psi^{(31)}(u_{34})
+  \frac{3}{2} \chi^2 \text{sym}_{123} [ s^{(31)}_{2000}( \tilde u_{123}, u_{34})] \\
&&
+ \frac{3}{2} \chi \tilde{u}_{13}  {\sf R}'''(u_{34})
s^{(22)}_{020}( \tilde u_{12}, 0, u_{34})
 + 3 \chi^2 r''( \tilde u_{12}) (
s^{(31)}_{1100}( \tilde u_{113}, u_{34}) -
s^{(31)}_{1100}( \tilde u_{123}, u_{34}) )
\\
&& + 3 \chi {\sf R}''(u_{34})
s^{(31)}_{1001}( \tilde u_{123}, u_{34}) ) +
\frac{3}{2} {\sf R}''(u_{34}) s^{(4)}_{1100}( \tilde u_{1123}) \\
&&
- 5 \chi^2 \tilde u_{12}^2 r''(u_{13}) \psi^{(31) \prime}(u_{34})
- \chi \tilde u_{12}^2 \tilde u_{13} ( \psi^{(31) \prime \prime}(u_{34})
{\sf R}''(u_{34}) + 3 \psi^{(31) \prime}(u_{34}) {\sf R}'''(u_{34}) ) \\
&&
+ \frac{3}{2} \chi^2 \phi'(u_{34})  \tilde{u}_{13} r''( \tilde u_{12}) +
\frac{3}{4}  {\sf R}''(u_{34}) s^{(3)}_{200}(\tilde u_{123}) +
\frac{5}{4} \chi^2 r''( \tilde u_{12}) s^{(21)}_{(20)}(\tilde u_{13},u_{34}) \\
&&
+ 9 \chi^3 \tilde u_{12}^2 \tilde u_{13} \phi'(u_{34}) \phi''(u_{34}) +
3 \chi^2 \tilde u_{13} s^{(3)}_{110}(\tilde u_{112}) \phi'(u_{34}) \\
&&  -
\frac{3}{2} \chi^2 s^{(21)}_{20}(0,u_{34}) s^{(3)}_{200}(\tilde u_{123})
- 3 \chi^2 s^{(21)}_{20}(\tilde u_{12},u_{34})
( s^{(3)}_{110}(\tilde u_{123}) - s^{(3)}_{110}(\tilde u_{113}) ) \\
&& + \gamma ( 3 \chi^2 \tilde u_{13} r''(\tilde u_{12})^2
\phi'(u_{34}) + \tilde u_{1} \tilde u_{2} \tilde u_{3} (- 12 {\sf
R}''(u_{34}) {\sf R}'''(u_{34}) \phi''(u_{34}) - 18
({\sf R}'''(u_{34}))^2 \phi'(u_{34})) \\
&& - 3  \chi^2 r''(\tilde u_{12}) r''(\tilde u_{13})
s^{(21)}_{20}(0,u_{34}) + 3  \chi^2 r''(\tilde u_{13})^2
s^{(21)}_{(20)}(\tilde u_{12},u_{34}) \\
&& + 3 \chi^2 (2 r''(\tilde u_{12}) r''(\tilde u_{13})  - r''(\tilde u_{13}) r''(\tilde u_{23}) )
s^{(21)}_{(20)}(\tilde u_{12},u_{34}) \\
&& + 6  r''(\tilde u_{12}) {\sf R}''(u_{34}) (
s^{(3)}_{110}(\tilde u_{113}) - s^{(3)}_{110}(\tilde u_{123}) ) )
+ 3 \gamma \,{\rm sym} [ r''(\tilde{u}_{12})r''(\tilde{u}_{13})]
{\sf
  R}''(u_{34}) \\
&& + \gamma' \big( 9 \, {\rm sym} [
(r''(\tilde{u}_{12}))^2r''(\tilde{u}_{13})] {\sf
  R}''(u_{34}) - 3
r''(\tilde{u}_{12})r''(\tilde{u}_{13})r''(\tilde{u}_{23}) {\sf
  R}''(u_{34}) \\
&& + {\rm sym}( \tilde{u}_{13}\tilde{u}_{12} r''(\tilde{u}_{12})) \frac{{\sf
  R}''(u_{34}) ({\sf
  R}'''(u_{34}) )^2}{\chi^2} -3 \, {\rm sym}[ \tilde{u}_{12}\tilde{u}_{23}^2 ] \frac{{\sf
  R}''(u_{34}) ({\sf
  R}'''(u_{34}) )^3}{\chi^3} \big).
\end{eqnarray}

Upon taking the limit $\tilde u_{13}$ large the whole equation
should grow linearly in $\tilde u_{13}$ and we should obtain an
equation of the form:
\begin{eqnarray}
\tilde u_{13} [ \tilde u_{12}^2 f(u_{34}) + g(\tilde
u_{12},u_{34}) ] + |\tilde u_{13}| [ \tilde u_{12}^2 \tilde
f(u_{34}) + \tilde g(\tilde u_{12},u_{34}) ] = 0
\end{eqnarray}
The parts $f$ and $\tilde f$ must match the small $u_{13}$ limit
of the outer equation discussed previously (and the corresponding
non analytic piece should vanish). The parts $g$ and $\tilde g$
should match the small $\tilde u_{13}$ limit of PBL211. All the
obtained equations should be independent. We will not present
these equations in detail here.

One notes that again the equation for $s^{(31)}$ is linear. It has
a slightly different structure from the previous PBL ones, there
being an additional first derivative term w.r.t. the TBL
variables. This should not prevent it being formally solvable. The
inhomogeneous piece can be simplified using the equations
discussed above, the one obtained derived above for $\psi^{(31)}$,
and various other BL equations. We will not give further details
here.

\subsection{PBL 22}

Considering the ERG equation (\ref{ergS4}) for $S^{(4)}$ in the
regime where $u_1$ is close to $u_2$ and $u_3$ close to $u_4$ and
inserting the various PBL forms from (\ref{regimatch}) as
appropriate, we find at leading order $O(\tilde T_l^3)$:
\begin{eqnarray}
  \label{eppbl22}
  && 0 =
  r''({{\tilde{u}}_{34}})\,
  (s_{020}^{(22)}({{\tilde{u}}_{12}},{{\tilde{u}}_{34}},{u_{13}})-
  s_{020}^{(22)}({{\tilde{u}}_{12}},0,{u_{13}}))+
 r''({{\tilde{u}}_{12}})\,(
 s_{200}^{(22)}({{\tilde{u}}_{12}},{{\tilde{u}}_{34}},{u_{13}})
 -s_{200}^{(22)}(0,{{\tilde{u}}_{34}},{u_{13}}) ) \\
 &&  +
s_{020}^{(22)}({{\tilde{u}}_{12}},{{\tilde{u}}_{34}},{u_{13}})+
s_{200}^{(22)}({{\tilde{u}}_{12}},{{\tilde{u}}_{34}},{u_{13}})]
\end{eqnarray}
The result is rather simple as many terms do not contribute to
this order. The rescaling term gives zero to order $T^3$ because
there is no cubic polynomial allowed by symmetry. The feeding $T
RRR$ and $RRRR$ term also yields zero to this order. For the $T
RRR$ term one checks that the $r''(\tilde u_{12}) r''(\tilde
u_{34})$ term cannot appear and for the $RRRR$ term one checks
that although it appears there are powerful cancellations between
the various resulting terms. The other terms result from Taylor
expansions in $\tilde u_{12}$ and $\tilde u_{34}$ not to high
enough order to produce anything not antisymmetric (one needs at
least $\tilde u_{12}^2 \tilde u_{34}^2$ but this is higher order
in $T$).

We can solve this generally, remembering that the l.h.s. can be
gauge, suppressing the dependence on $u_{13}$ (which acts only as
a parameter). The left hand side can be taken as an arbitrary
function of $\tilde{u}_{12}$ plus the same function of
$\tilde{u}_{34}$ (since the right hand side is symmetric in the
two variables).  Doing this, one finds the general solution
\begin{eqnarray}
s_{02}^{(22)}({{\tilde{u}}_{12}},{{\tilde{u}}_{34}},{u_{13}}) -
s_{02}^{(22)}({{\tilde{u}}_{12}},0,{u_{13}})
= \frac{g(\tilde{u}_{12},\tilde{u}_{34})}{1+ r''(\tilde{u}_{34})} +
\frac{a(\tilde{u}_{12})+a(\tilde{u}_{34})}{1+ r''(\tilde{u}_{34})},
\end{eqnarray}
where $a(x)$ is arbitrary and $g(x,y)=-g(y,x)$ is antisymmetric but
otherwise arbitrary.  The most trivial but likely physical solution
corresponds to $a(x)=g(x,y)=0$:
\begin{eqnarray}
  s^{(22)}({{\tilde{u}}_{12}},{{\tilde{u}}_{34}},{u_{13}}) =
  \tilde{u}_{12}^2 \tilde{u}_{34}^2 g^{(22)}({u_{13}}).
\end{eqnarray}
This matches properly provided that $\tilde
\sigma(\tilde{u}_{12},u_{13})=0$, i.e. that $s^{211}$ is
sufficiently smooth. We have not checked whether the (here
neglected) fifth cumulant will produce some additional non trivial
feeding term to $s^{(22)}$. This does not however modify the
conclusion that $s^{(22)}$ can be obtained formally in terms of
higher PLB cumulant functions.

\section{beta function to higher orders in R}
\label{sec:higher-orders-}

The iteration to four loop order reads:
\begin{eqnarray}
&& S_0^{(5)}= \gamma_P \gamma_5 (R'' R'' R'' R'' R'') \\
&& S_1^{(4)}= \gamma_Q ((R'' S_1^{(4)\prime \prime}) + \gamma (R''
R'' S_1^{(3) \prime \prime}) + [S_0^{(5) \prime \prime}] ) \\
&& S_2^{(3)}= \gamma_S ( (R'' S_1^{(3) \prime \prime}) + [S_1^{(4)\prime \prime}] ) \\
&& \beta_2 = [S_2^{(3) \prime \prime}]
\end{eqnarray}

The only term needed not yet determined is the $R^5$ feeding term
into the fifth cumulant. The general expression for the feeding
term of the $p$-th cumulant is \cite{ergchauve}:
\begin{eqnarray}
&& c'_p p! (Tr W^p - \sum_{a a_1 .. a_p} \tilde R''_{a a_1}..
\tilde R''_{a
a_p}) \\
&& W_{ab} = \delta_{ab} \sum_c \tilde R''_{ac} - \tilde R''_{ab}
\end{eqnarray}
where $R_{ab}=R(u_{ab})$, the last term being the $p+1$-th replica
term which appear in the trace and must thus be subtracted. Here
$c'_p=c_p=1/(2p)$ for Wilson and $c'_p=1/2^p$ which comes from
rescaling for the ERG. As can also be extracted from Appendix A,
the feeding term into the fifth cumulant equation (not written)
reads:
\begin{eqnarray}
&& 12 \gamma_5 \text{sym} \big[ 5 {\sf R}''(u_{12}) {\sf
R}''(u_{13}){\sf R}''(u_{14}) ({\sf R}''(u_{12}) - {\sf
R}''(u_{23})) ({\sf R}''(u_{15}) + {\sf R}''(u_{25}))
\\
&& + (5 {\sf R}''(u_{14}) - {\sf R}''(u_{45})) {\sf
R}''(u_{15}){\sf R}''(u_{12}){\sf R}''(u_{23}){\sf R}''(u_{34})
\big] \nonumber
\end{eqnarray}
with $\gamma_5=1$ for Wilson and $\gamma_5=5/16$ for the ERG.

At each step of the calculation, upon applying the $[..]_{12}$ we
check carefully that no ambiguity arises. In general, upon
inserting the non analytic expansion for $R''(u_1-u_2)$ in powers
of $|u_1-u_2|$, a very large number of terms proportional to
$\text{sgn}(u_1-u_2)$ arise. They however amazingly cancel when
the limit $u_2 \to u_1$ is taken. In fact this serves as a useful
and non trivial check at each stage of the calculation.

The final result for the four loop beta function $\partial_l
\tilde R = \beta(R)$ is:

\begin{eqnarray}
&& \beta_{\text{4 loop}}(R)=  \\
&& c_{10} {\sf R}'''(u)^4 {\sf R}''''(u) + c_{11} {\sf R}'''(0+)^4
{\sf R}''''(u) + (c_{11}-c_{10}) {\sf R}'''(0+)^2 {\sf R}'''(u)^2
{\sf R}''''(0+) - (c_{10}+c_{11}) {\sf R}'''(0+)^2 {\sf R}'''(u)^2
{\sf R}''''(u) \nonumber \\
&& c_{20} {\sf R}''(u) {\sf R}'''(u)^2 {\sf R}''''(u)^2 + c_{21}
{\sf R}''(u) {\sf R}'''(0^+)^2 {\sf R}''''(0^+)^2 + c_{22} {\sf
R}''(u) {\sf R}'''(0^+)^2 {\sf R}''''(0^+) {\sf R}''''(u) \nonumber \\
&& - ( c_{20}+c_{21}+c_{22}) {\sf R}''(u) {\sf R}'''(0^+)^2 {\sf
R}''''(u)^2 \nonumber \\
&& + c_{30} {\sf R}''(u) {\sf R}'''(u)^3 {\sf R}'''''(u) + c_{31}
{\sf R}''(u) {\sf R}'''(0+)^2 {\sf R}'''(u){\sf R}'''''(u) -
(c_{30} + c_{31}) {\sf R}''(u) {\sf R}'''(0+)^3 {\sf R}'''''(0^+)
\nonumber \\
&& + c_{40} {\sf R}''(u)^2 {\sf R}''''(u)^3 + c_{50} {\sf
R}''(u)^2 {\sf R}'''(u)
 {\sf R}''''(u) {\sf R}'''''(u) + c_{60}
{\sf R}''(u)^2 {\sf R}'''(u)^2 {\sf R}''''''(u) + c_{61}
{\sf R}''(u)^2 {\sf R}'''(0+)^2 {\sf R}''''''(u) \nonumber \\
&& + c_{70} {\sf R}''(u)^3 {\sf R}'''''(u)^2 + c_{80} {\sf
R}''(u)^3 {\sf R}''''(u) {\sf R}''''''(u) + c_{90} {\sf R}''(u)^3
{\sf R}'''(u) {\sf R}'''''''(u) \nonumber
\end{eqnarray}
The various terms have been grouped together with their associated
anomalous (''counter'') terms. The coefficient $c_{ij}$ are found
to satisfy the constraint that the supercusp vanishes for each
group $i$, i.e. the associated anomalous terms successfully
substract the $|u|$ piece of the main term (indexed $j=0$). Terms
with even powers of $R''(u)$ do not need anomalous term,
nevertheless one notes the existence of the $c_{61}$ term. The
table of coefficients $c_{ij}$ is found to be:
\begin{eqnarray}
&& c_{10}=
372\,{{\gamma}_5}\,{{\gamma}_P}\,{{\gamma}_Q}\,{{\gamma}_S} +
354\,\gamma'\,{{{\gamma}_Q}}^2\,{{\gamma}_S} +
  248\,{\gamma}^2\,{{\gamma}_Q}\,{{{\gamma}_S}}^2 +
  156\,\gamma'\,{{\gamma}_Q}\,{{{\gamma}_S}}^2
  + 138\,\gamma\,{{{\gamma}_S}}^3 \\
  && c_{11}= -12\,{{\gamma}_5}\,{{\gamma}_P}\,{{\gamma}_Q}\,{{\gamma}_S}
  - 18\,\gamma'\,{{{\gamma}_Q}}^2\,{{\gamma}_S} -
  8\,{\gamma}^2\,{{\gamma}_Q}\,{{{\gamma}_S}}^2
  - 6\,\gamma'\,{{\gamma}_Q}\,{{{\gamma}_S}}^2
  - 12\,\gamma\,{{{\gamma}_S}}^3 \\
  && c_{20}= 672\,{{\gamma}_5}\,{{\gamma}_P}\,{{\gamma}_Q}\,{{\gamma}_S}
  + 664\,\gamma'\,{{{\gamma}_Q}}^2\,{{\gamma}_S} +
  488\,{\gamma}^2\,{{\gamma}_Q}\,{{{\gamma}_S}}^2
  + 364\,\gamma'\,{{\gamma}_Q}\,{{{\gamma}_S}}^2
  + 346\,\gamma\,{{{\gamma}_S}}^3 \\
  && c_{21}=-624\,{{\gamma}_5}\,{{\gamma}_P}\,{{\gamma}_Q}\,{{\gamma}_S}
  - 588\,\gamma'\,{{{\gamma}_Q}}^2\,{{\gamma}_S} -
  432\,{\gamma}^2\,{{\gamma}_Q}\,{{{\gamma}_S}}^2
  - 300\,\gamma'\,{{\gamma}_Q}\,{{{\gamma}_S}}^2
  - 264\,\gamma\,{{{\gamma}_S}}^3 \\
  && c_{22}=-24\,\gamma'\,{{{\gamma}_Q}}^2\,{{\gamma}_S}
  - 24\,\gamma'\,{{\gamma}_Q}\,{{{\gamma}_S}}^2
  - 32\,\gamma\,{{{\gamma}_S}}^3 \\
  && c_{30} = 192\,{{\gamma}_5}\,{{\gamma}_P}\,{{\gamma}_Q}\,{{\gamma}_S}
  + 220\,\gamma'\,{{{\gamma}_Q}}^2\,{{\gamma}_S} +
  192\,{\gamma}^2\,{{\gamma}_Q}\,{{{\gamma}_S}}^2
  + 148\,\gamma'\,{{\gamma}_Q}\,{{{\gamma}_S}}^2
  + 166\,\gamma\,{{{\gamma}_S}}^3 \\
  && c_{31}=-28\,\gamma'\,{{{\gamma}_Q}}^2\,{{\gamma}_S}
  - 32\,{\gamma}^2\,{{\gamma}_Q}\,{{{\gamma}_S}}^2 -
  28\,\gamma'\,{{\gamma}_Q}\,{{{\gamma}_S}}^2
  - 46\,\gamma\,{{{\gamma}_S}}^3 \\
  && c_{40} = 48\,{{\gamma}_5}\,{{\gamma}_P}\,{{\gamma}_Q}\,{{\gamma}_S}
  + 64\,\gamma'\,{{{\gamma}_Q}}^2\,{{\gamma}_S} +
  48\,{\gamma}^2\,{{\gamma}_Q}\,{{{\gamma}_S}}^2 + 52\,\gamma'\,{{\gamma}_Q}\,{{{\gamma}_S}}^2 + 52\,\gamma\,{{{\gamma}_S}}^3 \\
  && c_{50}= 96\,{{\gamma}_5}\,{{\gamma}_P}\,{{\gamma}_Q}\,{{\gamma}_S}
  + 184\,\gamma'\,{{{\gamma}_Q}}^2\,{{\gamma}_S} +
  144\,{\gamma}^2\,{{\gamma}_Q}\,{{{\gamma}_S}}^2 + 172\,\gamma'\,{{\gamma}_Q}\,{{{\gamma}_S}}^2 + 224\,\gamma\,{{{\gamma}_S}}^3 \\
  && c_{60}= 24\,\gamma'\,{{{\gamma}_Q}}^2\,{{\gamma}_S} + 32\,{\gamma}^2\,{{\gamma}_Q}\,{{{\gamma}_S}}^2 +
  24\,\gamma'\,{{\gamma}_Q}\,{{{\gamma}_S}}^2
  + 50\,\gamma\,{{{\gamma}_S}}^3 \\
  && c_{61}= -2 \gamma \gamma_S^3 \\
  && c_{70}= 4\,\gamma'\,{{{\gamma}_Q}}^2\,{{\gamma}_S} + 4\,\gamma'\,{{\gamma}_Q}\,{{{\gamma}_S}}^2
  + 12\,\gamma\,{{{\gamma}_S}}^3 \\
  && c_{80}= 4\,\gamma'\,{{{\gamma}_Q}}^2\,{{\gamma}_S} + 4\,\gamma'\,{{\gamma}_Q}\,{{{\gamma}_S}}^2
  + 16\,\gamma\,{{{\gamma}_S}}^3 \\
  && c_{90} = 4\,\gamma\,{{{\gamma}_S}}^3
\end{eqnarray}

We have also computed the correction to the fourth cumulant,
yielding:
\begin{eqnarray}
f_4= \frac{6}{d-4 \theta + 4 \zeta}  \chi^{-4} {\sf R}'''(0^+)^4 (
\gamma' + 28 \gamma' \gamma_Q {\sf R}''''(0^+) +
    16 ( 2 \gamma_5 \gamma_P + \gamma^2 \gamma_S ) {\sf R}''''(0^+)
    )
\end{eqnarray}
up to $O(R^6)$ terms (strictly for $\zeta=0$).

\section{Above four dimensions}
\label{sec:above-four-dimens}

In this section we consider the behavior for $d>4$, and in particular
the fate of the TBL in this case. Four dimensions plays to a certain
extent the role of an upper critical dimension.  In particular, for
$d>4$ and a weak and smooth pinning potential, the Larkin
approximation (of Taylor expanding the pinning potential to obtain a
random force) is self-consistent in the sense that the deformations
$\overline{\langle (u(r) - u(0))^2\rangle}$ remain bounded and small.
The self-consistency of this argument suggests that under these
conditions there is an (almost) unique local minimum energy
conformation of the elastic medium, and consequently no glassy
behavior.  (We include the proviso ``almost'' here to indicate that
the self-consistent Larkin argument, which considers only the variance
of the manifold displacement, may incorrectly miss rare
samples/regions with anomalously strong pinning, arising from the
tails of the disorder distribution.) Nevertheless, clearly for strong
enough pinning -- or a weak but non-smooth pinning potential -- there
{\sl must} be multiple metastable states for the manifold.

To analyze the nature of these low-energy states and their
consequences for the boundary layer, we consider both the FRG and a
complementary $1/d$ expansion approach, both of which support the same
qualitative behavior.  For a weak, smooth, pinning potential, both
approaches indicate that the effective action remains analytic at
$T=0$, and consequently that the TBL is absent.  Since this implies
the variance of the curvature of the effective potential,
$\overline{(V''(u))^2}$ is bounded (and indeed small -- see below),
this indicates the absence of metastability.  For stronger disorder,
the effective potential does become non-analytic at $T=0$, and there
is thus a boundary layer for $T>0$.  Unlike for $d<4$, however, the
width of this boundary layer is $O(T)$ rather than $O(T/L^\theta)$.
Our interpretation of this result is that the metastable states for
$d>4$ occur at {\sl small scales}, and hence between states with
bounded energy differences from the ground state.  Physically, even
for strong disorder the manifold remains asymptotically flat on large
scales (with a bounded displacement larger than the transverse
correlation length of the pinning potential), because it is too costly
in elastic energy to distort the manifold over large distances.  The
metastable states thus correspond to small local distortions of the
manifold atop this generally flat background.

\subsection{FRG for $d>4$}
\label{sec:FRGdgt4}

We consider the Wilson FRG.  First, note that the
distinction between strong and weak pinning can be observed from
very simple considerations at $T=0$.  In particular, from
(\ref{Wilson2cum}), and ignoring higher cumulants (e.g.
$S^{(3)}$) for the moment, one finds by differentiating:
\begin{equation}
  \partial_l R^{(4)}(0) = \epsilon R^{(4)}(0) + 3 \left[
    R^{(4)}(0)\right]^2 + \tilde{T}_l R^{(6)}(0)\label{eq:derivRG}.
\end{equation}
For $d>4$, $\epsilon<0$, and for $R^{(4)}(0)< |\epsilon|/3$ (at
low $T$) it flows to zero, as do all higher derivatives and
cumulants. Taking $\zeta=\epsilon/2 (<0)$, only the quadratic
Gaussian Larkin force term survives under renormalization.  This
is the weak pinning state.  From (\ref{eq:derivRG}), however,
for stronger disorder (and low $T$), one moves outside the domain
of stability, and the disorder correlator becomes non-analytic (at
$T=0$) at some finite scale $l=l_1$ as for $d<4$.

In this case for $l>l_1$ we must return to general functional
analysis.  The detailed behavior of the FRG for scales of order $l_1$
is quite complex, both inside and outside the boundary layer. However,
considering the behavior outside the boundary layer (i.e. at $T_l=0$),
it is clear from (\ref{Wilson2cum}) that the full force-force
correlator $\tilde\Delta_l(u)$ does not have a stable fixed point, and
that ultimately the flow of $\tilde\Delta_l$ will be toward zero.
Higher cumulants also flow toward zero, and moreover do so more
rapidly with scale than does the second cumulant.  Hence there will be
another scale $l_2>l_1$, beyond which the entire function $\tilde
R(u)$ will become small, so that the $O(R^2\sim \Delta^2)$ terms can
be neglected.  Once this occurs, there is no longer any feedback to
continue to narrow the boundary layer, which will have some width set
by its narrowness at $l=l_2$, presumably of $O(\tilde{T}_{l_2})$.  In
particular, in the linearized FRG equation for $\tilde\Delta_l$, the
$\tilde{T}_l\tilde\Delta''_l$ term can then be neglected (presuming as
usual $\tilde{T}_l \ll 1$), and one has
\begin{equation}
  \tilde\Delta_l(u) \sim e^{-|\epsilon|l}\overline{\Delta}(u) \qquad
  {\rm for}\, l>l_2.
\end{equation}
Here to avoid confusion we have not made any rescaling, i.e. used
$\zeta=0$.  In the same scheme it is appropriate to take
$\theta=d-2$.  This behavior obtains hence {\sl uniformly} for all
$u$, indicating that indeed there is no further sharpening of the
boundary layer.  The physical interpretation is that the important
metastable states occur at finite scales $L \sim \Lambda^{-1}
e^{l_2}$, with consequent finite energy separations.

\subsection{Large $d$ Limit}

Next we consider the large dimension limit $d\rightarrow\infty$.  The
FRG analysis sketched above suggests that the qualitative physics is
independent of dimension for $d>4$, so we may expect the large $d$
approach to recover similar behavior. The results to leading order are
identical to those for a fully-connected ``mean field'' model, so we
will present the two in tandem.  For the large dimension analysis, it
is convenient to consider the model on a $d$-dimensional hypercubic
lattice.
\begin{eqnarray}
  \label{eq:modellarged}
   \beta H[\{u_i\}] & = & \frac{\tilde{K}}{2d} \sum_{\langle ij \rangle}
    (u_i - u_j)^2 + \sum_i \tilde{V}_i(u_i),
\end{eqnarray}
where $\tilde{K}=K/T$ and $\tilde{V}_i(u_i) = V_i(u_i)/T$.
We consider in particular the free energy (effective action)
$\Gamma_0[u_0]$ for the uniform (zero) mode of the displacement field
$u_0$, {\sl in a given disorder realization}, defined according to
\begin{eqnarray}
  \Gamma_0[u_0,\lambda]  & = & - \ln \int [du_i] e^{
    - \beta H[\{u_i\}]  + \sum_i \lambda (u_i -
    u_0)}, \\
  \partial_\lambda \Gamma_0[u_0,\lambda] & = & 0. \label{eq:legendre}
\end{eqnarray}
Here $\lambda$ is a Lagrange multiplier used to define
$\Gamma_0[u_0]=\Gamma_0[u_0,\lambda(u_0)]$ in the usual way via the
condition in (\ref{eq:legendre}).  One may also readily calculate
a variety of other similar quantities at large $d$, e.g. the full
effective action for all coordinates in a particular disorder
realization, or the effective action for the replicated model.  The
present choice seems simplest and the most physically transparent.

The large-$d$ limit of
$\Gamma_0[u_0]$ can be calculated using the efficient method of
Georges and Yedidia,\cite{GY}\ which gives the result, to leading
(zeroth) order in $1/d$,
\begin{equation}
\Gamma_0[u_0]  =  \sum_i (g_i(\lambda) + \lambda u_0) -
\frac{\tilde{K}} {d} \sum_{\langle ij \rangle} g'_i(\lambda)
g'_j(\lambda), \label{eq:gamma0a}
\end{equation}
where $\lambda(u_0)$ is determined by the condition $\sum_i
(g'_i(\lambda)+u_0) = 0$ and
\begin{eqnarray}
  g_i(x) = -\ln \int du e^{ - \tilde{K} u^2 - \tilde{V}_i(u) + x
    u}.\label{eq:gdef}
\end{eqnarray}
is proportional to the free energy of the ``toy model'' with potential
$V_i(u)$.  To leading order in $1/d$ the last term in
(\ref{eq:gamma0a}) can be simplified to $\frac{1}{d}\sum_{\langle
  ij\rangle} g_i'(\lambda)g_j'(\lambda) \approx N u_0^2$, where $N$ is
the number of sites.  With this replacement, the result is identical
to the exact effective action for the fully connected mean-field model,
\begin{equation}
  \beta H[\{u_i\}] = \frac{\tilde{K}}{2 N} \sum_{i j}
  (u_i - u_j)^2 + \sum_i \tilde{V}_i(u_i).
\end{equation}
The effective action for this model is easily and exactly obtained by mean-field
theory.  One finds
\begin{equation}
 \Gamma_{FC,0}[u_0]= \Gamma_0[u_0] = \sum_i g_i(\lambda) + N \lambda
 u_0 -  N  \tilde{K}  u_0^2, \label{eq:fcgamma}
\end{equation}
with the same condition (\ref{eq:legendre}) determining $\lambda$
as above.  Using the fully-connected form in (\ref{eq:fcgamma}),
one can then generally show that the renormalized random force obeys
\begin{equation} \label{eq:lambdacondition}
  \frac{\Gamma_{FC,0}'(u_0)}{N\tilde{K}} = \tilde{\lambda} - 2u_0,\label{eq:gammafcpresult}
\end{equation}
where $\tilde{\lambda} = \lambda/\tilde{K}$.

We now use this result to evaluate the two-point function of the
``renormalized pinning potential'' defined by $\Gamma_0[u_0]$
(note that because we consider the uniform mode, there is no mean
elastic contribution to the energy for a displacement $u_0$).
For concreteness, we focus on the periodic CDW case, and take
$\tilde{V}_i(u_i)=V(u_i - \beta_i)/T$, with $\beta_i$ uniformly
distributed between $0$ and $2\pi$.

\subsubsection{Renormalized disorder correlator for weak pinning}

Consider first the case of a weak and smooth potential, (e.g.
$V(u)=a\cos u$ for $a\ll K$) with $|V(u)|,|V'(u)|\cdots \ll K$.  In
this case, we expect from the FRG analysis, that the renormalized
disorder will remain analytic at $T=0$, and hence that there will be
no TBL.  Thus we concentrate on zero temperature, for which
\begin{equation}
  g_i(\lambda) = \min_{u} \left[ \tilde{K}u^2 + \tilde{V}_i(u) -
    \lambda u\right].
\end{equation}
This can be evaluated perturbatively in $V_i(u)$ to give
\begin{equation}
  g_i(\lambda) \approx -\tilde{K}\frac{\tilde{\lambda}^2}{4}+
  \tilde{V}_i({\scriptstyle\frac{\tilde\lambda}{2}})-
  \frac{(\tilde{V}_i'({\scriptstyle\frac{\tilde\lambda}{2}}))^2}{4\tilde{K}^2}+O(\tilde{V}_i^3).
\end{equation}
Then (\ref{eq:lambdacondition}) gives perturbatively
\begin{equation}
  \tilde\lambda \approx 2u_0 + \frac{1}{N}\sum_i
  \frac{1}{\tilde{K}} \tilde{V}_i'(u_0) - \frac{1}{N}\sum_i
  \frac{1}{\tilde{K}^2} \tilde{V}_i'(u_0)\tilde{V}_i''(u_0) + \frac{1}{N^2}\sum_{ij}
  \frac{1}{2\tilde{K}^2} \tilde{V}_i'(u_0) \tilde{V}_j''(u_0) +
  O(\tilde{V}^3/\tilde{K}^3).\label{eq:lambdaperturb}
\end{equation}
It is straightforward to obtain the force-force correlator by
directly averaging using
Eqs.~(\ref{eq:gammafcpresult},\ref{eq:lambdaperturb}). One finds
\begin{equation}
  \overline{(\Gamma_0'(u_0) - \Gamma_0'(u_0'))^2 } = N \tilde{K}^2
  \left[ \Delta_1(u_0-u_0') + \Delta_2(u_0-u_0') +
    O(V^4/K^2)\right], \label{eq:ffsmooth}
\end{equation}
with
\begin{eqnarray}
  \Delta_1(u_0) & = & \int_0^{2\pi}\! \frac{d\beta}{2\pi}\,
  \frac{V'(\beta)V'(u_0+\beta)}{K^2}, \\
  \Delta_2(u_0) & = & \int_0^{2\pi}\! \frac{d\beta}{2\pi}\,
  \frac{[V'(\beta)]^2[V''(\beta+u_0)+ V''(\beta-u_0)]}{K^3}.
\end{eqnarray}
Provided the potential $V(u_0)$ is smooth (the existence of all
derivatives is a sufficient condition), the force-force correlator in
(\ref{eq:ffsmooth}) is analytic at zero temperature to all orders in
$V/K$.  Since it is analytic at zero temperature, there is a smooth
limit as $T\rightarrow 0$, without any boundary layer.  This can also
be explicitly verified by a direct calculation of $g_i(\lambda)$
perturbatively in $V$ from (\ref{eq:gdef}).  This describes the
weak-disorder behavior in large $d$.

\subsubsection{Renormalized disorder correlator for strong pinning}

For stronger disorder, or for a non-smooth pinning potential, the
above analysis does not obtain.  We now consider the behavior
explicitly for such a case, taking for concreteness the Villain potential:
\begin{eqnarray}
&&  e^{- \tilde{V}_i(u)} = \sum_n e^{- \frac{\tilde{a}}{2}  (u -
\beta_i - 2 \pi n)^2 },
\end{eqnarray}
with $\beta$ uniformly distributed between $0 < \beta < 2 \pi$.  In
the low temperature limit with $a=\tilde{a}/\tilde{K}$ fixed as
$T\rightarrow 0$, the single site ``free energy'' is then given
approximately by
\begin{eqnarray}
&&  e^{- g_i(\lambda)} \sim \sum_n e^{- \tilde{K}
g_n(\beta_i,\tilde{\lambda}) }, \\
&& g_n(\beta,\tilde{\lambda}) = \frac{a}{2} (\beta + 2 \pi n)^2 -
\frac{1}{2(a+2)} (\tilde{\lambda} + a  (\beta + 2 \pi n))^2.
\end{eqnarray}
Consider first the zero temperature limit, for which this simplifies
to
\begin{eqnarray}
  g_i(\lambda) = \min_n g_n(\beta_i,\tilde{\lambda}).
\end{eqnarray}
The minimum occurs for a given $n$ when $\tilde{\lambda}_{n-1,n}
< \tilde{\lambda} < \tilde{\lambda}_{n,n+1}$, where
\begin{eqnarray}
\tilde{\lambda}_{n,n+1} = 2 \beta + (2 n +1) 2 \pi
\end{eqnarray}
which is the point where $g_n=g_{n+1}$. The derivative of $g$ has a
jump at these values with:
\begin{eqnarray}
&& g_i'(\lambda,T=0) = \sum_n \frac{-1}{a+2} (\tilde{\lambda} + a
(\beta_i + 2 \pi n) ) \theta( 2 \beta_i + (2 n -1) 2 \pi <
\tilde{\lambda} < 2 \beta_i + (2 n +1) 2 \pi).
\end{eqnarray}
Hence $g'_i$ at $T=0$ consists of alternating linear segments with
slope $-1/(a+2)$ , and jumps of magnitude
$g'(\tilde{\lambda}_{n,n+1}+0^+) - g'(\tilde{\lambda}_{n,n+1}-0^+) = 2
\pi a/(a+2)$.  At low but non-zero temperature, the jumps in
$g'_i(\lambda)$ are rounded over a region of width $\delta \lambda
\sim \tilde{K}^{-1}=T/K$.

To obtain the second cumulant of the effective potential we use that
$\Gamma_0'(u_0) = \lambda$ which gives:
\begin{eqnarray}
&& \overline{(\Gamma_0'(u_0) - \Gamma_0'(u_0'))^2 } = N^2
\tilde{K}^2 \overline{(\tilde{\lambda}(u_0) -
\tilde{\lambda}(u_0') - 2 (u_0 - u_0'))^2 },
\end{eqnarray}
where as usual $\lambda$ is determined from
\begin{eqnarray}
&& u_0 = U(\lambda) = -  \frac{1}{N} \sum_i g_i'(\lambda_i).
\label{phi}
\end{eqnarray}
For simplicity (though it is not essential), we now study the limit
$a=\infty$ where $g'_i(\lambda)$ is flat and jumps exactly by $2 \pi$
at each shock.  Note that the infinite $a$ limit amounts to making the
model discrete, since each coordinate $u_i$ is then constrained to sit
exactly at one of the minima of the Villain potential,
$u_i=\beta_i+2\pi n_i$. This model is still non-trivial since $K$ is
still necessary to hold all particles near to one another (i.e. to
keep $n_i-n_{i'}$ small). The function $U(\lambda)$ for this case is
shown schematically in Fig. \ref{stairfig}
\begin{figure}
  \centerline{\fig{6cm}{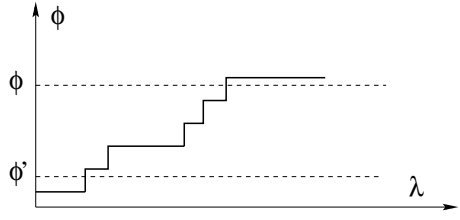}} \caption{The function
    $U(\lambda)$ defined in the text for the $N$-site $d=\infty$
    Villain model for the simplest limit $a=\infty$.  At zero
    temperature, it consists of $N$ ``shocks'' (step discontinuities)
    in each $4\pi$ interval of $\lambda$, each of magnitude $2\pi/N$.  At
    $T>0$, these steps are smoothed by an amount $\delta\lambda \sim
    \tilde{K}^{-1} =T/K$.
    \label{stairfig}}
\end{figure}
Graphically the solution to (\ref{phi}) is obtained by intersecting a
horizontal line with the function $u_0(\lambda)$ for a given disorder
realization. We consider first the behavior at $T=0$ and show that it
indeed exhibits the expected absolute value cusp in the force-force
correlation function.  $ $From the figure it is clear that (at $T=0$)
the solution to (\ref{phi}) is always such that $\lambda$ is equal one
of the shock values $\lambda = 2 \beta_i + (2 n_i +1) 2 \pi)$, with
$n_i$ an integer. Let us, for a fixed $N$, order the shocks by
increasing values of $\lambda$, i.e. $\lambda_m<\lambda_{m'}$ for
$m<m'$.

To obtain the two point statistics of the $\lambda(u_0)$ random
function we intersect with two lines $U(\lambda) = u_0$ and
$U(\lambda') = u_0'$. For given values of $u_0$ and $u_0'$ and a given
configuration of disorder, the solution is $U(\lambda_m) = u_0$ and
$U(\lambda_{m'}) = u_0'$ with the above-mentioned ordering. Clearly
from the figure one has, since $U(0)=\frac{1}{N}\sum_i \beta_i$,
\begin{eqnarray}
&&  m - m' = n = \big[ \frac{N u_0 - \sum_i \beta_i }{2 \pi}
\big] - \big[ \frac{N u_0' - \sum_i \beta_i }{2 \pi} \big],
\end{eqnarray}
where $[x]$ indicates the greatest integer less than or equal to $x$.
This involves the ``center of mass'' of the $\beta_i$ variables. On
the other hand we know that the $\lambda_m$ are of the form $2 \beta_i
+ (2 n_i +1) 2 \pi)$ for some $\beta_i, n_i$ but the differences
$\lambda_m - \lambda_{m'}=2\beta_i-2\beta_{i'}$ depend only on the
differences $\beta_i - \beta_{i'}$. Thus we can use that the
differences are statistically uncorrelated with the center of mass.
This is true because the distribution of the $\beta_i$ is uniform.
Thus we may consider for fixed $n=m-m'$ the distribution over the
differences.

This problem is formulated as follows. One chooses $N$
independent variables $\theta_i=2\beta_i$ on the interval $[0,4 \pi]$
with uniform distribution. Since the quantities of interest depend
only upon the differences, we may arbitrarily fix the first one at
zero, and order them as $\theta_0 = 0 <\theta_1<\theta_2<...<
\theta_{N-1} < \theta_N= 4 \pi$. The distribution of $\psi=\theta_n -
\theta_0$ is the object of interest.  Switching to difference
variables $\delta_k = \theta_k - \theta_{k-1}$, $k=1\ldots N$, this
probability reads
\begin{eqnarray}
  \label{eq:diffint}
  && P_n(\psi) = Z_{N} \int \prod_{i=1}^{N} d\delta_i \delta(
\sum_{i=1}^{N} \delta_i - 4 \pi)
\delta( \sum_{i=1}^{n} \delta_i - \psi),
\end{eqnarray}
where $Z_N$ is a normalization constant.  The first delta function here
expresses the constraint that the sum of the difference variables gives
$\theta_N-\theta_0=4\pi$.  Standard techniques (using a Fourier
representation of the delta functions) give the normalized result
\begin{eqnarray}
&& P_n(\psi) = \frac{1}{(4 \pi)^{N-1}} \frac{\Gamma(N)}{\Gamma(n)
\Gamma(N-n)} \psi^{n-1} (4 \pi -\psi)^{N-n-1}
\end{eqnarray}
Thus, finally, the probability distribution of
$\lambda_m-\lambda_{m'}$ is obtained by the sum
\begin{eqnarray}
&& P(\lambda(u_0) - \lambda(u_0') = \psi) = \sum_n P_n(\psi)
{\rm Prob}\Big\{ \big[ \frac{N u_0 - \sum_i \beta_i }{2 \pi} \big] - \big[
\frac{N u_0' - \sum_i \beta_i }{2 \pi} \big] = n\Big\}.
\end{eqnarray}
Clearly the fluctuations of $n$ are $O(1)$, while $n$ itself is
$O(N)$.  Hence, in the large system limit $N\gg 1$ the center of mass
disorder is negligible and we can take
\begin{eqnarray}
&&  x = n/N = (u_0 - u_0')/(2 \pi).
\end{eqnarray}
Then, using the Sterling formula, one has
\begin{eqnarray}
&& P(\psi) \sim \exp(2 \pi N ( u_0 \ln (\frac{\psi}{2 u_0}) +
(2 \pi - u_0) \ln(\frac{4 \pi - \psi}{2(2 \pi - u_0)}))
\end{eqnarray}
where the denominators have been chosen so that it is approximately
normalized, and we have taken $u'_0=0$ without loss of generality
since $P$ is translationally invariant.  Due to $N$ in the
exponential, it is strongly peaked. We write $\psi = 2 u_0 + f$ to
bring out the physical $f$, representing the fluctuation of the force
difference per site (in units of $\tilde K$). Then
\begin{eqnarray}
&& P_n(\psi) \sim \exp\left[2 \pi N \left( u_0 \ln (1 + \frac{f}{2 u_0}) +
(2 \pi - u_0) \ln(1 - \frac{f}{2(2 \pi - u_0)})\right)\right] \\
&& \sim \exp\left[ - N \left(\frac{f^2}{2 D} + A f^3 \right)\right],  \\
&& D(u_0) = \frac{1}{\pi^2} u_0 ( 2 \pi - u_0), \\
&& A(u_0) = - \frac{\pi^2}{3} \frac{\pi - u_0}{ u_0^2 ( 2 \pi -
u_0)^2 }.
\end{eqnarray}
The force correlator is finally obtained as:
\begin{eqnarray}
&& \overline{(\Gamma_0'(u_0) - \Gamma_0'(u_0'))^2 } = N^2
\tilde{K}^2 \overline{(\tilde{\lambda}(u_0) -
\tilde{\lambda}(u_0') - 2 (u_0 - u_0'))^2 } = \tilde{K}^2 N
u_0 ( 2 \pi - u_0)/\pi^2.
\end{eqnarray}
Indeed, as expected, this form has the linear cusp at small $u_0$.
Higher cumulants are readily extracted by further expansion around the
saddle point $f=0$, e.g.
\begin{eqnarray}
  && \overline{(\Gamma_0'(u_0) - \Gamma_0'(u_0'))^3 } = - 24 N
  \tilde{K}^3 A(u_0) D(u_0)^3.
\end{eqnarray}
This also exhibits the non-analytic behavior expected at $T=0$.

Turning finally to $T>0$, we note that shock contributing to
$U(\lambda)$ in Fig.~\ref{stairfig} is smoothed on the scale
$\delta\lambda \sim T/K$.  For large $N \gg T/K$ each shock will
overlap many others over this range.  Therefore the $T>0$ solution is
expected to be fully analytic in $u_0$, and rounded over a scale $\sim
T/K$.  Thus one finds agreement with the na\"ive FRG treatment for
$d>4$ of Sec.~\ref{sec:FRGdgt4}: the thermal boundary layer width does
not diverge with scale (in this case $N$ since we consider the center
of mass) but remains finite.

\section{Issues in extensions to $N>1$ or $d>0$}
\label{sec:diff-extend-erg}

\subsection{more than one component}
\label{sec:some-n1-calculation}

In this Appendix we attempt the calculation of the two loop
$\beta$-function for $d=0$ and $N>1$, by the direct method of
expansion in powers of $R$ of Section \ref{expansion-power-R} and
show that it fails.

In the $N$ component model the zero temperature exact RG equations
truncated to order $R^3$ read:
\begin{eqnarray}
&& - m \partial_m \tilde R(u) = (\epsilon - 4 \zeta  + \zeta u^i
\partial_{u^i} ) \tilde R(u) + \frac{1}{2} {\sf R}''_{ij}(u)^2
+ \partial_{u_1^i} \partial_{u_2^i} \tilde S(u_1, u_2, 0)|_{u_1=u_2=u} \\
&& - m \partial_m  \tilde S(u_1, u_2, u_3) = (-2 + 2 \epsilon - 6
\zeta  + \zeta u^i_a \partial_{u^i_a} ) \tilde S(u_1, u_2, u_3)
\\
&& + \gamma \big ( 3 \text{sym} [ {\sf R}''_{ij}(u_{12}) {\sf
R}''_{jk}(u_{12}) {\sf R}''_{ki}(u_{13}) ] - {\sf
R}''_{ij}(u_{12}) {\sf R}''_{jk}(u_{23}) {\sf R}''_{ki}(u_{31})
\big )
\end{eqnarray}
where summation over repeated indices is implicit, arguments are
now $N$ components vectors $u^i$, $i=1,..N$. Considering the model
with $O(N)$ symmetry, we define the functions:
\begin{eqnarray}
&& \tilde R(u) = N b(u^2/N) \\
&& h(x) = b(x^2)
\end{eqnarray}
for any $N$. To compute the beta function by the direct method
requires the calculation of the feedback of $S$ into $R$, denoted
as:
\begin{eqnarray}
&& \beta_{2 loop} = \gamma_S \gamma [ Tr (R'' R'' R'')'']
\end{eqnarray}
 where
we use the schematic notations defined in Section
\ref{expansion-power-R} and the $Tr$ denote the trace to be
performed in $O(N)$ space (and $\zeta$ is set to zero for now).

Performing the necessary contractions, and assuming that $h(x)$
has a well defined Taylor expansion near zero with $h'(x)=0$ we
find the fixed point equation for $h$:
\begin{eqnarray} \label{eq:betaN}
&& 0 = \epsilon h + \frac{1}{2 N} \bar h''(x)^2 + \frac{N-1}{2 N}
\frac{\bar{h}'(x)^2}{x^2} \\
&& + 4 \gamma( \frac{1}{2 N^2} \bar{h}''(x)^2 ( h'''(x)^2 -
h'''(0)^2) + \frac{(N-1)}{N^2} [ \frac{\bar{h}'(x)^3}{x^5} -
\frac{3}{2} \frac{\bar{h}'(x)^2 \bar{h}''(x)}{x^4} + \frac{1}{2}
\frac{\bar{h}''(x)^3}{x^2} ] \nonumber  \\
&& - \frac{(N-1)}{8 N^2} h'''(0)^2 [ (N+3) \frac{\bar h'(x)}{x} +
h''(x) ]) \nonumber
\end{eqnarray}
where $\bar h'(x)=h'(x)-h''(0) x$. The first line contains the
rescaling term and the one loop $O(h^2)$ terms, while the second
and third line is the result of the calculation to $O(h^3)$ of the
above-mentioned feedback. Extension for finite $\zeta$ using
convolutions as in (\ref{exactbeta0}) is straightforward. The
non-anomalous terms, present in an analytic theory, have the same
structure as obtained in previous similar field theory
calculations. The anomalous terms all contain $h'''(0)$
which plays the role of $\tilde R'''(0)$ for $N=1$. In fact, for
$N=1$, $h(x)$ identifies to $\tilde R(x)$ and one recovers the
beta function obtained in (\ref{exactbeta0}). One nice feature of
the calculation leading to the above is that when taking vectors
$u_1$ and $u_2$ close one can check explicitly that terms
containing $(\hat u_{12} \cdot \hat u_{13})^2$ vanish, thus the
result does not depend on how the points are brought together,
i.e. it is unambiguous.

However, the above is not an acceptable $\beta$-function. Indeed
one finds that the behaviour at small argument of the r.h.s. of
(\ref{eq:betaN}) is:
\begin{eqnarray}
&& - \frac{\gamma}{4} \frac{N^2-1}{N^2} h'''(0)^3 x + O(x^2)
\end{eqnarray}
This singularity (supercusp) occurring in the feeding term at two
loop is inconsistent with the working hypothesis (that the
behavior be similar to the one loop result which starts as $u^2$),
necessary for this direct $R$ expansion method to work. Thus, for
$N>1$ one of the assumptions leading to this expansion must break
down.

One also notes that a two loop term survives for $N=\infty$, which
is surprising in view of the results found in (). It reads: $ -
\frac{\gamma}{2} h'''(0)^2 \frac{\bar h'(x)}{x} = - \gamma
h'''(0)^2 b'(x^2)$. It is easy to trace it to the expansion of the
leading order term:
\begin{eqnarray}
&& S = 4 \gamma_S \gamma N b'(\frac{u_{ab}^2}{N})
b'(\frac{u_{ab}^2}{N})( b'(\frac{u_{ac}^2}{N}) +
b'(\frac{u_{bc}^2}{N}) ) + 0(1)
\end{eqnarray}
expanded to order $u_{ab}^2$ keeping only the leading order in
$N$.

The fact that the most naive extension of the non-analytic
expansion in power of $R$ fails here signals that the matching
problem must be reexamined for $N>1$. Work is in progress in that
direction.

\subsection{higher dimensions}

It is instructive to consider how the naive one loop zero
temperature FRG is encapsulated in the ERG.  Since it requires
keeping only terms up to second order in $R(u)$, it is
straightforward to write the corresponding multilocal terms used
in this formalism. We require up to bilocal order for the second
cumulant, which is a term in the effective action of the form
\begin{eqnarray}
&& V^{(2)}_l = \frac{1}{2 T^2} \int_{x_1,x_2} \sum_{a b}
R(u_{x_1}^{ab},u_{x_2}^{ab},x_{12}) .
\end{eqnarray}
Naively truncating to this order, the rescaled local and bilocal
disorder correlators obey the equations
\begin{eqnarray}
&& \partial_l \tilde R(u) = (d - 2 \theta + \zeta u \partial_u ) \tilde R(u) \label{eq:localERG}\\
&& + \tilde T \partial g(x=0) \tilde R''(u) + 2 \tilde T
\int_x \partial g(x) \tilde R_{11}(u,u,x)
+ \frac{1}{2} (\int_x \partial g(x) g(x)) (\tilde R''(u)^2
- 2 \tilde R''(0) \tilde R''(u))\nonumber
\end{eqnarray}
and
\begin{eqnarray}
&& \partial_l \tilde R(u_1,u_2,x) =
(2 d - 2 \theta + \zeta u_i \partial_{u_i} + x_i \partial_{x_i})
\tilde R(u_1,u_2,x) + F(u_1,u_2,x) - \delta^d(x)
\int_x F(u_1,u_2,x) \label{eq:bilocalERG}\\
&& F(u_1,u_2,x) =\tilde T \partial g(x=0) ( \tilde R_{20}(u_1,u_2,x)
+ \tilde R_{02}(u_1,u_2,x) )
+ 2 \tilde T \partial g(x) \tilde R_{11}(u_1,u_2,x)  \nonumber \\
&& +  \frac{1}{2} \partial g(x) g(x) {\sf R}''(u_1) {\sf
R}''(u_2) .
\end{eqnarray}
We use the same rescaling factor $1/4$ as in the text for second
cumulant (and an extra $m^d$ for the bilocal term). The projector
in \ref{eq:bilocalERG} extracts the purely local part. Here the
Fourier transforms of $g(x),\partial g(x)$ are $ g(p) =
\frac{1}{p^2 + 1}$, $\partial g(p) = \frac{1}{(p^2 + 1)^2}$. Note
that if $\tilde{T}$ is taken to zero in (\ref{eq:localERG}), it
becomes exactly the naive one loop FRG equation of the Wilson
approach for $\tilde R(u)$, with $\int_x
\partial g(x) g(x) = 1/(32\pi^2)$ in $d=4$ (yielding the usual
universal number appearing at one loop). From $\tilde R(u)$ one
obtains, at least naively, the $q=0$ correlations to lowest order
in $\epsilon$.

Attempts to ``improve'' on this calculation however lead to
difficulties.  These occur already in considering the two point
correlation function at non-zero momentum, which is determined
from $\tilde R_{11}(0,0,q)$.  Naively solving
(\ref{eq:bilocalERG}) in the zero temperature limit gives the
outer solution:
\begin{eqnarray}
&& \tilde R_{outer}(u_1,u_2,q) = \phi(q) {\sf R}''(u_1) \tilde
{\sf R}''(u_2) ,
\end{eqnarray}
which, we know from considerations in the present paper can at
best hold only for $u_i \sim O(1)$. The momentum integral is found
to be:
\begin{eqnarray}
&& \phi(q)  = \int_0^\infty dl' e^{\epsilon l'} (h(q e^{- l'}) - h(0)),
\end{eqnarray}
and $h(q) = \frac{1}{2} \int_k \frac{1}{(k^2 +1)^2} \frac{1}{(k+q)^2
  +1}$.

Taking two derivatives gives $\tilde R_{11}(u_1,u_2,q) = \phi(q)
\tilde R'''(u_1) \tilde R'''(u_2)$.  Unlike similar local $(\tilde
R'''(u))^2$ terms, this does not have an unambiguous limit for
small $u_1,u_2$.  This is indeed a sign of difficulty for the
matching.  In particular, the naive Larkin term in the bilocal
TBL,  $\tilde R_{TBL}(u_1,u_2,q)\sim \tilde{T}_l^2 f_{2,2}(q)
\tilde u_1\tilde u_2$, does not match the non-analytic outer
solution at small argument $\tilde R_{outer}(u_1,u_2,q) \sim
\phi(q) (R'''(0^+))^2 |u_1||u_2|$. Thus the matching problem is
more complex in this case and requires further analysis. Without
such a prescription, the zero temperature beta function becomes
itself ambiguous at $O(R^3)$, since at this order there is a
feeding term in the local equation of the form $\sim {\sf
R}''(u){\sf R}_{11}(u,u,\cdot)$, an expression which naively
starts as $|u|$ at small $u$, which seems problematic. Work is in
progress to tackle these issues.

\end{widetext}


\begin{thebibliography}{10}
\bibitem{footnote2}
the way it is connected is itself defined by the
connected replica average. We do not need to specify it
here, and detailed examples are studied in Section.
\bibitem{footnotehat}
it is also useful to define a more symmetric probability
distribution $\hat P(x_1,x_2,\epsilon=\epsilon_2-\epsilon_1)$ that
the two lowest minima are located at $x_1$,$x_2$ with energy
difference $\epsilon$ of arbitrary sign, with $\hat
P(x_1,x_2,\epsilon)=\theta(\epsilon) P(x_1,x_2,\epsilon) +
\theta(-\epsilon) P(x_2,x_1,- \epsilon)$ being normalized to $2$.
The RG equation (\ref{eq:RGpdf}) is established, strictly, for
$\hat P$ for any $\epsilon$, but clearly also applies to $P$ for
any $\epsilon
>0$. The symmetry of $D(x_1,x_2)=P(x_1,x_2,\epsilon=0^+)$ then
follows if $\hat P$ is continuous (which is implied by the RG
equation for $x_1 \neq x_2$).
\bibitem{f4} the condition exactly for $d=4$ becomes $\tilde{T}_l
  \ll l^2$ and all other considerations are very similar
\bibitem{f3}
Although as an extra precaution one may want to
  distinguish $T \to 0^+$ (studied here, $lim_{T \to 0} lim_{l \to
    \infty}$) and $T=0$ (absolute minimum of the energy $lim_{l \to
    \infty} lim_{T \to 0}$), we will denote $T=0$ everywhere below for
  simplicity (also, at this stage we do not see any reason why the two
  should differ).
\bibitem{f2}
in the simplest version of the droplet theory one assumes that $p$
is uncorrelated from $x_1,x_2$. This restriction can be lifted
although it would change the amplitudes
\end{thebibliography}

\end{document}